\pdfoutput=1
\documentclass[journal=enfuem,manuscript=article,hyperref]{achemso}

\usepackage{hyperref}

\usepackage{achemso}
\usepackage[version=3]{mhchem} 
\usepackage{latexsym,amsmath,amssymb}
\usepackage{bm}

\usepackage{booktabs,multicol,multirow}

\usepackage{graphicx}
\usepackage{caption}
\usepackage{subcaption}

\usepackage{url,doi}

\usepackage{mathtools}
\usepackage{exscale, relsize}
\usepackage[retainorgcmds]{IEEEtrantools}

\usepackage{nicefrac}

\usepackage{listings}

\usepackage{cancel}
\usepackage[super]{nth}

\usepackage{chemfig}

\usepackage[T1]{fontenc}
\usepackage[english]{babel}
\usepackage{csquotes}
\usepackage{textcomp}

\usepackage{siunitx}
\DeclareSIUnit\atm{atm}
\DeclareSIUnit\rpm{rpm}
\DeclareSIUnit{\calorie}{cal}
\DeclareSIUnit{\Calorie}{\kilo\calorie}

\graphicspath{{./figures/}}

\usepackage[textsize=footnotesize,textwidth=2cm]{todonotes}

\author{Xin Hui}
\email{huixin@buaa.edu.cn}
\affiliation[Beihang University]{School of Energy and Power Engineering, Beihang University, Beijing, China}

\author{Kyle E.\ Niemeyer}
\affiliation[Oregon State University]
{School of Mechanical, Industrial, and Manufacturing Engineering, Oregon State University, Corvallis, OR, USA}

\author{Kyle B.\ Brady}
\altaffiliation{Innovative Scientific Solutions Inc., Dayton, OH, USA}
\author{Chih-Jen Sung}
\affiliation[University of Connecticut]
{Department of Mechanical Engineering, University of Connecticut, Storrs, CT, USA}

\title[]
{Reduced chemistry for butanol isomers at engine-relevant conditions}

\begin{document}

\begin{abstract}
\noindent
Butanol has received significant research attention as the second-generation biofuel in the past few years.
In the present study, skeletal mechanisms for four butanol isomers were generated from two widely accepted, well-validated detailed chemical kinetic models for the butanol isomers.
The detailed models were reduced using a two-stage approach consisting of the directed relation graph with error propagation and sensitivity analysis.
During the reduction process, issues encountered with pressure-dependent reactions formulated using the logarithmic pressure interpolation approach were discussed, with recommendations made to avoid ambiguity in its future implementation in mechanism development.
The performances of the skeletal mechanisms generated here were compared with those of detailed mechanisms in simulations of autoignition delay times, laminar flame speeds, and perfectly stirred reactor temperature response curves and extinction residence times, over a wide range of pressures, temperatures, and equivalence ratios.
Good agreement was observed between the detailed and skeletal mechanisms, demonstrating the adequacy of the resulting reduced chemistry for all the butanol isomers in predicting global combustion phenomena.
The skeletal mechanisms also closely predicted the time-histories of fuel mass fractions in homogeneous compression-ignition engine simulations.
Finally, the performances of the butanol isomers were compared with that of a gasoline surrogate with an anti-knock index of 87 in a homogeneous compression-ignition engine simulation.
The gasoline surrogate consumed faster than any of the butanol isomers, and \textit{tert}-butanol had the slowest fuel consumption rate; \textit{n}-butanol and isobutanol came closest to matching the gasoline, but the two literature chemical kinetic models predicted different orderings.
\end{abstract}

\section{Introduction}
\label{sec:intro}

Interest in renewable energy has grown significantly in the last decade, driven primarily by unstable oil prices and the environmental costs associated with fossil fuels.
Alcohol biofuels, renewable fuels produced from biological sources, have attracted significant research interest because they may offer significant benefits in terms of reduced emissions, lowered lifecycle carbon footprint, improved price stability, and more distributed production facilities.
Moreover, as a result of the reduced chemical complexity---relative to petroleum distillates---and the accompanying reduction in fuel variability associated with alcohol biofuels, detailed modeling of complete combustion systems becomes significantly more tractable, aiding the development of novel engine designs.
However, while novel engine designs and alternative fuels promise improved efficiencies and better emissions performance, to a great extent their success depends on a comprehensive understanding of fuel kinetics.
Robust chemical kinetic models are therefore needed that can provide accurate and efficient predictions of combustion performance across a wide range of engine relevant conditions.

\subsection{Butanol}

Butanol has many advantages as a biofuel over ethanol, including a higher heating value, reduced corrosiveness and susceptibility to water contamination, better engine performance, and a wider range of feedstocks~\cite{Nigam:2011aa,Bergthorson:2015dg}.
For these reasons, butanol is under consideration to replace ethanol as an alternative fuel to gasoline and diesel.
Extensive efforts have been made in recent years to study the combustion performance of the four butanol isomers in various well-defined fundamental and engine experiments.
Numerous fundamental studies have investigated a variety of combustion characteristics of the butanol isomers, including homogeneous autoignition delays~\cite{Heufer:2011jh,Weber:2011fv,Stranic:2012jl,Weber:2013hs,Moss:2008bva}, counterflow ignition temperatures~\cite{Brady:2016dw,Brady:2015fk,Liu2011995}, laminar flame speeds~\cite{Liu2011995,Sarathy:2009js,Broustail:2011ez,Gu:2010bo,Veloo:2011fr,Wu:2013et}, flame extinction limits~\cite{Agathou:2012ey,Lefkowitz:2012do}, and species evolutions~\cite{Sarathy:2009js,Lefkowitz:2012do}.
Sarathy et al.~\cite{Sarathy:2014iq} reviewed fundamental experimental studies of alcohols, including the butanol isomers, and both Bergthorson and Thomson~\cite{Bergthorson:2015dg} and No~\cite{No:2016cq} reviewed the characteristics of butanol combustion in engines.

\begin{figure}[htbp]
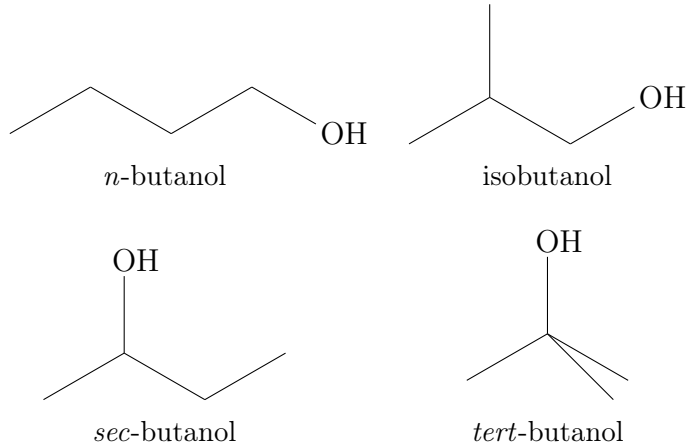

   \centering
   \begin{subfigure}[b]{0.25\textwidth}
    \centering
        \chemfig{[:30]--[:-30]--[:-30]OH}
        \caption*{\textit{n}-butanol}
   \end{subfigure}
   \quad\quad
   \begin{subfigure}[b]{0.25\textwidth}
   \centering
        \chemfig{-[:30](-[:90])-[:-30]-[:30]OH}
        \caption*{isobutanol}
   \end{subfigure}
   \\ \vspace{1em}
   \begin{subfigure}[b]{0.25\textwidth}
   \centering
        \chemfig{-[:30](-[:90]OH)-[:-30]-[:30]}
        \caption*{\textit{sec}-butanol}
   \end{subfigure}
   \quad\quad
   \begin{subfigure}[b]{0.25\textwidth}
   \centering
        \chemfig{-[:30](-[:90]OH)(-[:-45])-[:-30]}
        \caption*{\textit{tert}-butanol}
   \end{subfigure}
   %
   \caption{Molecular structure of butanol isomers}
   \label{fig:butanol_isomers}
\end{figure}

Several of these studies compared combustion properties among the four butanol isomers and found differences in ignition propensity~\cite{Stranic:2012jl,Weber:2013hs,Moss:2008bva,Brady:2016dw} and flame propagation~\cite{Gu:2010bo,Veloo:2011fr,Wu:2013et}, demonstrating that the four isomers exhibit differing reactivities as a result of their different molecular structures, which Fig.~\ref{fig:butanol_isomers} depicts.
In addition, other studies focused on the combustion characteristics of \textit{n}-butanol in terms of ignition temperature~\cite{Brady:2015fk,Liu2011995}, flame propagation~\cite{Liu2011995,Sarathy:2009js,Broustail:2011ez}, and speciation~\cite{Sarathy:2009js}.
Lefkowitz et al.~\cite{Lefkowitz:2012do} explored flame extinction and speciation to characterize the global reactivity of \textit{tert}-butanol due to its distinctive characteristics compared with the other three butanol isomers.
The performances of butanol blends with gasoline or diesel have also been assessed in spark ignition (SI) engines~\cite{Szwaja:2010db,Gu:2012gh}, compression ignition (CI) engines~\cite{AlHasan:2008cm,Yao:2010es,Rakopoulos:2011cp}, and advanced homogeneous charge compression ignition (HCCI) engines~\cite{Saisirirat:2011ci,He:2013bo} in terms of exhaust temperature, thermal efficiency, autoignition timing, and emissions.
Aside from the work of Al-Hasan and Al-Momany~\cite{AlHasan:2008cm} on isobutanol, the majority of these studies focused on blends of \textit{n}-butanol with conventional petroleum-derived fuels at various blending ratios.
Recently, He et al.~\cite{He:2014gl} also evaluated the behavior of pure \textit{n}-butanol in an HCCI engine in terms of combustion performance and emissions at engine speeds of \SIlist{1200;1500}{rpm}.

In addition to this experimental work, much progress has been made in developing kinetic models that describe the chemical kinetics of the butanol isomers; see Sarathy et al.~\cite{Sarathy:2014iq} for a detailed review of these efforts.
Moss et al.~\cite{Moss:2008bva} developed the first detailed chemical kinetic model for four butanol isomers based on the high temperature ignition delay data measured in a shock tube.
Later, Dagaut et al.~\cite{Dagaut:2009kj} proposed a detailed chemical kinetic model for \textit{n}-butanol based on jet stirred reactor data.
Sarathy et al.~\cite{Sarathy:2009js} then improved this model based on laminar flame speed and non-premixed flame speciation data, developing a detailed model for \textit{n}-butanol consisting of 118 species and 878 reactions.
More recently, Sarathy et al.~\cite{Sarathy:2012fj} proposed a detailed chemical kinetic model for \textit{n}-, iso, \textit{sec}-, and \textit{tert}-butanols.
The model describes low- and high-temperature oxidation of butanol isomers with 426 species and 2335 reactions, and Sarathy et al.~\cite{Sarathy:2012fj} validated its performance with atmospheric pressure laminar flame speeds, low-pressure flame species profiles, intermediate-temperature shock tube ignition delay times for \textit{n}-butanol, high-temperature shock tube ignition delay times for all four isomers, rapid-compression machine ignition delay times for all four isomers at low-to-intermediate temperatures, and jet-stirred reactor species profiles for selected isomers.
Vasu and Sarathy~\cite{Vasu:2013jc} updated the model of Sarathy et al.~\cite{Sarathy:2012fj} in the high-temperature regime for \textit{n}-butanol oxidation using rate constants measured in a single-pulse shock tube~\cite{RosadoReyes:2012fl}.
The updated high-temperature model of Vasu and Sarathy~\cite{Vasu:2013jc} consists of 284 species and 1892 reactions, and agrees better with experimental \textit{n}-butanol data compared with the original model of Sarathy et al.~\cite{Sarathy:2012fj}.

Frassoldati et al.~\cite{Frassoldati:2012jn} also proposed a detailed chemical kinetic model for all four butanol isomers, with 317 species and \num{12353} reactions.
This model built on the work of Grana et al.~\cite{Grana:2010gk} and emphasizes the major decomposition and oxidation pathways of each respective butanol isomer in the high-temperature, low-pressure combustion regime.
Frassoldati et al.~\cite{Frassoldati:2012jn} validated their detailed model using data from pyrolysis, shock-tube autoignition, and both premixed and non-premixed flame experiments.

Van Geem et al.~\cite{VanGeem:2010ca} proposed a detailed chemical kinetic model for \textit{n}-, \textit{sec}-, and \textit{tert}-butanol, and validated it using shock-tube ignition delay times, pyrolysis data, and flame species profiles.
Later, Merchant et al.~\cite{Merchant:2013kz} studied the pyrolysis and combustion of isobutanol and added its chemical pathways to the model of Van Geem et al.~\cite{VanGeem:2010ca}.
The combined detailed model of Merchant et al.~\cite{Merchant:2013kz} for all four butanol isomers consists of 372 species and 8723 reactions.

\subsection{Need for model reduction}

While the predictions of the aforementioned butanol models largely match fundamental experimental validation data, their large sizes pose a significant challenge to their implementation in practical engine simulations.
The computational cost of detailed chemistry in such simulations scales cubically with the number of species (in the worst case)~\cite{Lu:2009gh}.
The resulting large number of governing differential equations for the set of species are nonlinearly coupled, and exhibit vastly different time scales that render them mathematically stiff~\cite{Lu:2009gh}.
The large size and chemical stiffness of detailed chemical kinetic models therefore limit their application to large-scale simulations, and as such there is a growing need to reduce the size of these models while retaining their predictive capabilities.

A large number of model reduction methodologies have been developed in the past decade to counter the trend of increased numbers of species and reactions, as reviewed by Lu and Law~\cite{Lu:2009gh} and more recently by Tur\'{a}nyi and Tomlin~\cite{Turanyi:2014aa}.
Among the various methods, the direct relation graph (DRG) approach of Lu and Law~\cite{Lu:2005ce,Lu:2006bb,Lu:2006gi,Lu:2008bi} has received significant attention due to its effectiveness and efficiency.
The DRG method built on the earlier approach of Bendtsen et al.~\cite{Bendtsen:2001vh} for representing reaction pathways with weighted directed graphs, but instead uses the graph to quantify the importance of species.
The weights of graph edges represent species interaction coefficients, which estimate the error induced in the overall production rate of one species by the removal of the other, and are determined with normalized contributions to the overall production rates.
Low-valued graph edges then indicate unimportant relationships between species, and are trimmed; the DRG method produces a skeletal mechanism after trimming the graph in this way then identifying which species remain connected to certain important target species.

DRG is often used as the first step of a multistage reduction to quickly reduce a large detailed mechanism to a smaller skeletal mechanism.
However, DRG only considers direct interactions between species and thus assumes the worst-case scenario for error propagation, and can generate non-minimal skeletal mechanisms.
Pepiot-Desjardins and Pitsch~\cite{Pepiot-Desjardins:2008} proposed a more aggressive treatment that considers error propagation along the graph pathways: DRG with error propagation (DRGEP).
The DRGEP method can generate smaller mechanisms compared with DRG alone~\cite{Pepiot-Desjardins:2008,Niemeyer:2010bt}, while maintaining the fidelity of the resulting skeletal mechanism to the original detailed description.
Such skeletal mechanisms can be further reduced by various techniques such as sensitivity analysis (SA)~\cite{Niemeyer:2010bt,Niemeyer:2014,Niemeyer:2015wq} and path flux analysis~\cite{Sun:2010jf}.
In addition, techniques based on time-scale analysis can be used to both effectively decrease mechanism size and reduce stiffness, such as invoking quasi-steady state approximations~\cite{Bodenstein:1913tc,Chapman:1913dx} and\slash or using computational singular perturbation methods~\cite{Lam:1988wc,Lam:1993ub,Lam:1994ws,Lu:2001ve}.

Given the potential application of butanol isomers in internal combustion engines, it is necessary to reduce butanol mechanism sizes for use in realistic engine simulations---without sacrificing chemical fidelity relative to the underlying detailed mechanism.
Therefore, our first objective in this paper is to generate skeletal mechanisms for butanol isomers from available comprehensive models based on the DRGEPSA method, as described by Niemeyer et al.~\cite{Niemeyer:2010bt,Niemeyer:2011fe,Niemeyer:2014,Niemeyer:2015wq}.
We then compare intermediate and final skeletal mechanisms, resulting from the DRGEP and DRGEPSA (i.e., DRGEP followed by SA) methods, respectively, with their parent detailed mechanisms in terms of ignition delay times, laminar flame speeds, and perfectly stirred reactor (PSR) temperature response curves and extinction turning points, to assess the validity of the reduction methodology.
In addition, we investigate the potential of replacing gasoline with butanol by comparing the engine performance of the butanol isomers and a gasoline surrogate.
Finally, we discuss our overall conclusions and contributions of the paper.

\section{Methodology}
\label{S:method}

In this section, we describe the baseline chemical kinetic models, the reduction procedure, and validation of the resulting skeletal mechanisms.
The reduction procedure used here is based on the approach of Niemeyer and coworkers~\cite{Niemeyer:2010bt,Niemeyer:2011fe,Niemeyer:2014,Niemeyer:2015wq}, and a complete description can be found in those works.
However, we provide an overview of the method here for completeness, and discuss a few notable changes.

\subsection{Chemical kinetic models}
\label{sec:mechanisms}

We selected two detailed chemical kinetic models from the literature to generate skeletal mechanisms for the four butanol isomers: the mechanisms of Sarathy and coworkers~\cite{Sarathy:2012fj,Vasu:2013jc} and Merchant et al.~\cite{Merchant:2013kz}, hereafter referred to as the Sarathy mechanism (284 species, 1892 reactions) and Merchant mechanism (372 species, 8322 reactions), respectively. Note that one reaction has been removed from the original Merchant mechanism due to an issue of negative rate coefficient in a pressure-dependent reaction, which will be further explained in due course.
Both mechanisms include chemical pathways for all four butanol isomers: \textit{n}-, iso, \textit{sec}-, and \textit{tert}-butanols.
As discussed previously, Frassoldati et al.~\cite{Frassoldati:2012jn} developed another mechanism for all four butanol isomers.
However, this model contains reactions with non-integer stoichiometric coefficients that make it incompatible with the current CHEMKIN-based reduction code, and thus it is not included in the present study.

\begin{figure}[htbp]
   \centering
   \begin{subfigure}[b]{0.48\textwidth}
        \includegraphics[width=\textwidth]{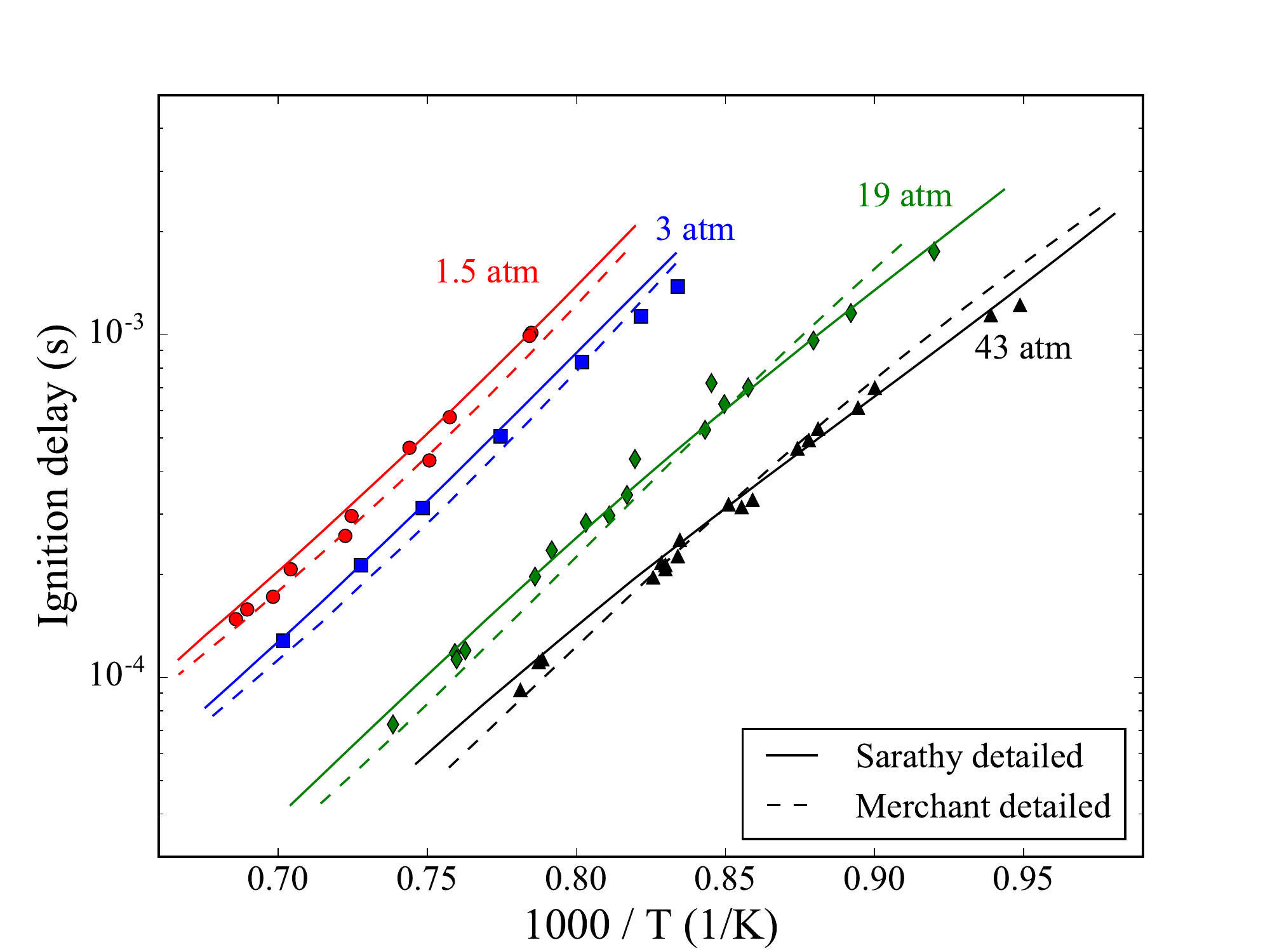}
        \caption{\textit{n}-butanol}
        \label{fig:nbutanol_ign_exp_comp}
    \end{subfigure}
    ~
    \begin{subfigure}[b]{0.48\textwidth}
        \includegraphics[width=\textwidth]{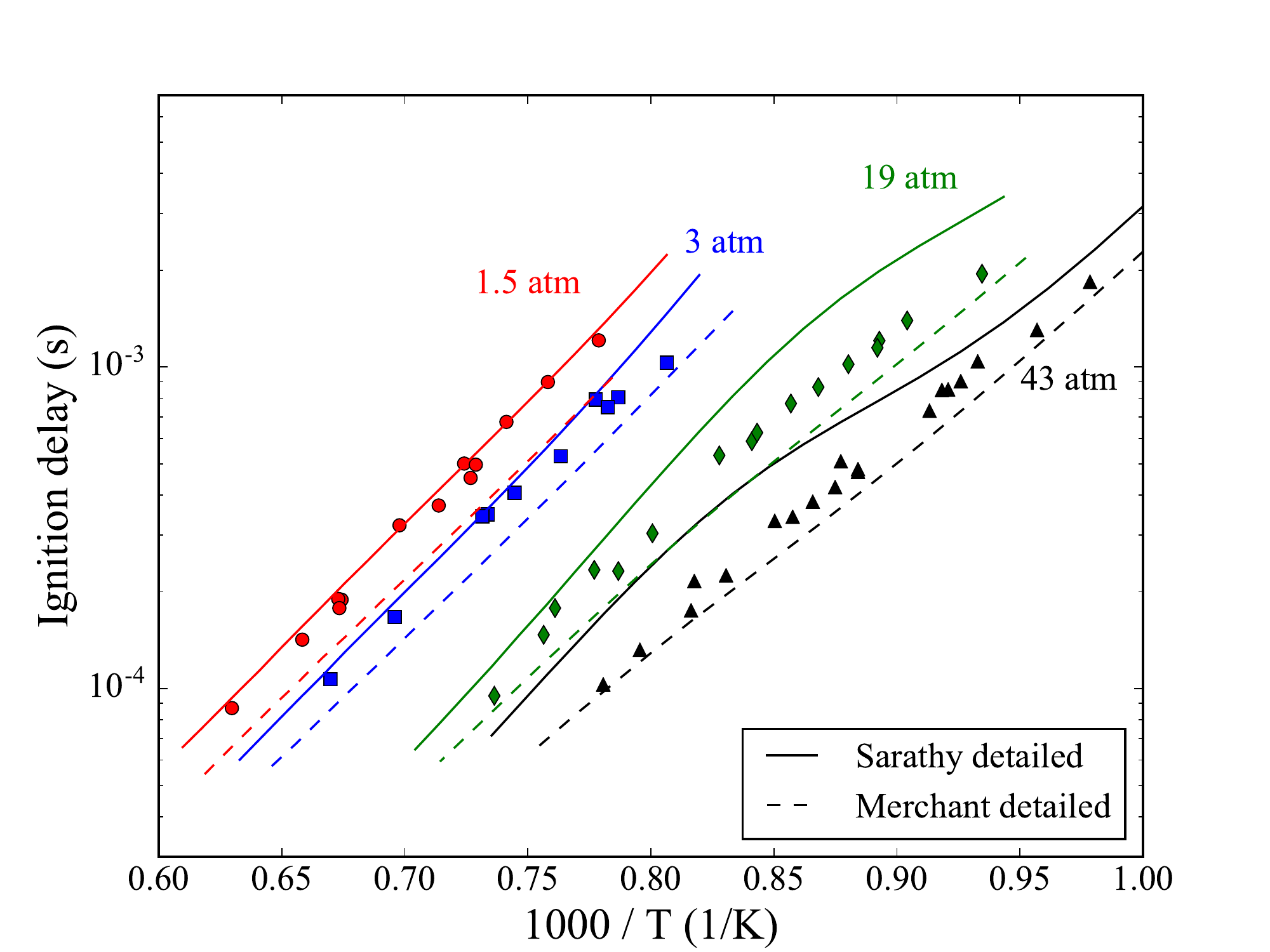}
        \caption{isobutanol}
        \label{fig:isobutanol_ign_exp_comp}
    \end{subfigure}
    \\
    \begin{subfigure}[b]{0.48\textwidth}
        \includegraphics[width=\textwidth]{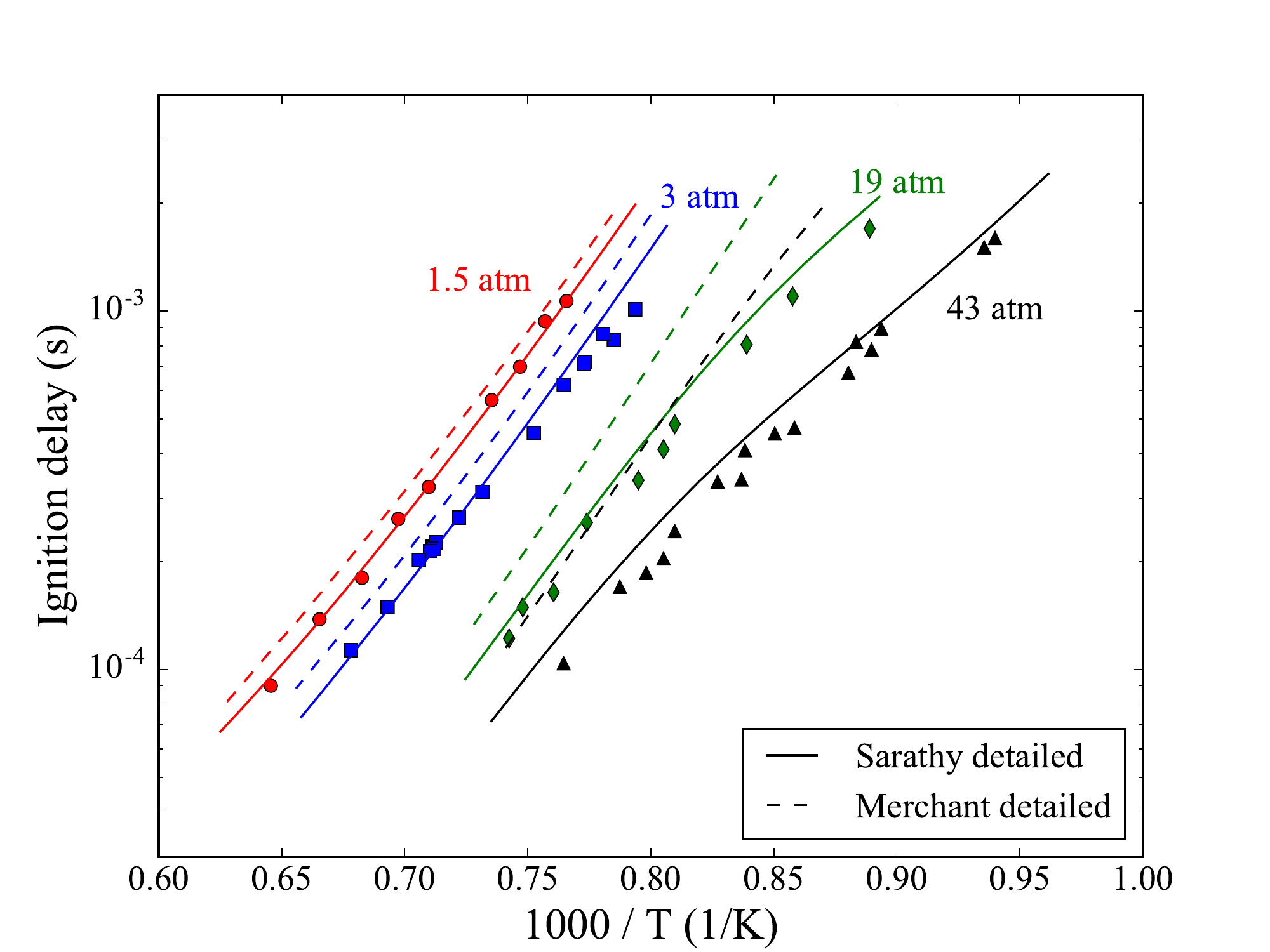}
        \caption{\textit{sec}-butanol}
        \label{fig:secbutanol_ign_exp_comp}
    \end{subfigure}
    ~
    \begin{subfigure}[b]{0.48\textwidth}
        \includegraphics[width=\textwidth]{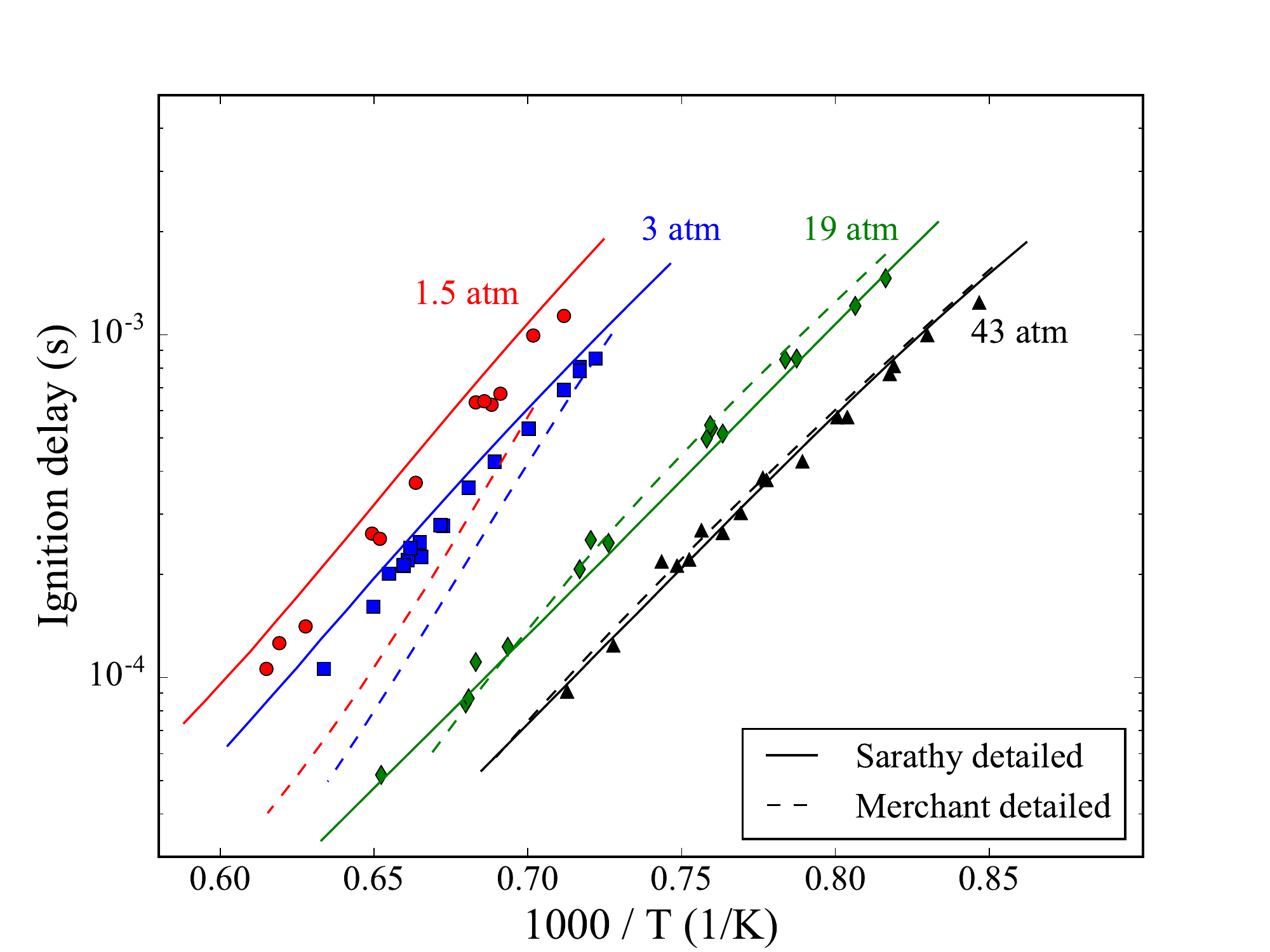}
        \caption{\textit{tert}-butanol}
        \label{fig:tertbutanol_ign_exp_comp}
    \end{subfigure}
    \caption{Comparison of butanol isomer ignition delay times between experiments (symbols) and simulations (lines) using detailed Sarathy and Merchant detailed mechanisms for pressures of \SIlist{1.5;3;19;43}{\atm}; a range of temperatures, and equivalence ratios of \numrange{0.6}{1.0} in mixtures containing \SI{4}{\percent} \ce{O2} diluted in argon. Experimental data are taken from Stranic et al.~\cite{Stranic:2012jl}.}
   \label{fig:ignition_validation}
\end{figure}

\begin{figure}[htbp]
   \centering
      \begin{subfigure}[b]{0.48\textwidth}
        \includegraphics[width=\textwidth]{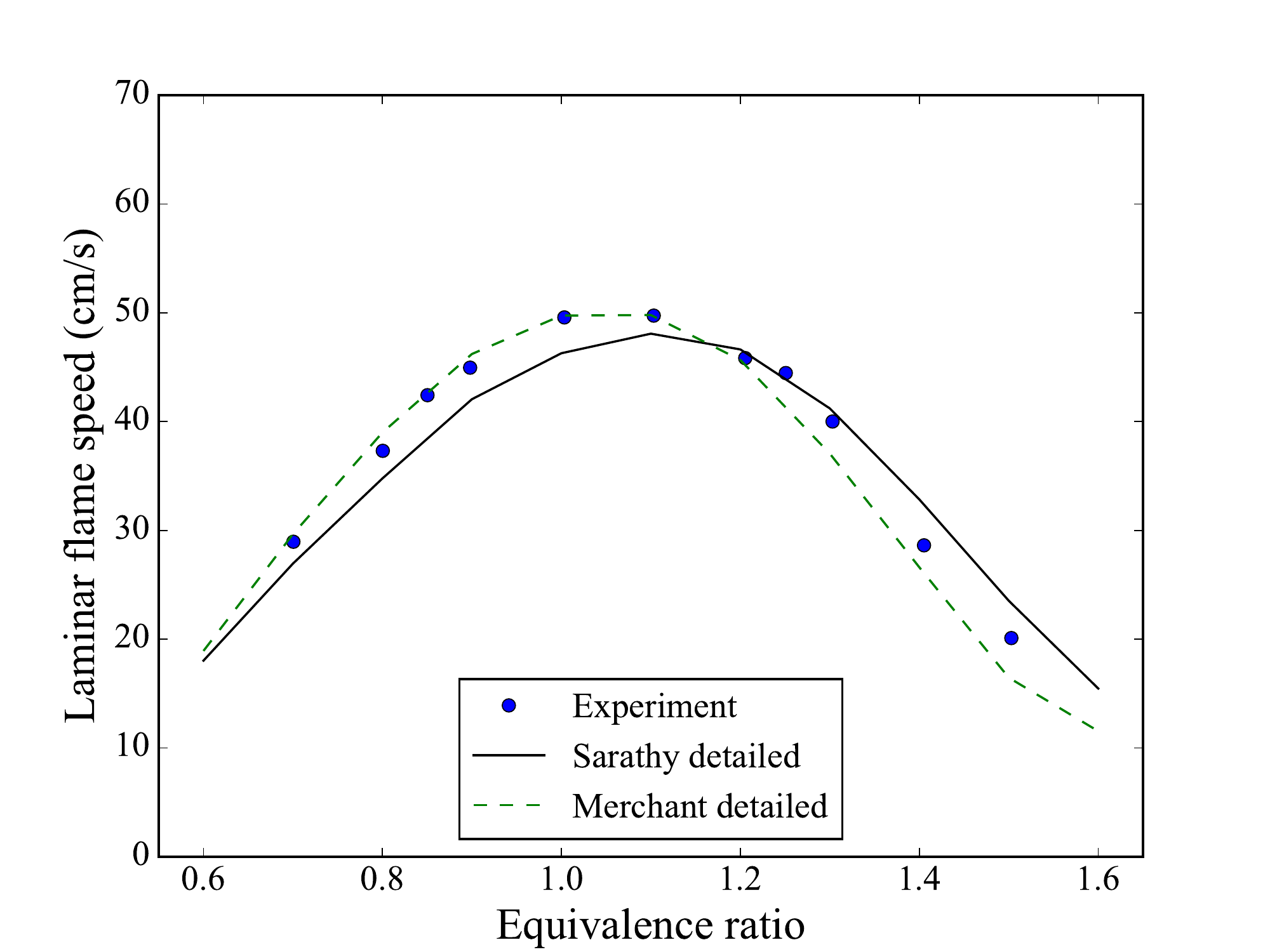}
        \caption{\textit{n}-butanol}
        \label{fig:nbutanol_flamespeed_exp_comp}
    \end{subfigure}
    ~
    \begin{subfigure}[b]{0.48\textwidth}
        \includegraphics[width=\textwidth]{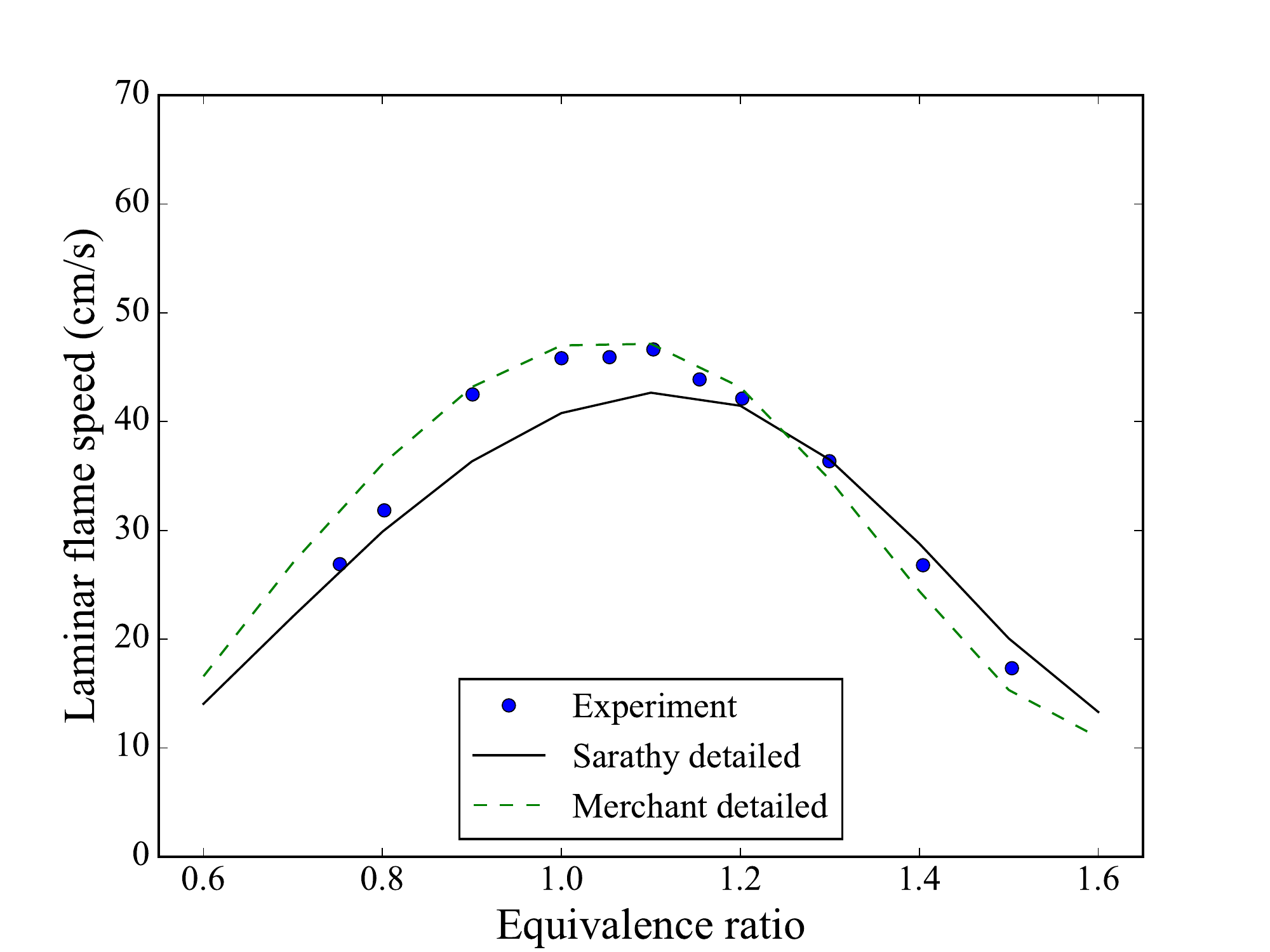}
        \caption{isobutanol}
        \label{fig:isobutanol_flamespeed_exp_comp}
    \end{subfigure}
    \\
    \begin{subfigure}[b]{0.48\textwidth}
        \includegraphics[width=\textwidth]{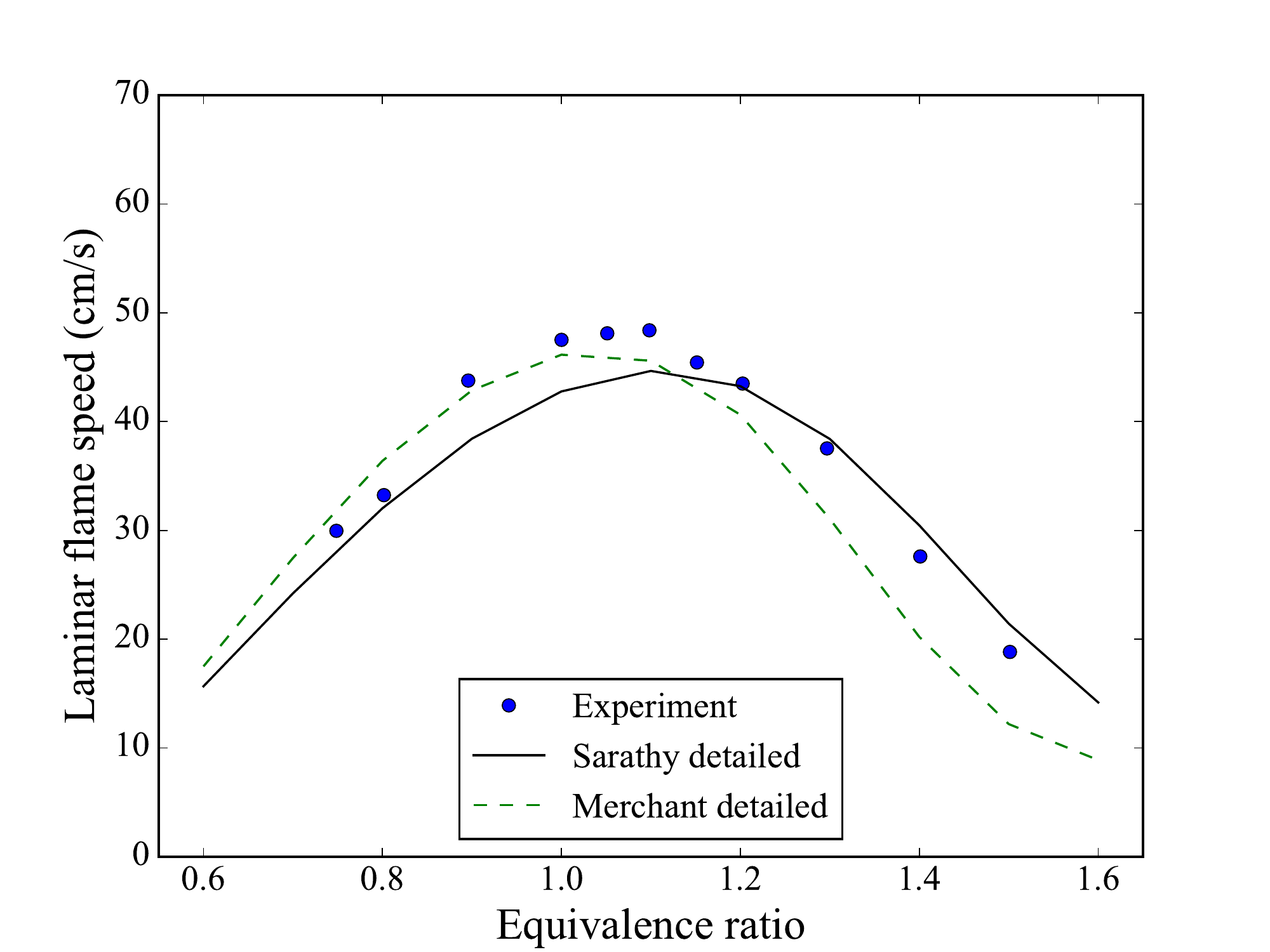}
        \caption{\textit{sec}-butanol}
        \label{fig:secbutanol_flamespeed_exp_comp}
    \end{subfigure}
    ~
    \begin{subfigure}[b]{0.48\textwidth}
        \includegraphics[width=\textwidth]{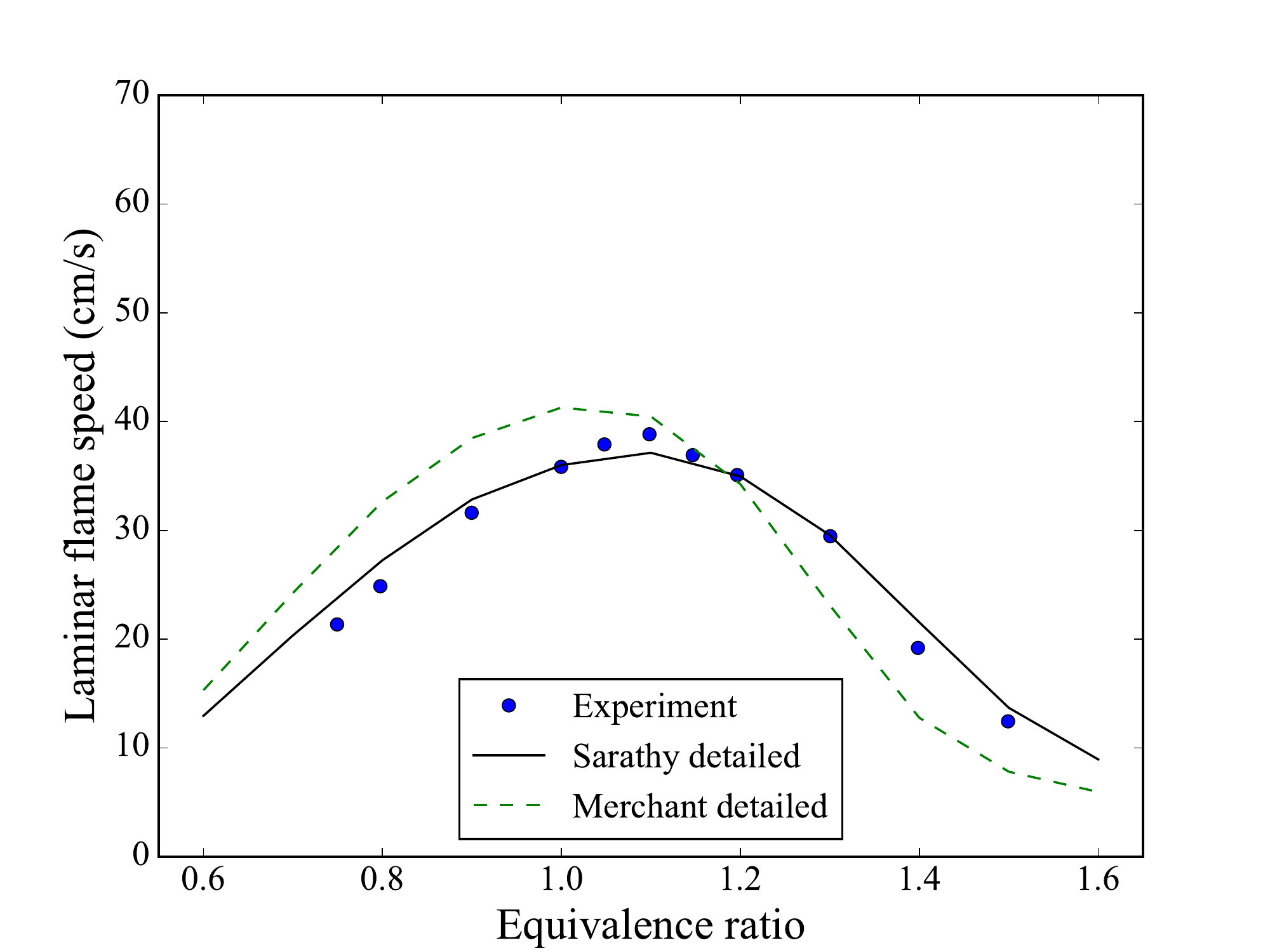}
        \caption{\textit{tert}-butanol}
        \label{fig:tertbutanol_flamespeed_exp_comp}
    \end{subfigure}
   \caption{Comparison of butanol isomer laminar flame speeds between experiments (symbols) and simulations (lines) using detailed Sarathy and Merchant detailed mechanisms for a range of equivalence ratios in air at \SI{1}{\atm} and an unburned mixture temperature of \SI{434}{\kelvin}. Experimental data are taken from Veloo et al.~\cite{Veloo:2011fr}.}
   \label{fig:validation_laminar_flame}
\end{figure}

Figures~\ref{fig:ignition_validation} and \ref{fig:validation_laminar_flame} compare the performances of the detailed Sarathy and Merchant mechanisms with experimentally measured ignition delay times~\cite{Stranic:2012jl} and laminar flame speeds~\cite{Veloo:2011fr}.
As seen in Fig.~\ref{fig:ignition_validation}, the Sarathy mechanism agrees with ignition delay data for all the isomers except for isobutanol at high pressures of \SIlist{19;40}{\atm}, while the Merchant mechanism agrees with most of the experimental data except for \textit{sec}-butanol at high pressures \SIlist{19;40}{\atm} and \textit{tert}-butanol at \SIlist{1.5;3}{\atm}.
The laminar flame speed comparison in Fig.~\ref{fig:validation_laminar_flame} shows that both the Sarathy and Merchant mechanisms approximately match the experimental data, though some discrepancies exist between the two mechanisms, including the equivalence ratio location of peak laminar flame speed.
The respective original publications provide additional validation against other experimental data for the Sarathy and Merchant mechanisms.

\subsection{Reduction procedure}

The reduction of the detailed mechanisms was performed at two levels: DRGEP and full DRGEPSA.
The first level applies the DRGEP method to quantify the importance of each species to the predetermined target species through a graph-based representation of species interdependence within the reaction system.
Species are removed when their importance values fall below a cutoff threshold, which is determined iteratively based on a user-specified error limit.
After a skeletal mechanism is generated by DRGEP, greedy sensitivity analysis further removes certain ``limbo'' species whose overall importance coefficients fall between the cutoff threshold and a specified upper threshold, while remaining within the specified error limit.
Additional details of the reduction methodology can be found in the works of Niemeyer and coworkers~\cite{Niemeyer:2010bt,Niemeyer:2011fe,Niemeyer:2014,Niemeyer:2015wq}.
This study used version 2.3.0 of MARS (Mechanism Automatic Reduction Software)~\cite{MARS:2.3} to perform the reduction, which added support for the newer logarithmic and Chebyshev pressure-dependent reaction formulations (discussed in more detail later) compared to the version used by Niemeyer and Sung\cite{Niemeyer:2015wq}.

The reduction procedure used autoignition and PSR simulation data to both generate thermochemical state data (e.g., to calculate DRGEP species interaction coefficients) and to evaluate trial skeletal mechanism performance.
The reactants butanol, \ce{O2}, and \ce{N2} were chosen as the target species in the reduction process, following the practice of Niemeyer and coworkers in prior studies using DRGEP~\cite{Niemeyer:2010bt,Niemeyer:2011fe,Niemeyer:2014,Niemeyer:2015wq}.
Since we are primarily interested in SI or CI engines operating at high temperatures, the reduction validations are limited to temperatures above \SI{1000}{\kelvin}.
Specifically, the autoignition simulations cover initial temperatures of \SIrange{1000}{1800}{\kelvin}, pressures of \SIrange{1}{40}{\atm}, and equivalence ratios of \numrange{0.5}{1.5}.
PSR simulations cover the same ranges of pressure and equivalence ratio, with an inlet temperature of \SI{400}{\kelvin}.
The error limit level for DRGEP was set to \SI{10}{\percent}, while the upper threshold value for the SA phase was set to 0.1.
MARS determines the DRGEP cutoff threshold value iteratively based on the \SI{10}{\percent} error limit, starting at \num{1e-3} and increasing until it reaches the error limit.
Niemeyer and Sung~\cite{Niemeyer:2014,Niemeyer:2015wq} suggested using these values to achieve appreciable mechanism reduction without compromising accuracy.
MARS evaluates skeletal mechanism error using both ignition delay times and PSR temperatures at three points along the upper branch of the response curve with respect to residence time: the extinction turning point, near a residence time of \SI{0.1}{\second}, and the logarithmic midpoint between the first two points~\cite{Niemeyer:2015wq}.

\begin{table}[htbp]
   \centering
   \begin{tabular}{@{} lccccc @{}}
      \toprule
    & & \multicolumn{2}{c}{DRGEP} & \multicolumn{2}{c}{DRGEPSA} \\
    & $\epsilon_{\text{EP}}$ & Species & Reactions & Species & Reactions \\
      \midrule
      \textit{n}-butanol & \num{2.3e-2} & 86 & 637 & 61 & 409 \\
      isobutanol & \num{2.4e-2} & 85 & 574 & 58 & 385 \\
      \textit{sec}-butanol & \num{1.5e-2} & 95 & 687 & 62 & 425 \\
      \textit{tert}-butanol & \num{2.0e-2} & 81 & 536 & 64 & 445 \\
      \bottomrule
   \end{tabular}
   \caption{Summary of results from skeletal reduction of the Sarathy mechanism for butanol isomers, with 284 species and 1892 reactions.
   $\epsilon_{\text{EP}}$ represents the DRGEP cutoff threshold.}
   \label{tab:sarathy_skeletal}
\end{table}


\begin{table}[htbp]
   \centering
   \begin{tabular}{@{} lccccc @{}}
      \toprule
    & & \multicolumn{2}{c}{DRGEP} & \multicolumn{2}{c}{DRGEPSA} \\
    & $\epsilon_{\text{EP}}$ & Species & Reactions & Species & Reactions \\
      \midrule
      \textit{n}-butanol & \num{1.4e-2} & 102 & 1904 & 68 & 1046 \\
      isobutanol & \num{2.0e-3} & 153 & 3405 & 81 & 1709 \\
      \textit{sec}-butanol & \num{1.1e-2} & 108 & 2197 & 69 & 1357 \\
      \textit{tert}-butanol & \num{7.0e-3} & 117 & 1755 & 78 & 1114 \\
      \bottomrule
   \end{tabular}
   \caption{Summary of results from skeletal reduction of the Merchant mechanism for butanol isomers, with 372 species and 8322 reactions.
   $\epsilon_{\text{EP}}$ represents the DRGEP cutoff threshold.}
   \label{tab:merchant_skeletal}
\end{table}

Tables~\ref{tab:sarathy_skeletal} and \ref{tab:merchant_skeletal} show the sizes of the resulting skeletal mechanisms for the Sarathy and Merchant mechanisms, respectively.
The DRGEP method removes most of the unimportant species for all butanol isomers, resulting in skeletal mechanisms less than one third the size of the respective detailed parent, except in the case of isobutanol in the Merchant model, whose skeletal mechanism is ~\SI{50}{\percent} larger than those of the other three isomers.
As the Merchant mechanism was recently optimized for isobutanol, it contains more detailed chemical pathways for this isomer.
Further application of SA removes an additional \SIrange{21}{36}{\percent} of remaining species---except in the case of Merchant isobutanol, where \SI{47}{\percent} are removed---to generate the final skeletal mechanisms.

\subsubsection{Pressure-dependent reactions}

In addition to the canonical Arrhenius dependence of reaction rates on temperature, the reaction rates of certain reactions are also dependent on pressure.
Such reactions include, for example, dissociation reactions, isomerization reactions, radical-radical recombination reactions, and elimination reactions.
Pressure dependence is typically expressed as unimolecular\slash recombination fall-off reactions and chemically activated bimolecular reactions.
In general, the rates of unimolecular\slash recombination fall-off reactions increase with increasing pressure, while the rates of chemically activated bimolecular reactions decrease with increasing pressure.
To capture the ``fall-off'' behavior of these reactions, their reaction rate coefficients are usually described by modified Arrhenius expressions utilizing low- and high-pressure limit constants and a fall-off factor that smoothly connects the limiting rate coefficients between the fall-off regimes.
The Lindemann, Troe, and SRI formulations of Lindemann et al.~\cite{lindemann1922discussion}, Gilbert et al.~\cite{gilbert1983theory}, and Stewart et al.~\cite{stewart1989pressure}, respectively, provide analytical expressions for the fall-off factor and have been quite successful at describing most pressure-dependent reactions.

However, for more complex reactions with multiple energy wells and products, the fall-off behavior cannot be satisfactorily fitted using a single Arrhenius expression.
More accurate formulations based on logarithmic interpolation (i.e., pressure-log)~\cite{chemkin:2012,cantera} and Chebyshev polynomials~\cite{venkatesh1997parameterization,Venkatesh:2000gj,chemkin:2012,cantera} have been proposed and used in more recent chemical kinetic models, such as the Sarathy and Merchant mechanisms employed in the present study.
Caution needs to be exercised when implementing the pressure-log and Chebyshev reactions due to their complex formulations, and care needs to be taken with regards to their validity range to prevent unjustified extrapolation outside their specified limits.
This is particularly true in the case of Chebyshev polynomials, as their formulation is mathematically constrained to  the stated pressure and temperature limits.
Beyond these general cautions, however, in the process of our mechanism reduction several issues were encountered for certain pressure-log reactions, and we would like to clarify them in the remainder of this section in an effort to motivate a more consistent method of applying such pressure-log reactions in future studies.

In a pressure-log reaction, the reaction rate constant $k_i$($T,P_i$) at a specified pressure $P_i$ and a temperature $T$ is given by
\begin{equation}
k_i (T,P_i) = A_i T^{b_i} \exp{(-E_i / RT)}\;,
\label{ki}
\end{equation}
where $A_i$, $b_i$, and $E_i$ are the pre-exponential factor, temperature exponent, and activation energy, respectively, given for a specified pressure $P_i$ in a pressure-log reaction.
Usually, several rate expressions at different specified pressures will be given to cover a range of pressures.
The reaction rate constant $k(T,P)$ at any intermediate pressure $P_i < P < P_{i+1}$ can be computed by a log fitting method using $k_i (T,P_i)$ and $k_{i+1} (T,P_{i+1})$ as given by
\begin{equation}
\log k (T,P) = \log k_i (T,P_i) + \left( \log k_{i+1} (T,P_{i+1}) - \log k_i (T,P_i) \right) \frac{ \log P - \log P_i }{ \log P_{i+1} - \log P_i } \;.
\label{logk}
\end{equation}
When $P$ is outside the specified pressure range, CHEMKIN-PRO~\cite{chemkin:2012} uses the rate expression at the limit pressure to avoid an error, although the validity of this approach cannot be justified.

Many mechanisms also allow certain identical reactions to proceed at different rates; these reactions can be defined as ``declared duplicate'' reactions with separate sets of rate expressions.
For pressure-log reactions, such utility can be achieved by using multiple rate expressions at the same pressure in one reaction, in which case the rate constants used in Eq.~\eqref{logk} are the sum of all the rate constants given at that pressure.
Both methods are commonly used; however, for pressure-log reactions the formulation of duplicate reactions and the formulation of multiple rate expressions are not equivalent, and in fact can yield very different results.
We use the decomposition reaction of isobutanol (\ce{$i$BuOH <-> $i$C4H8 + H2O}) in the Merchant mechanism to illustrate this behavior.
This reaction is originally defined in the form of duplicate reactions in the Merchant mechanism, as shown below in CHEMKIN format (for the sake of brevity the rate expressions at only \SIlist{10;100}{\atm} are shown):

\begin{lstlisting}[basicstyle=\linespread{1.0}\ttfamily\small]
iBuOH = iC4H8 + H2O
! Pressure (atm)     A          b      E (cal/mol)
PLOG 10.0            3.13E+51  -10.82  96480
PLOG 100.0           4.24E+27  -4.0    80731.98
duplicate
iBuOH = iC4H8 + H2O
! Pressure (atm)     A          b      E (cal/mol)
PLOG 10.0            5.10E+16  -1.46   66679.9
PLOG 100.0           5.83E+09   0.61   63570
duplicate
\end{lstlisting}

The above duplicate pressure-log reactions can also be reformulated using multiple rate expressions into a single pressure-log reaction format:
\begin{lstlisting}[basicstyle=\linespread{1.0}\ttfamily\small]
iBuOH = iC4H8 + H2O
! Pressure (atm)     A          b      E (cal/mol)
PLOG 10.0            3.13E+51  -10.82  96480
PLOG 10.0            5.10E+16  -1.46   66679.9
PLOG 100.0           4.24E+27  -4.0    80731.98
PLOG 100.0           5.83E+09   0.61   63570
\end{lstlisting}

\begin{figure}[htbp]
   \centering
   \includegraphics[width=0.7\linewidth]{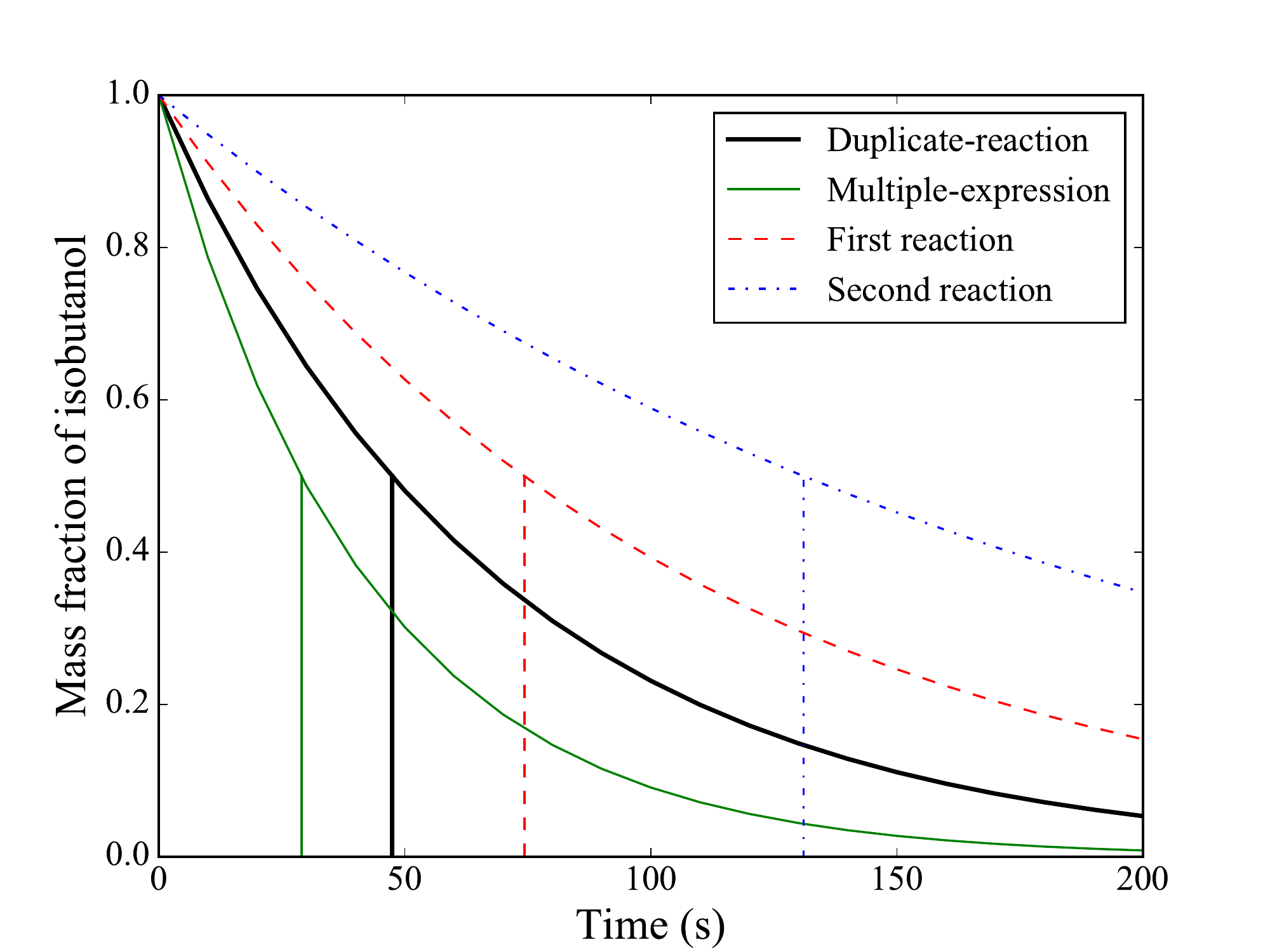}
   \caption{Comparison of half-life times of isobutanol for the reaction \ce{$i$BuOH <-> $i$C4H8 + H2O} implemented using the duplicate-reaction and multiple-expression pressure-log formulations, at a constant temperature of \SI{1000}{\kelvin} and pressure of \SI{40}{\atm}. The consumption of isobutanol due to the first and second reactions alone in the duplicate-reaction formulation are also shown for reference.}
   \label{fig:plog_isobutanol_halflife}
\end{figure}

Both formulations are syntactically valid, but each will cause isobutanol to decompose into isobutene and water at different rates.
Figure~\ref{fig:plog_isobutanol_halflife} compares the half-life times of isobutanol using the duplicate-reaction and multiple-expression formulations at a constant pressure of \SI{40}{\atm} and temperature of \SI{1000}{\kelvin}, illustrating the differences between the two approaches.
The decomposition of isobutanol can proceed at two different rates as represented by the first and second reactions in the duplicate-reaction formulation; however, when combined the duplicate-reaction formulation predicts a slower decay than the multiple-expression formulation, resulting in half-life times of \SI{47.5}{\second} and \SI{29.0}{\second} for the duplicate-reaction and multiple-expression formulations, respectively.
This difference occurs because the duplicate-reaction formulation calculates the net rate coefficient as the sum of the logarithmically interpolated rate coefficients, while the multiple-expression formulation sums the rate coefficients prior to interpolation.
As long as the difference between these two formulations is understood, the mechanism developer should intentionally choose the proper formulation for a given pressure-log reaction with multiple reaction rates.

Another issue can arise with negative pre-exponential factors in pressure-log reactions: duplicate-reaction formulations cannot allow a negative value, since the logarithm of a negative rate coefficient would be undefined.
In contrast, the single-reaction, multiple-expression formulation allows a negative pre-exponential factor as long as the sum of the rate constants at the same pressure remains positive.
As it turns out, this issue is not hypothetical: in the Merchant mechanism, one isobutanol reaction describing decomposition into a methyl radical and 2-hydroxypropyl (\ce{$i$BuOH <-> CH3 + C3H7O-2}) contains a negative pre-exponential factor in the duplicate pressure-log format, which forces the logarithm of a negative rate constant and thus causes a failure in our mechanism reduction process due to an undefined value.
Brady et al.~\cite{Brady:2016dw} observed the same problem in their study on forced ignition of the butanol isomers.
Perhaps more worryingly, when the same reaction is used in CHEMKIN-PRO~\cite{chemkin:2012}, the software proceeds with the calculation without showing any warning or error messages.
Cantera, an alternative open-source software~\cite{cantera}, stops with an error message.
Given the nature of the interpolation, it is unclear what allows CHEMKIN-PRO calculations to proceed.
In order to facilitate as broad a mechanism comparison as possible, this reaction was removed from the Merchant mechanism before our reduction process.
The removal of this isobutanol decomposition reaction is deemed valid for \textit{n}-, \textit{sec}-, and \textit{tert}-butanols since it has little effect on their oxidation pathways.
As for isobutanol, we assume that the pathway of isobutanol decomposing into methyl and 2-hydroxypropyl is not significant due to its high activation energy (\SI{132857}{\calorie\per\mole}) compared with other pathways via \ce{H}-abstract reactions and other decomposition reactions.
Calculations of ignition delays and laminar flame speeds without this problematic reaction agree well with experimental data as shown in Figs.~\ref{fig:ignition_validation} and \ref{fig:validation_laminar_flame}, suggesting that its removal does not adversely impact the mechanism performance.
Therefore, the reaction of \ce{$i$BuOH <-> CH3 + C3H7O-2} was excluded from the Merchant mechanism in our present study.

\section{Validation of skeletal mechanisms}
\label{S:validation}

\begin{figure}[htbp]
   \centering
   \begin{subfigure}[b]{0.48\textwidth}
        \includegraphics[width=\textwidth]{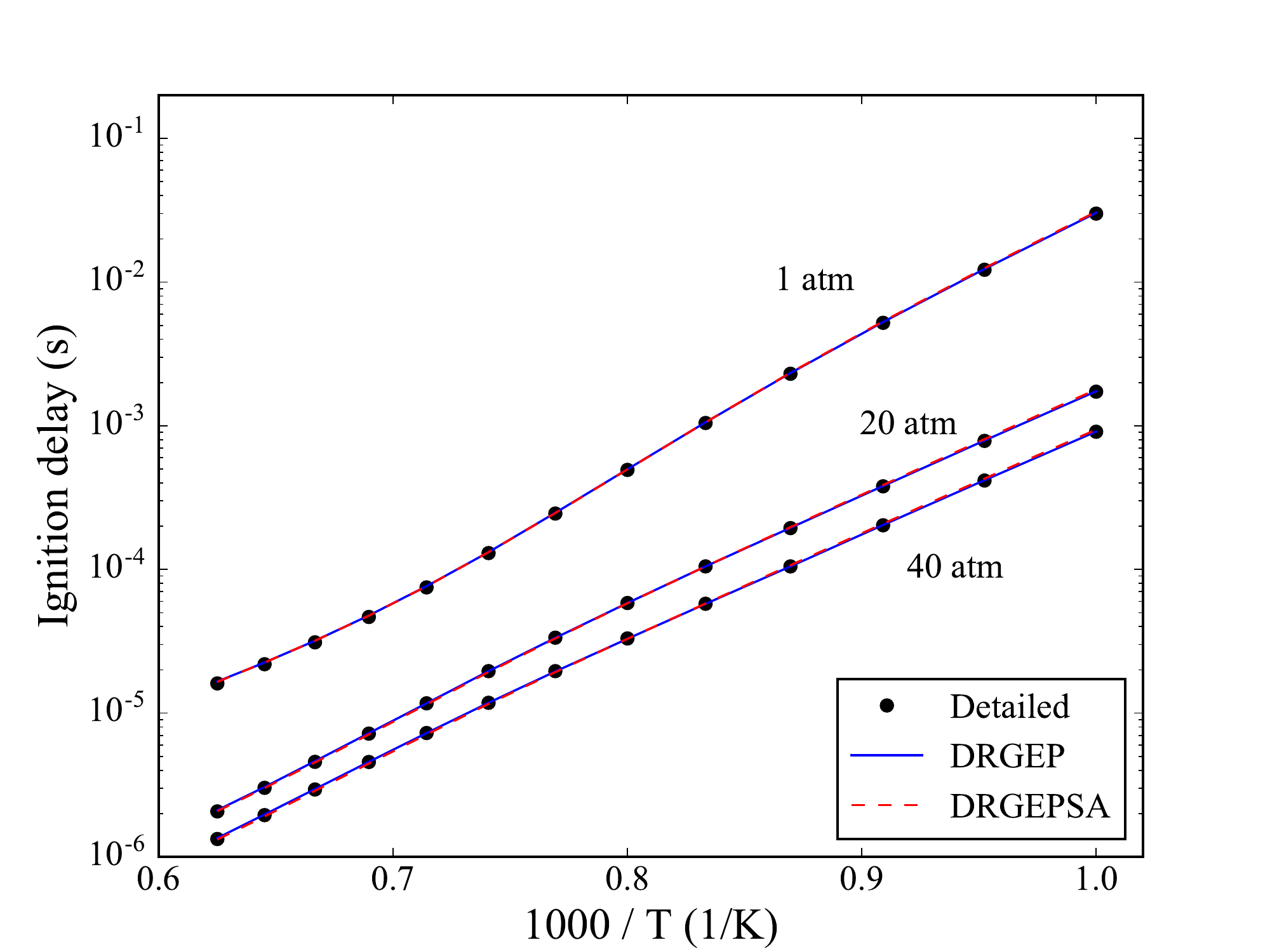}
        \caption{\textit{n}-butanol}
        \label{fig:Sarathy_nbutanol_skel_ign}
    \end{subfigure}
    ~
    \begin{subfigure}[b]{0.48\textwidth}
        \includegraphics[width=\textwidth]{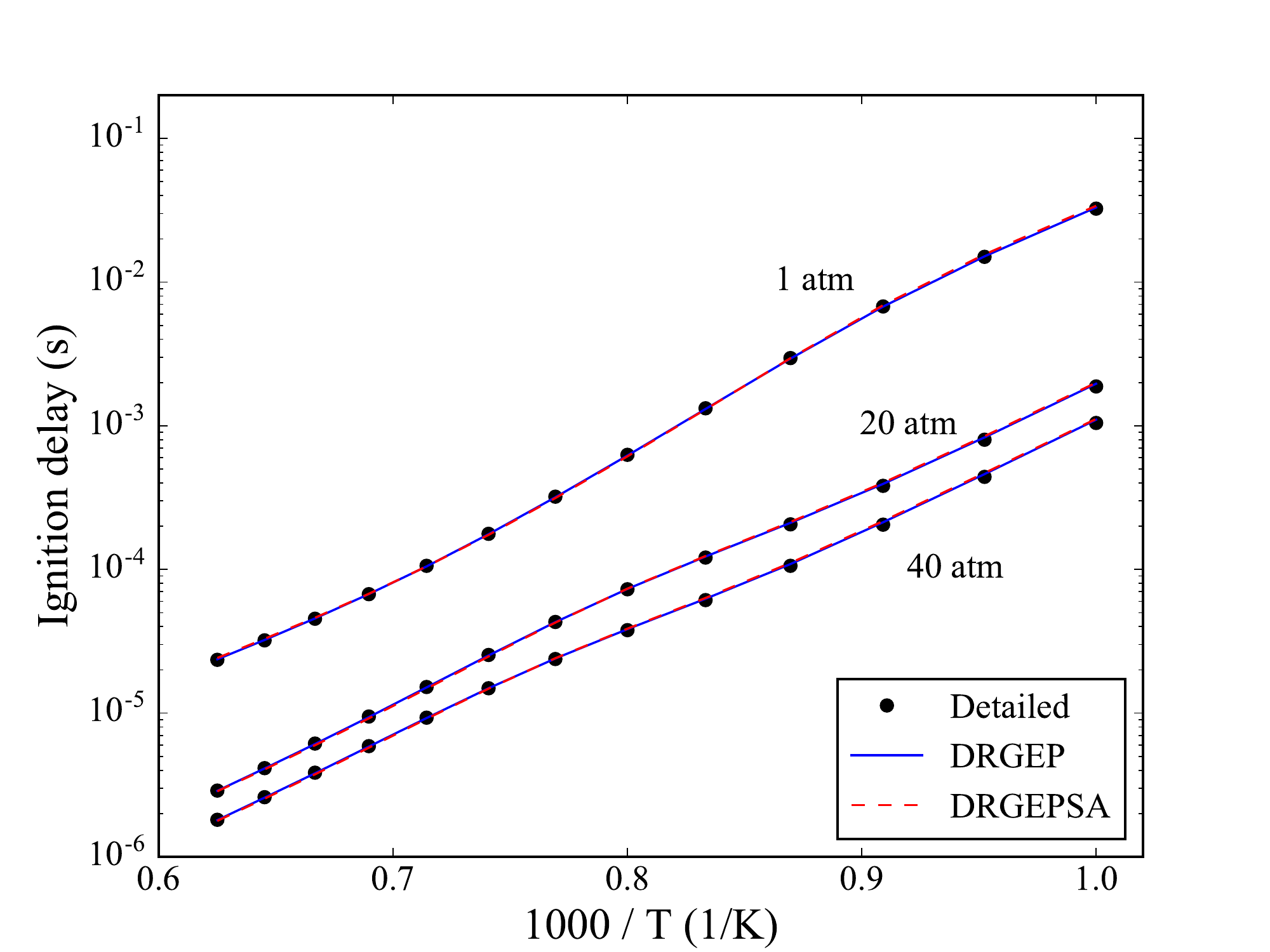}
        \caption{isobutanol}
        \label{fig:Sarathy_isobutanol_skel_ign}
    \end{subfigure}
    \\
    \begin{subfigure}[b]{0.48\textwidth}
        \includegraphics[width=\textwidth]{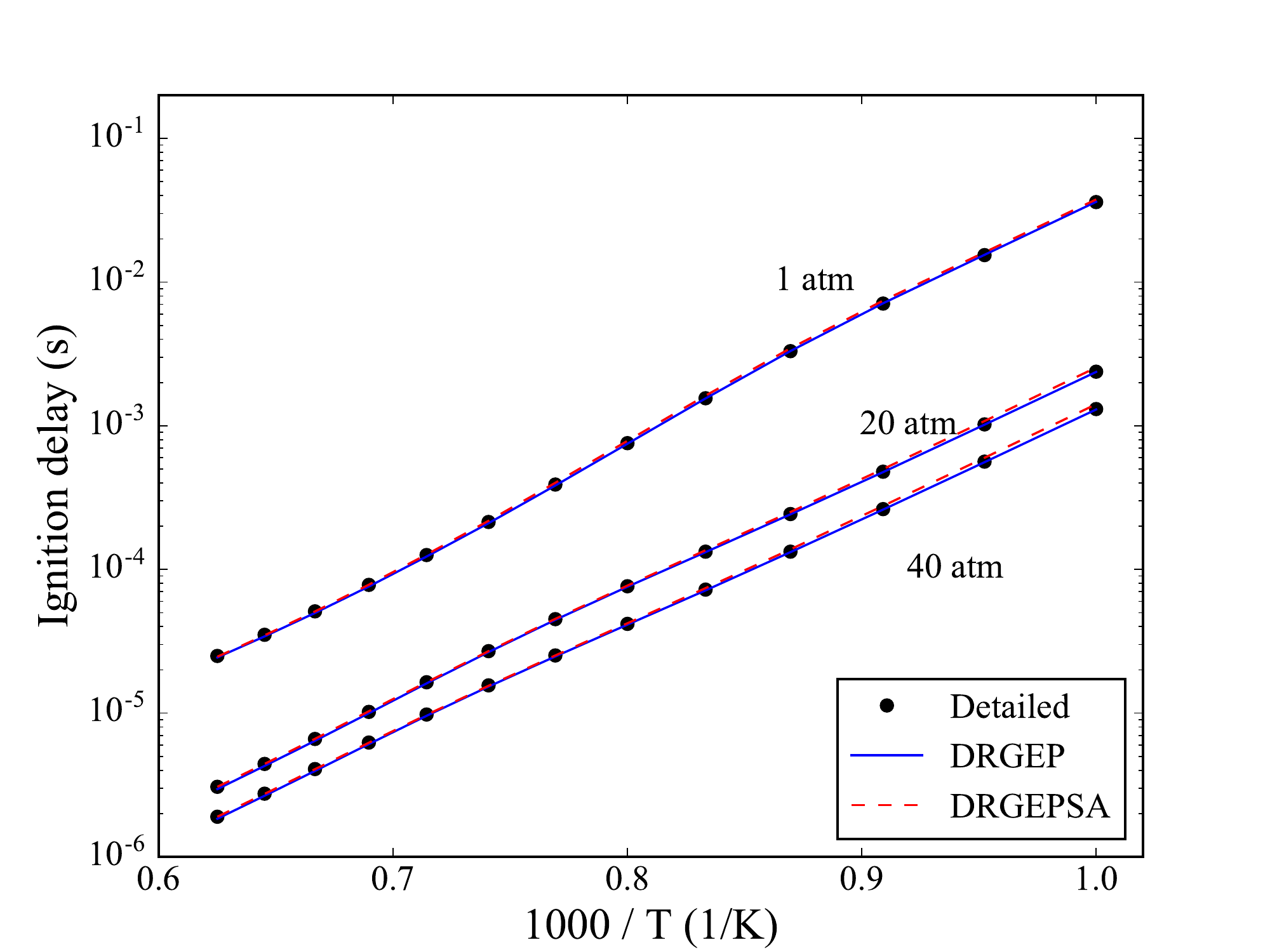}
        \caption{\textit{sec}-butanol}
        \label{fig:Sarathy_secbutanol_skel_ign}
    \end{subfigure}
    ~
    \begin{subfigure}[b]{0.48\textwidth}
        \includegraphics[width=\textwidth]{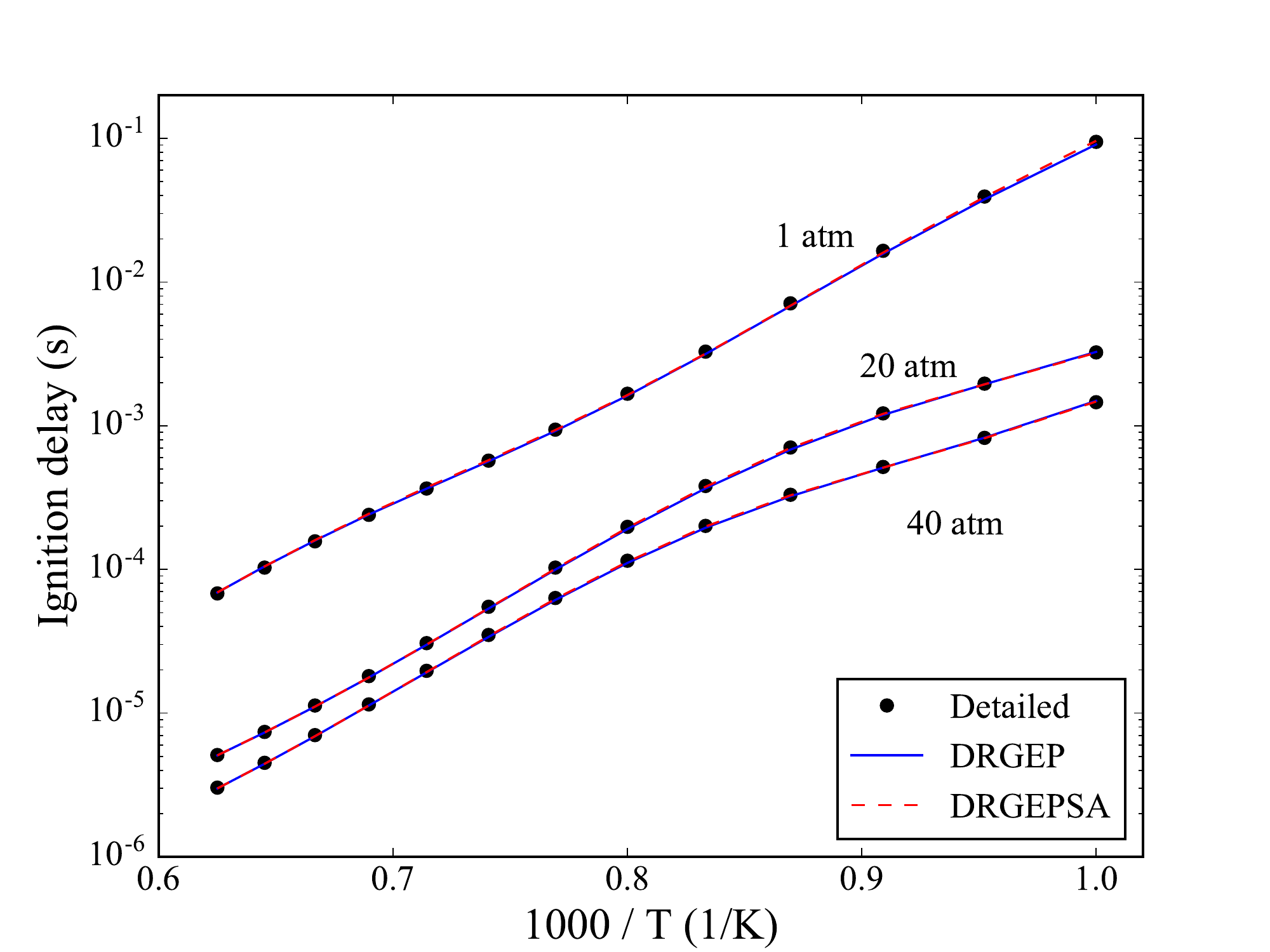}
        \caption{\textit{tert}-butanol}
        \label{fig:Sarathy_tertbutanol_skel_ign}
    \end{subfigure}
   \caption{Ignition delay times of butanol isomers using Sarathy detailed and skeletal DRGEP and DRGEPSA mechanisms for initial temperatures of \SIrange{1000}{1600}{\kelvin}; pressures of \SIlist{1;20;40}{\atm}; and an equivalence ratio of 1.0 in air.}
   \label{fig:Sarathy_ign_delay_comparison}
\end{figure}

\begin{figure}[htbp]
   \centering
   \begin{subfigure}[b]{0.48\textwidth}
        \includegraphics[width=\textwidth]{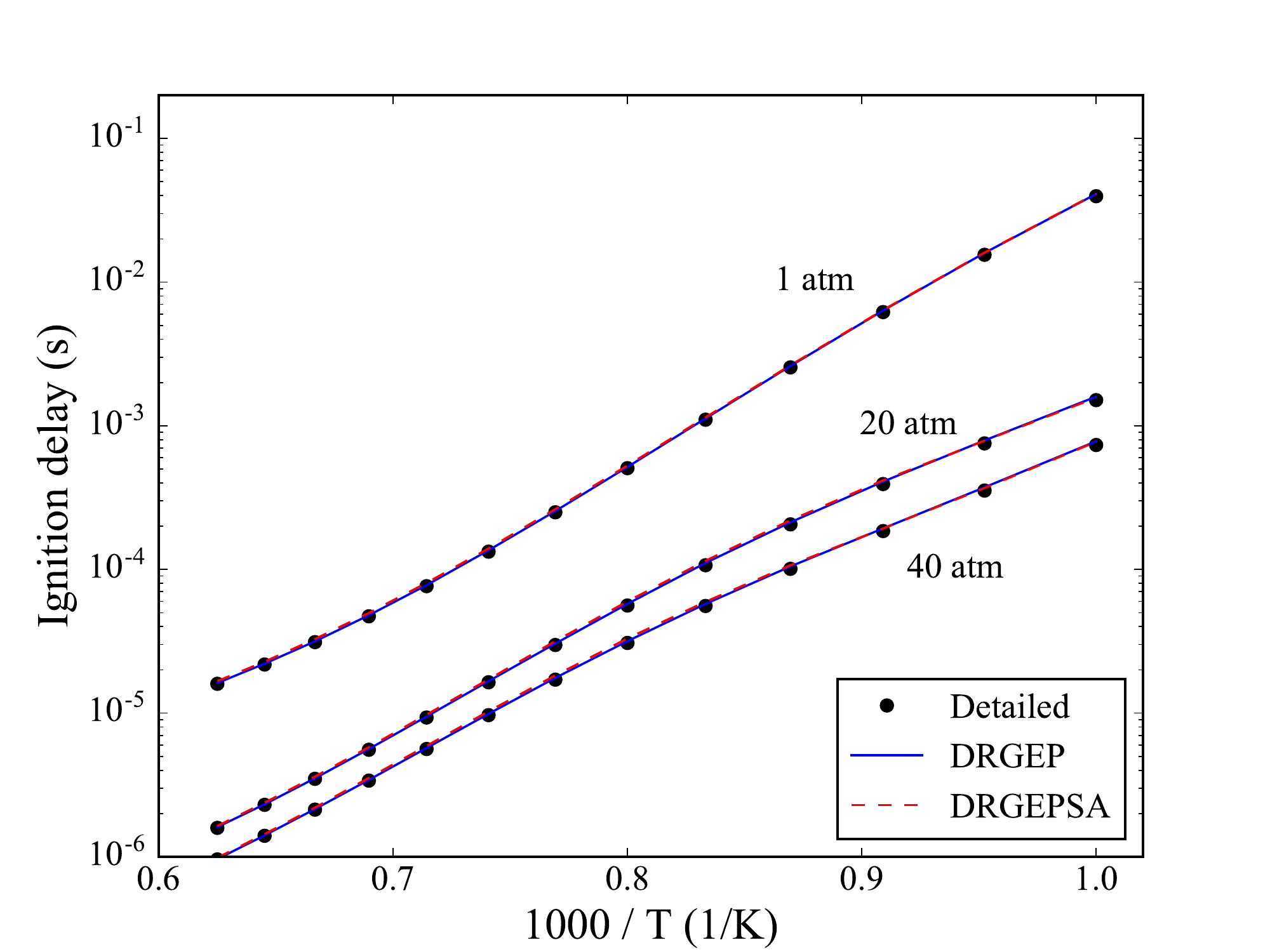}
        \caption{\textit{n}-butanol}
        \label{fig:Merchant_nbutanol_skel_ign}
    \end{subfigure}
    ~
    \begin{subfigure}[b]{0.48\textwidth}
        \includegraphics[width=\textwidth]{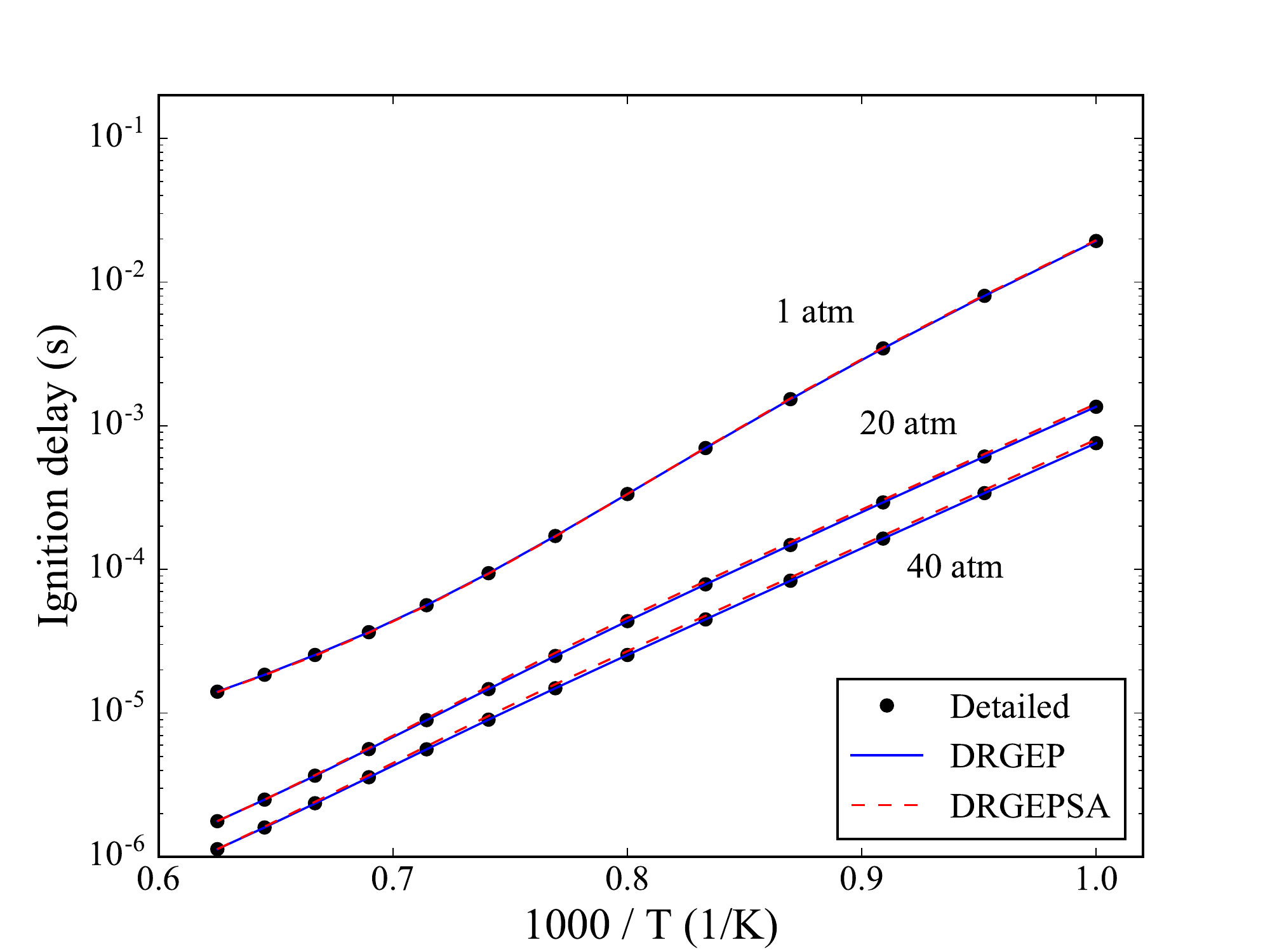}
        \caption{isobutanol}
        \label{fig:Merchant_isobutanol_skel_ign}
    \end{subfigure}
    \\
    \begin{subfigure}[b]{0.48\textwidth}
        \includegraphics[width=\textwidth]{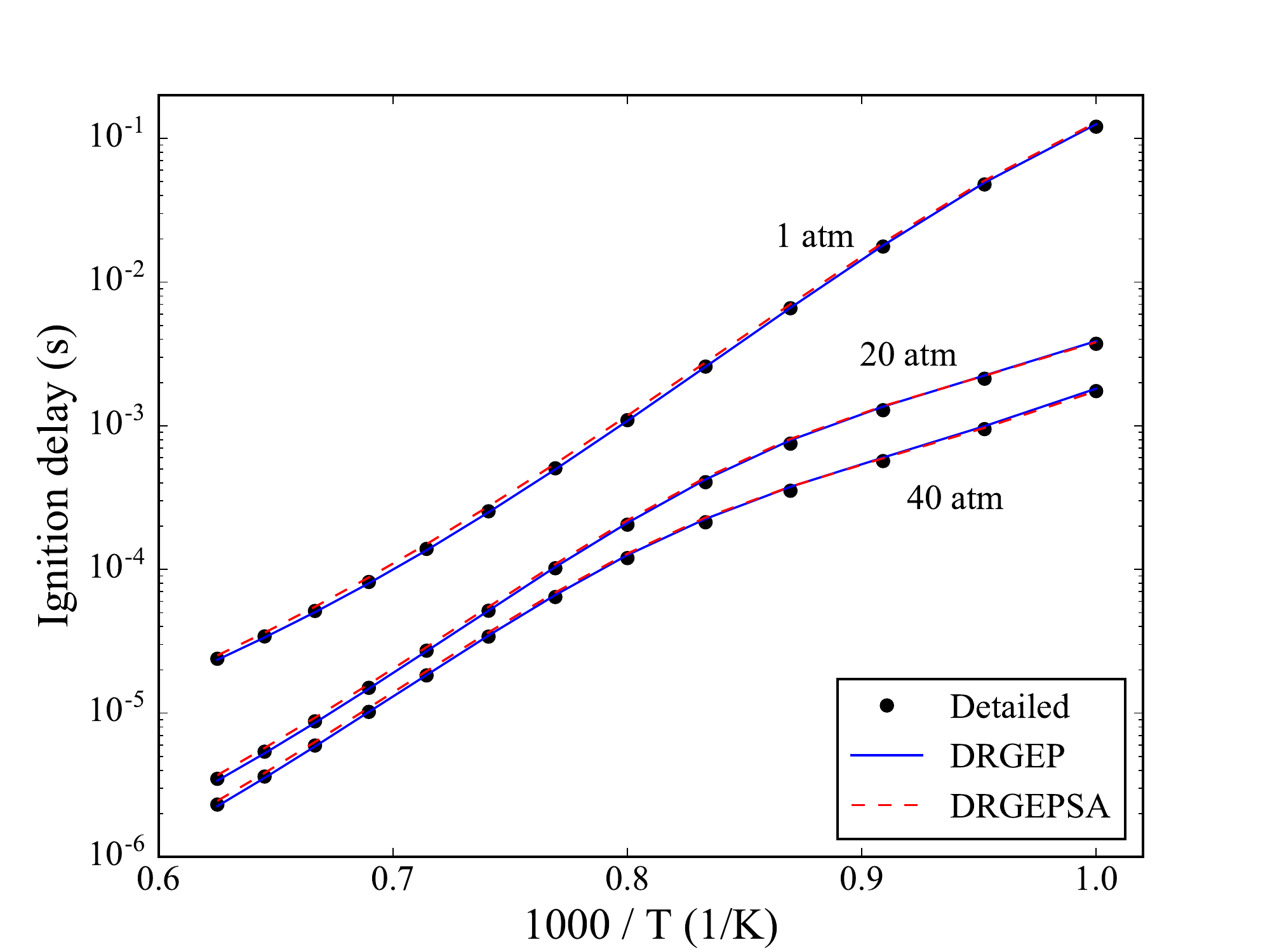}
        \caption{\textit{sec}-butanol}
        \label{fig:Merchant_secbutanol_skel_ign}
    \end{subfigure}
    ~
    \begin{subfigure}[b]{0.48\textwidth}
        \includegraphics[width=\textwidth]{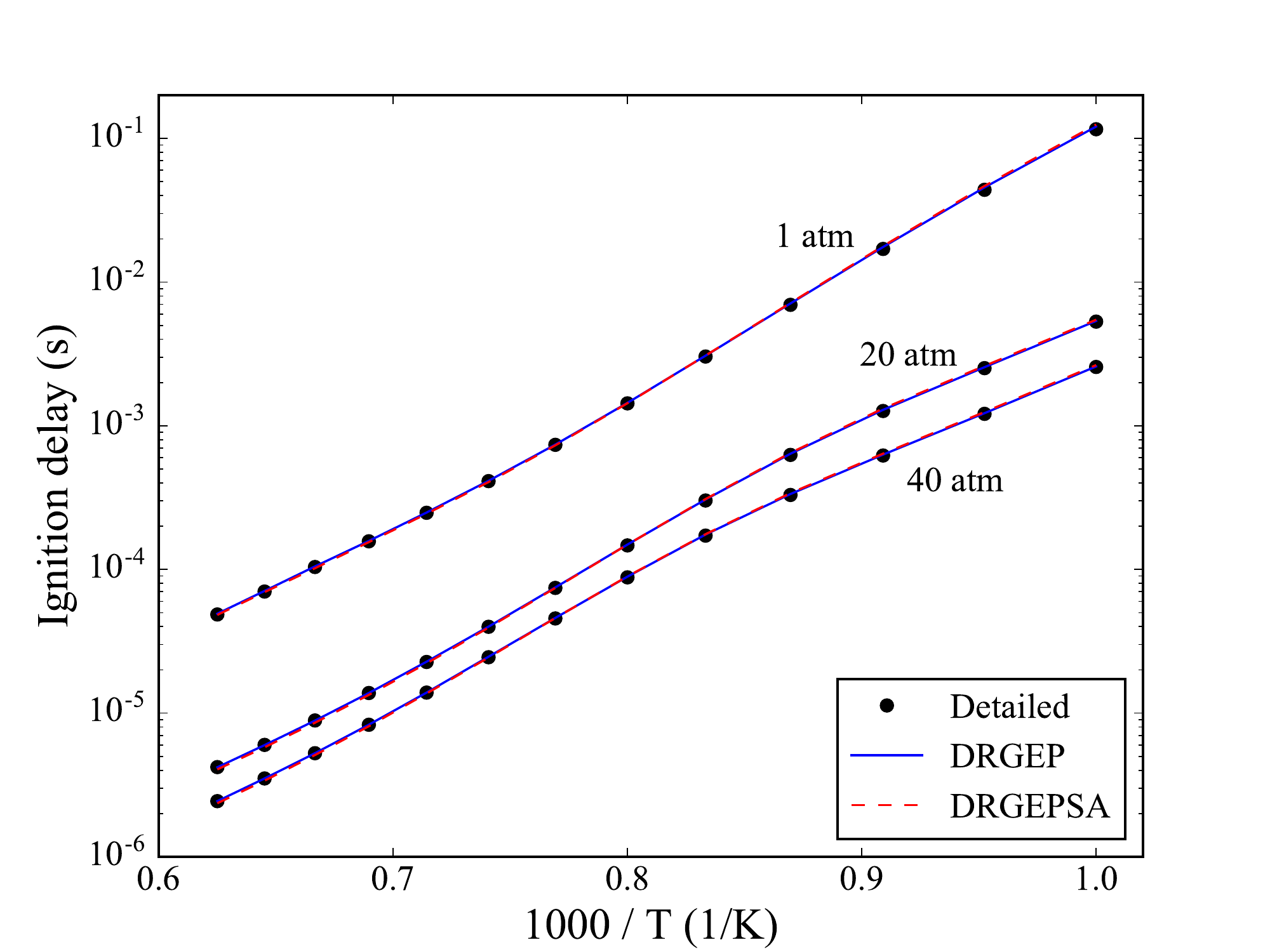}
        \caption{\textit{tert}-butanol}
        \label{fig:Merchant_tertbutanol_skel_ign}
    \end{subfigure}
   \caption{Ignition delay times of butanol isomers using Merchant detailed and skeletal DRGEP and DRGEPSA mechanisms for initial temperatures of \SIrange{1000}{1600}{\kelvin}; pressures of \SIlist{1;20;40}{\atm}; and an equivalence ratio of 1.0 in air.}
   \label{fig:Merchant_ign_delay_comparison}
\end{figure}

\begin{figure}[htbp]
    \centering
        \begin{subfigure}[b]{0.6\textwidth}
        \includegraphics[width=\textwidth]{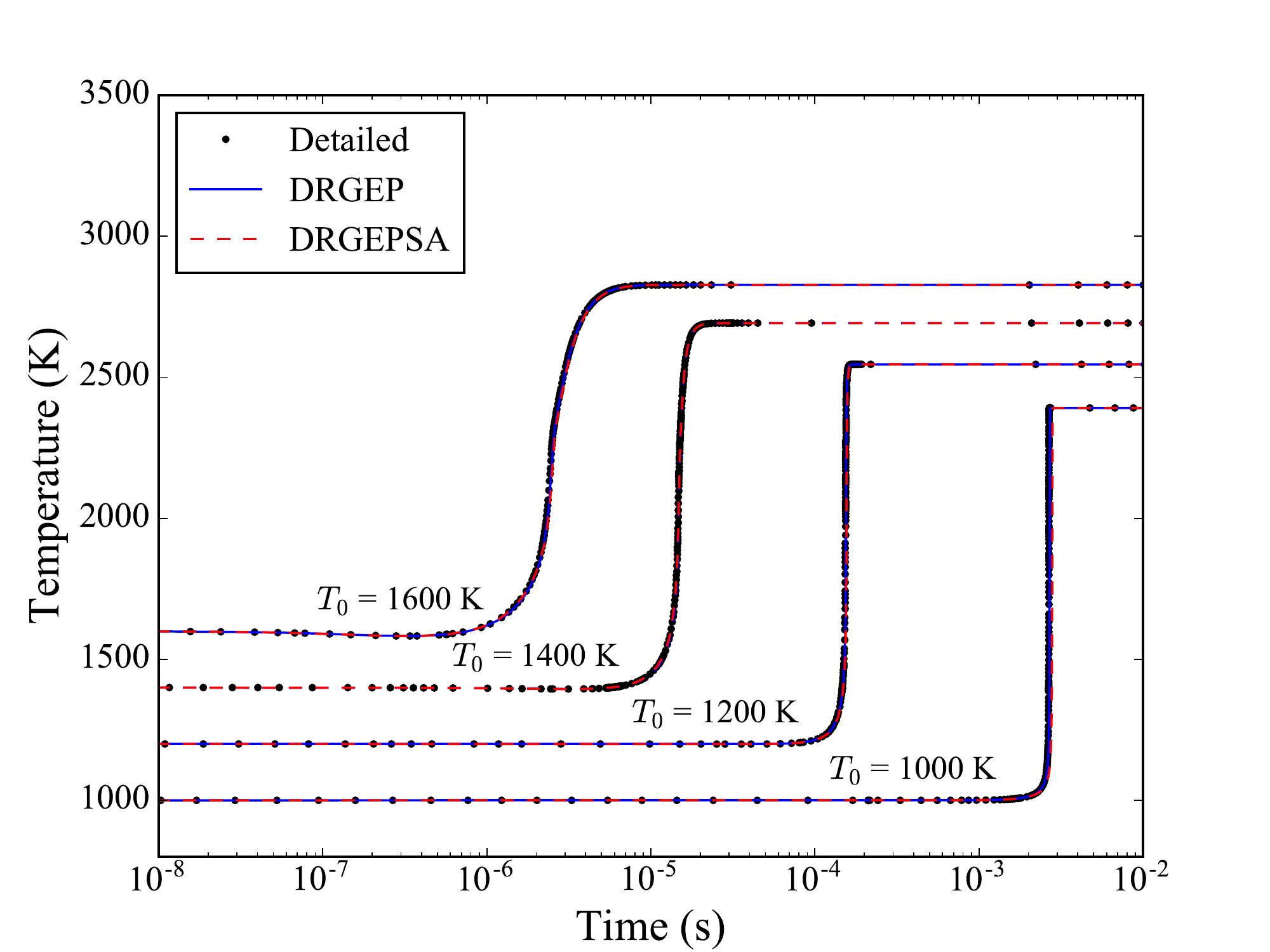}
        \caption{$\phi = 0.5$}
        \label{fig:Sarathy_temperature_profile_phi0.5}
    \end{subfigure}
    \\
    \begin{subfigure}[b]{0.6\textwidth}
        \includegraphics[width=\textwidth]{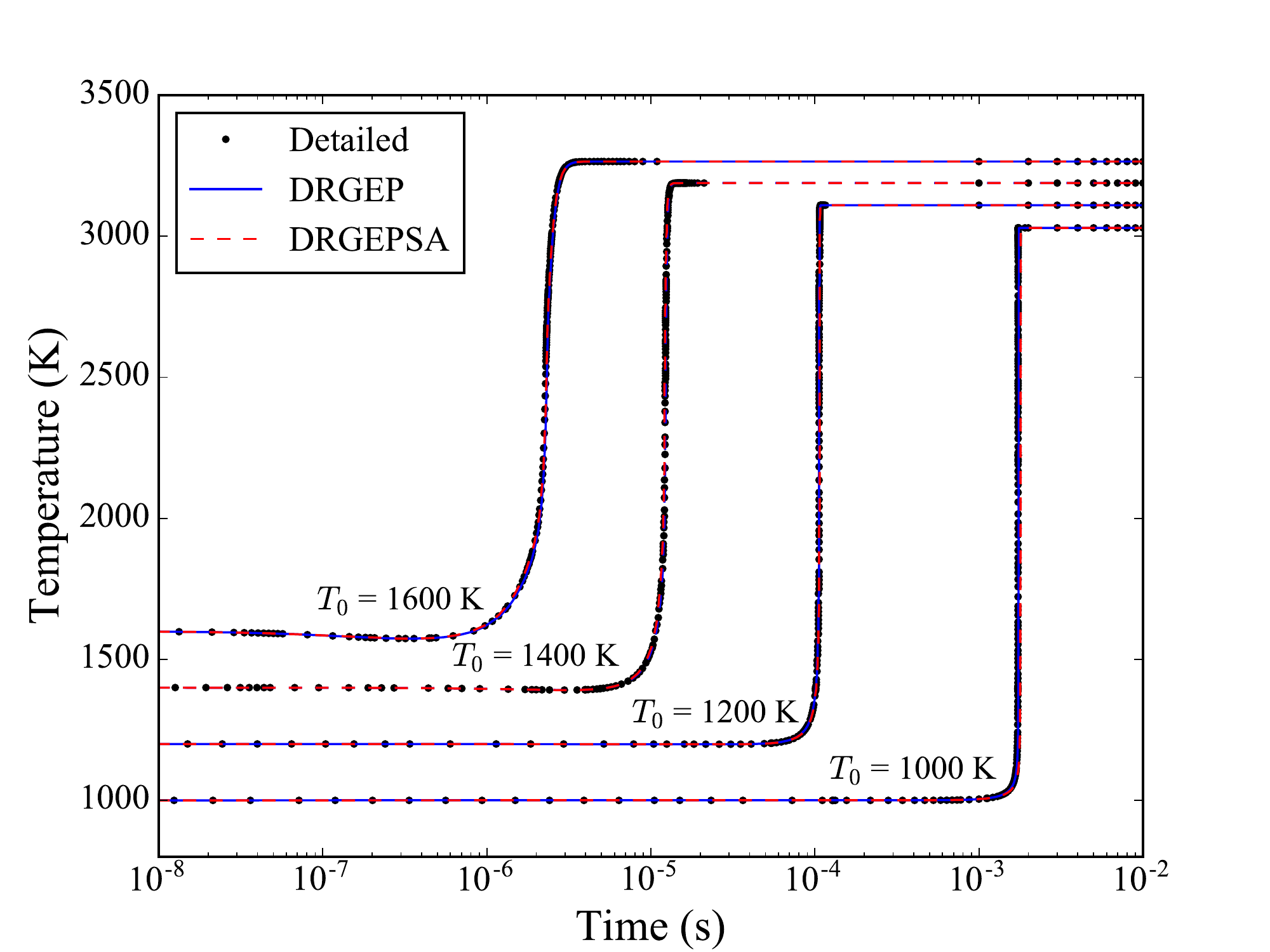}
        \caption{$\phi = 1.0$}
        \label{fig:Sarathy_temperature_profile_phi1.0}
    \end{subfigure}
    \caption{Comparison of autoignition temperature profiles for \textit{n}-butanol using Sarathy detailed and skeletal mechanisms at initial temperatures of $T_o$=\SIlist{1000;1200;1400;1600}{\kelvin}; an inlet pressure of \SI{20}{\atm}; and equivalence ratios of 1.0 and 0.5 in air.}
    \label{fig:Sarathy_temperature_profile}
\end{figure}

We first performed a validation of the skeletal mechanisms in autoignition simulations over a range of initial temperatures, pressures, and equivalence ratios.
Figures~\ref{fig:Sarathy_ign_delay_comparison} and \ref{fig:Merchant_ign_delay_comparison} show and compare the ignition delay times for the butanol isomers at stoichiometric conditions predicted by detailed and skeletal mechanisms for the Sarathy and Merchant mechanisms, respectively.
As the low-temperature chemistry is not considered in this study, the ignition simulations only consider high-temperature chemistry starting from \SI{1000}{\kelvin}.
The results show that both of the DRGEP and DRGEPSA skeletal mechanisms---indicated by the solid and dashed lines, respectively---accurately predict the ignition delay times of the corresponding detailed parent over the validated temperature and pressure ranges.
Comparisons of ignition delay predictions at equivalence ratios of 0.5 and 1.5, though not shown here, demonstrate similar agreement between skeletal and detailed mechanisms for both Sarathy and Merchant mechanisms.
Time evolutions of temperature are also compared to further validate the skeletal mechanisms.
Figure~\ref{fig:Sarathy_temperature_profile} compares the time evolutions of temperature for \textit{n}-butanol at equivalence ratios of 0.5 and 1.0 to further validate the skeletal mechanisms.
For all cases, the temperature profiles are indistinguishable over the entire ignition process.
Comparisons for other isomers also show good agreement, suggesting that the ignition kinetics are properly retained through the reduction process.

\begin{figure}[htbp]
   \centering
   \begin{subfigure}[b]{0.48\textwidth}
        \includegraphics[width=\textwidth]{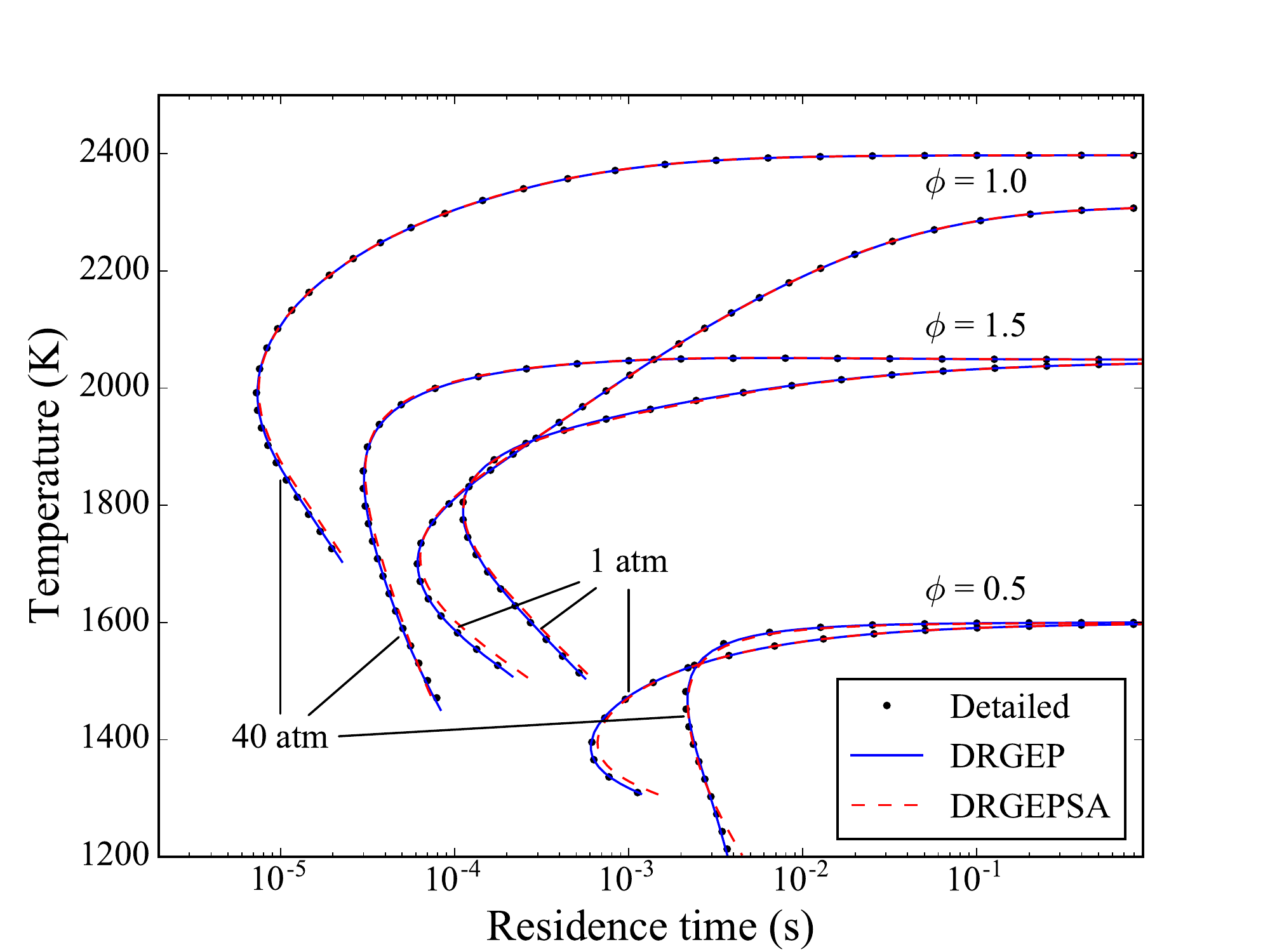}
        \caption{\textit{n}-butanol}
        \label{fig:Sarathy_nbutanol_skel_PSR}
    \end{subfigure}
    ~
    \begin{subfigure}[b]{0.48\textwidth}
        \includegraphics[width=\textwidth]{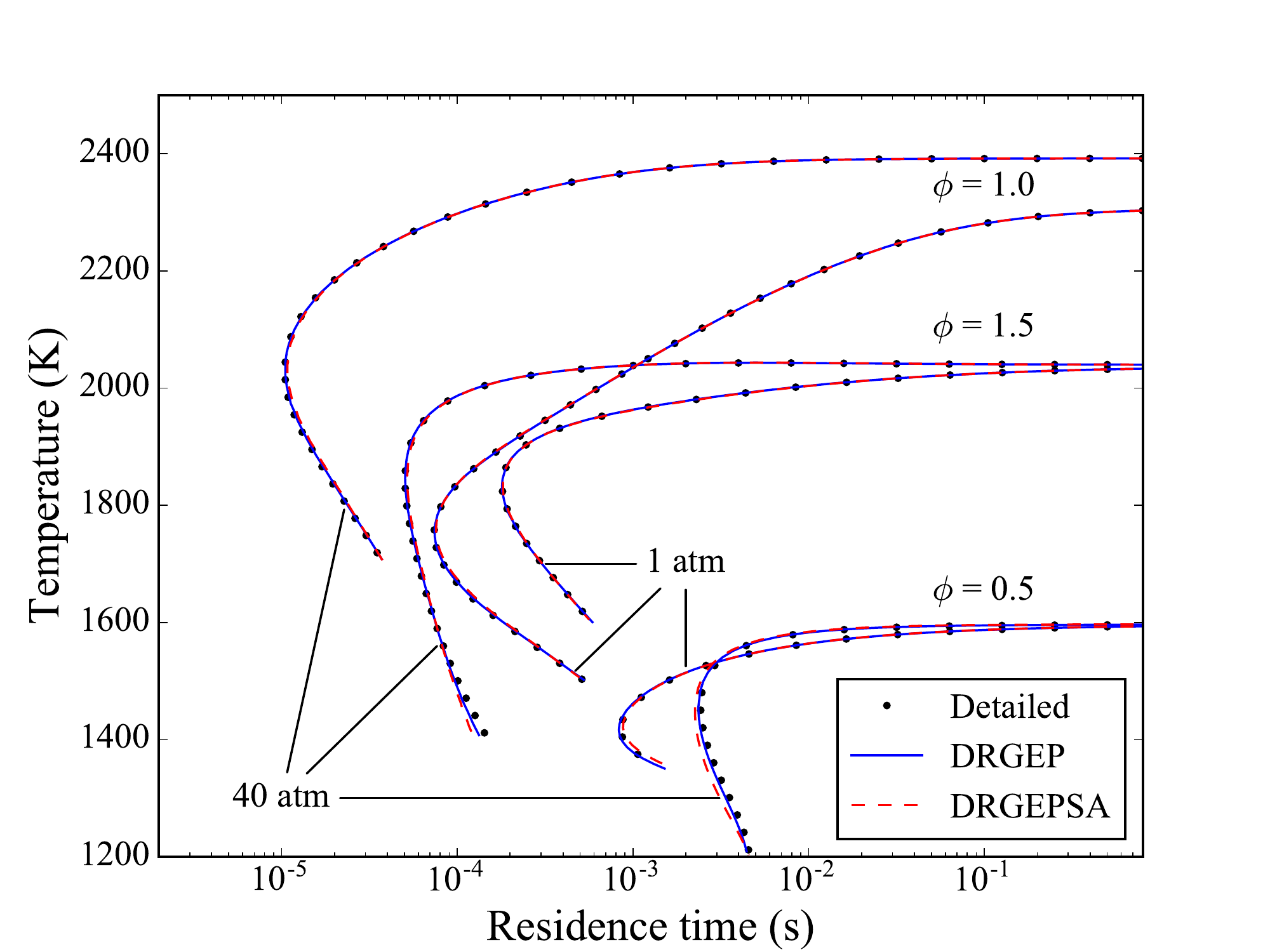}
        \caption{isobutanol}
        \label{fig:Sarathy_isobutanol_skel_PSR}
    \end{subfigure}
    \\
    \begin{subfigure}[b]{0.48\textwidth}
        \includegraphics[width=\textwidth]{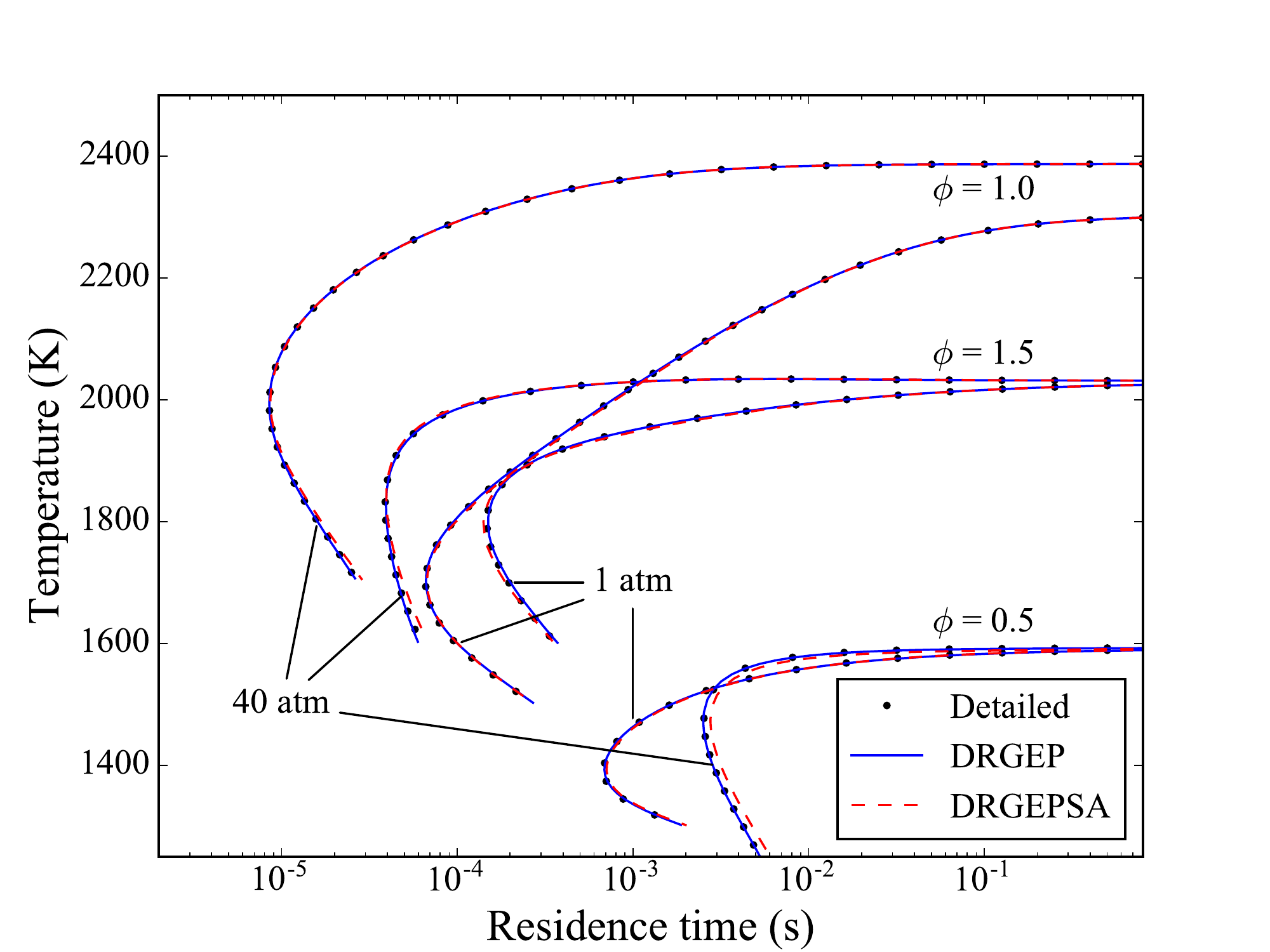}
        \caption{\textit{sec}-butanol}
        \label{fig:Sarathy_secbutanol_skel_PSR}
    \end{subfigure}
    ~
    \begin{subfigure}[b]{0.48\textwidth}
        \includegraphics[width=\textwidth]{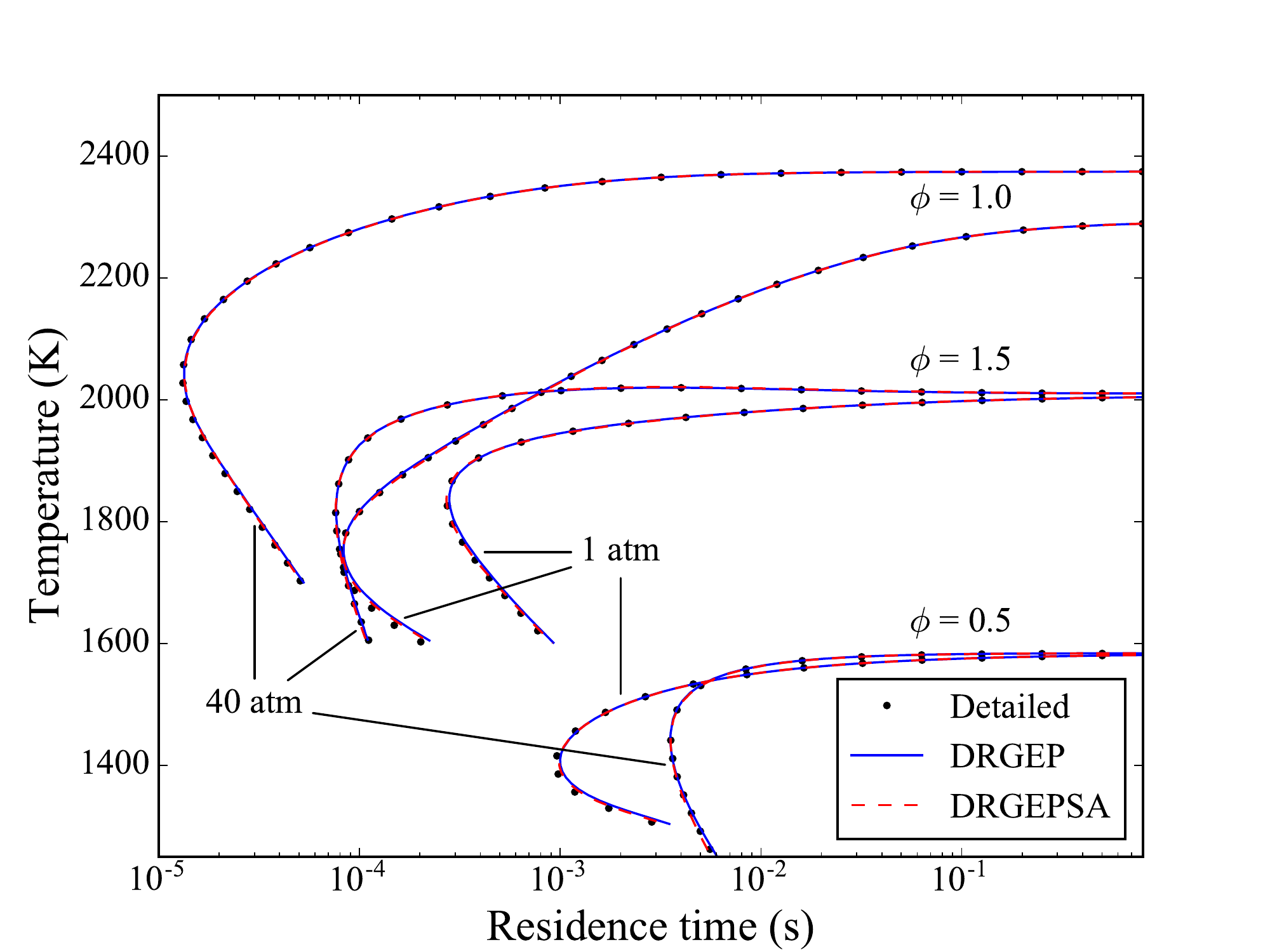}
        \caption{\textit{tert}-butanol}
        \label{fig:Sarathy_tertbutanol_skel_PSR}
    \end{subfigure}
   \caption{Comparison of PSR temperature response curves for butanol isomers using Sarathy detailed and skeletal DRGEP and DRGEPSA mechanisms at an inlet temperature of \SI{400}{\kelvin}, equivalence ratio of 1.0 in air, and pressures of \SIlist{1;40}{\atm}.}
   \label{fig:Sarathy_PSR}
\end{figure}

\begin{figure}[htbp]
   \centering
   \begin{subfigure}[b]{0.48\textwidth}
        \includegraphics[width=\textwidth]{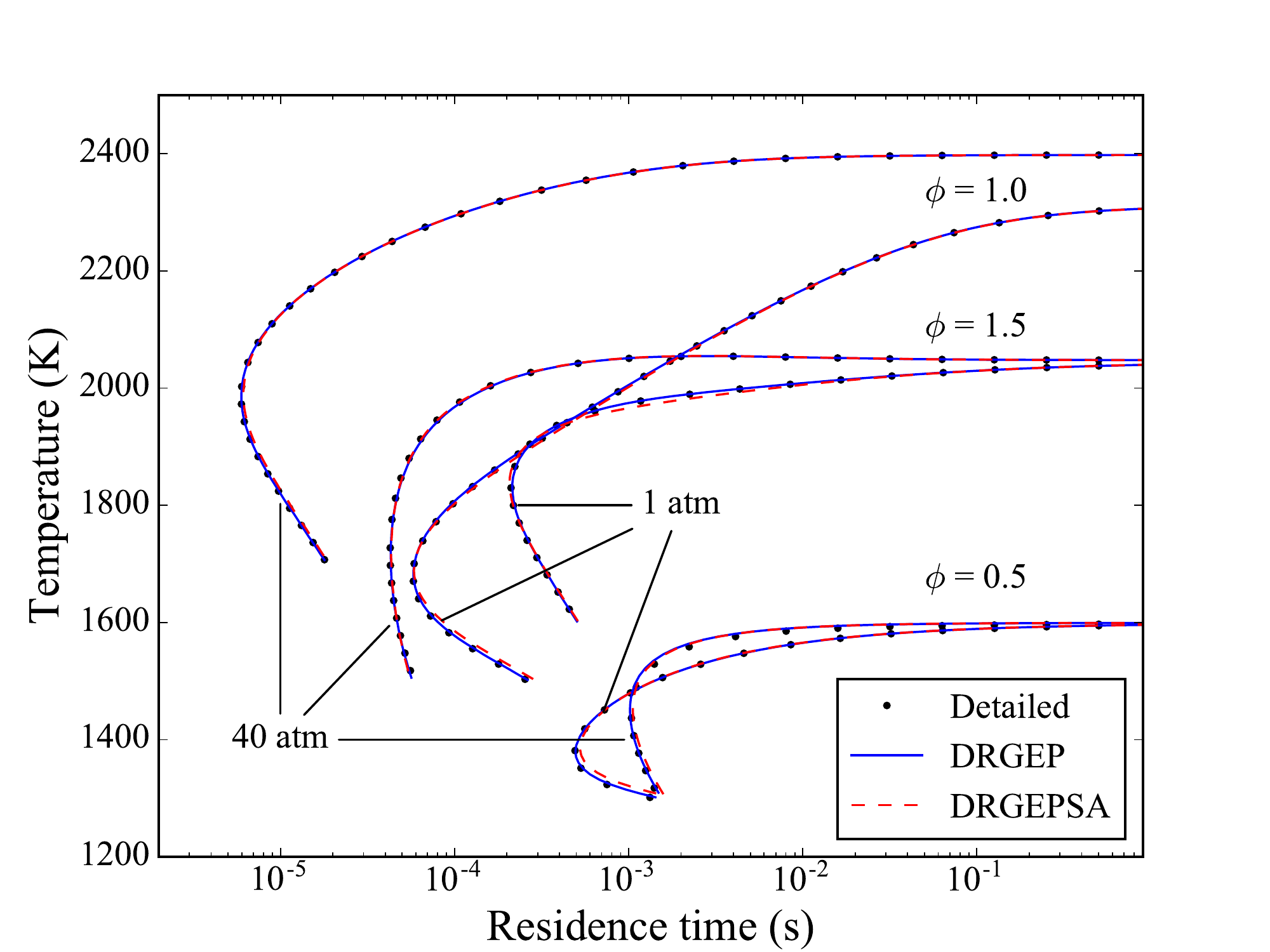}
        \caption{\textit{n}-butanol}
        \label{fig:Merchant_nbutanol_skel_PSR}
    \end{subfigure}
    ~
    \begin{subfigure}[b]{0.48\textwidth}
        \includegraphics[width=\textwidth]{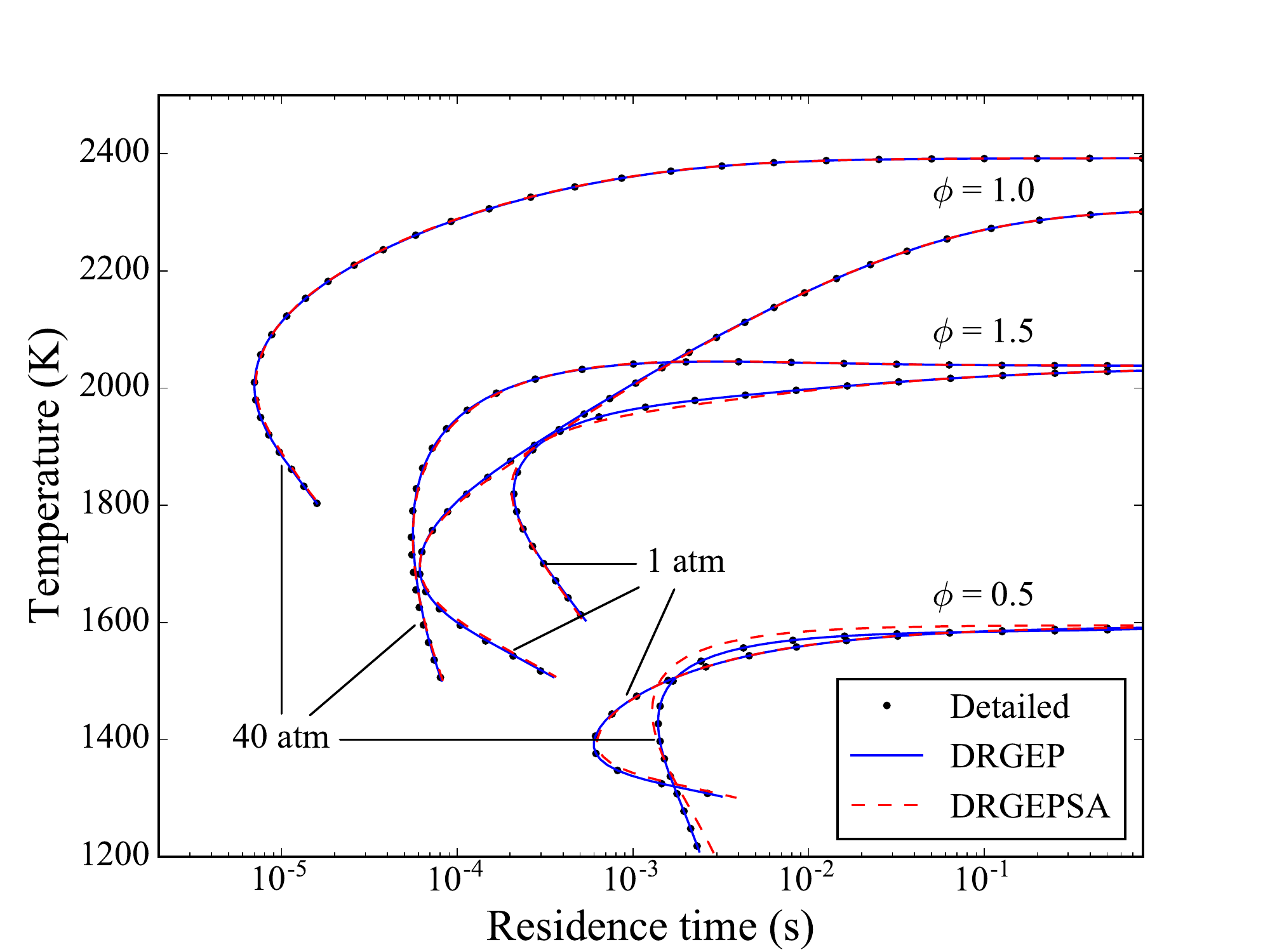}
        \caption{isobutanol}
        \label{fig:Merchant_isobutanol_skel_PSR}
    \end{subfigure}
    \\
    \begin{subfigure}[b]{0.48\textwidth}
        \includegraphics[width=\textwidth]{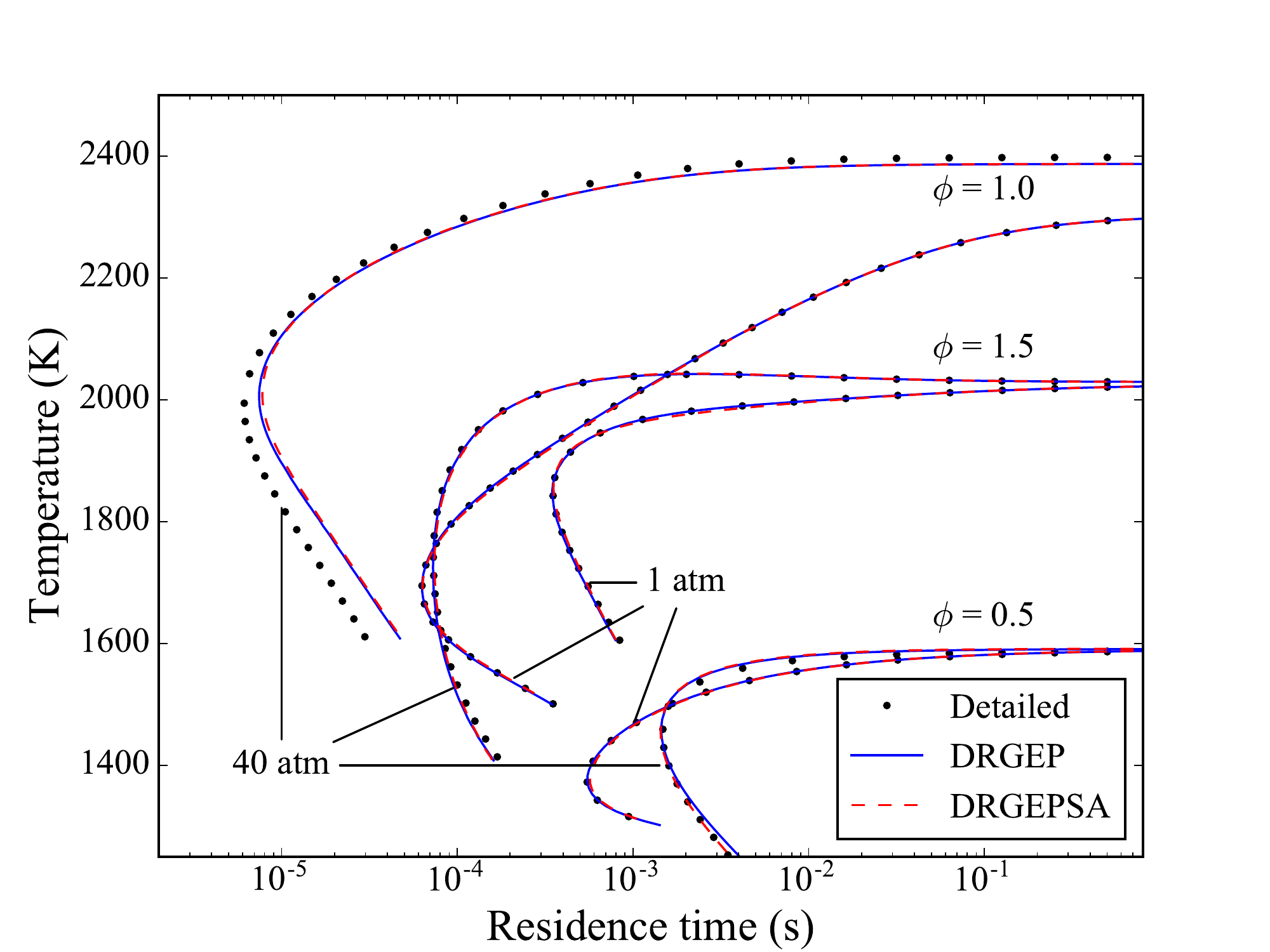}
        \caption{\textit{sec}-butanol}
        \label{fig:Merchant_secbutanol_skel_PSR}
    \end{subfigure}
    ~
    \begin{subfigure}[b]{0.48\textwidth}
        \includegraphics[width=\textwidth]{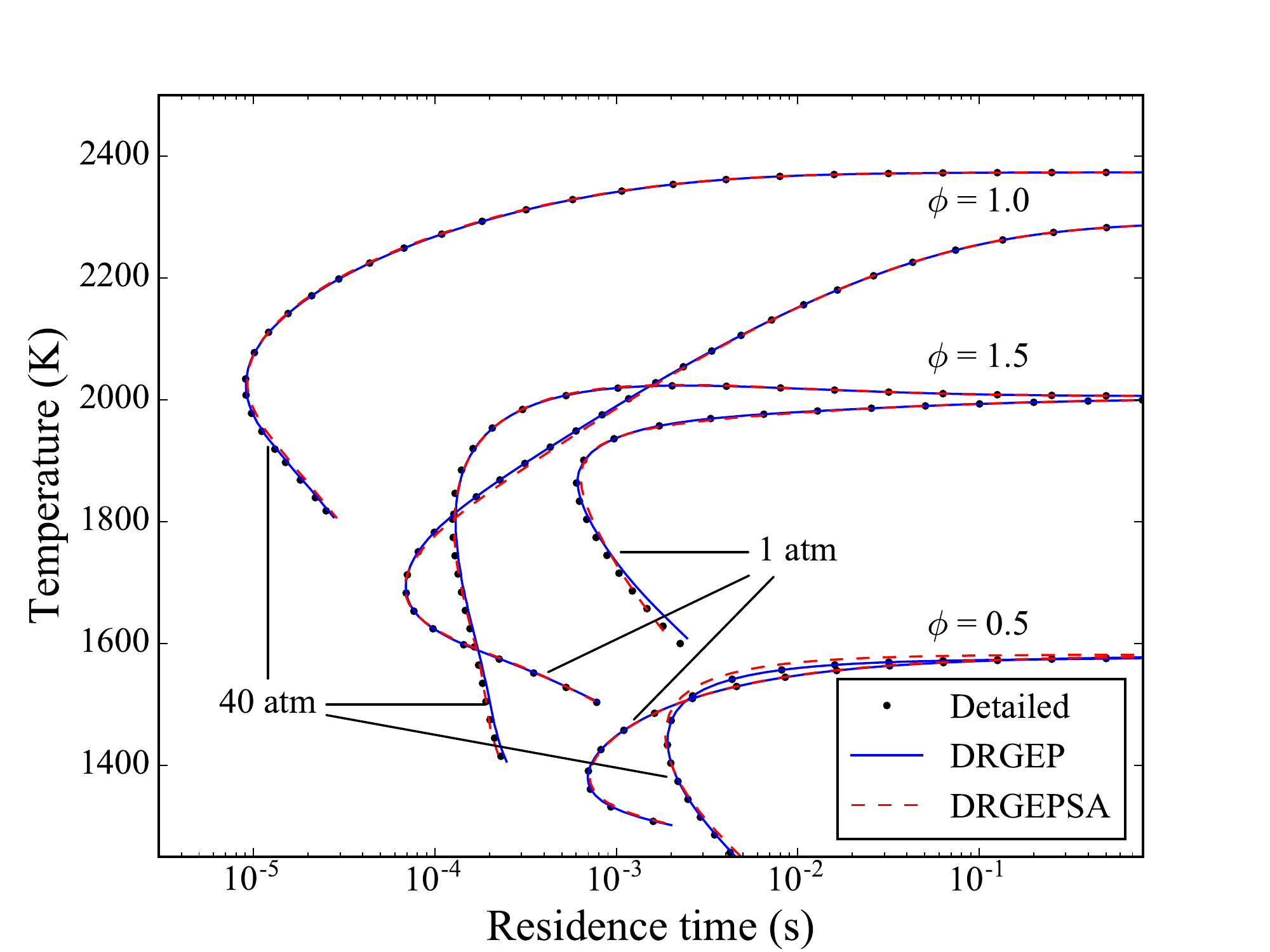}
        \caption{\textit{tert}-butanol}
        \label{fig:Merchant_tertbutanol_skel_PSR}
    \end{subfigure}
   \caption{Comparison of PSR temperature response curves for butanol isomers using Merchant detailed and skeletal DRGEP and DRGEPSA mechanisms at an inlet temperature of \SI{400}{\kelvin}, equivalence ratio of 1.0 in air, and pressures of \SIlist{1;40}{\atm}.}
   \label{fig:Merchant_PSR}
\end{figure}

Next, PSR simulations were used to validate the skeletal mechanisms in extinction phenomena.
Figures~\ref{fig:Sarathy_PSR} and \ref{fig:Merchant_PSR} show temperature response curves as a function of residence time at equivalence ratios of \SIlist{0.5;1.0;1.5} and pressures of \SIlist{1;40}{\atm} for the Sarathy and Merchant mechanisms, respectively.
The upper branch of the C-shaped temperature curve represents stable flame solutions, while the lower branch represents unstable flame solutions that are not experimentally accessible.
The turning point between the upper and lower branches represents the extinction limit, and the corresponding residence time is called the extinction residence time.
The skeletal mechanisms closely capture the temperature profiles predicted by the detailed mechanisms for all the isomers at lean, stoichiometric, and rich conditions, demonstrating the capability of the skeletal mechanisms in predicting the flame limit phenomenon.

\begin{figure}[htbp]
   \centering
   \begin{subfigure}[b]{0.48\textwidth}
        \includegraphics[width=\textwidth]{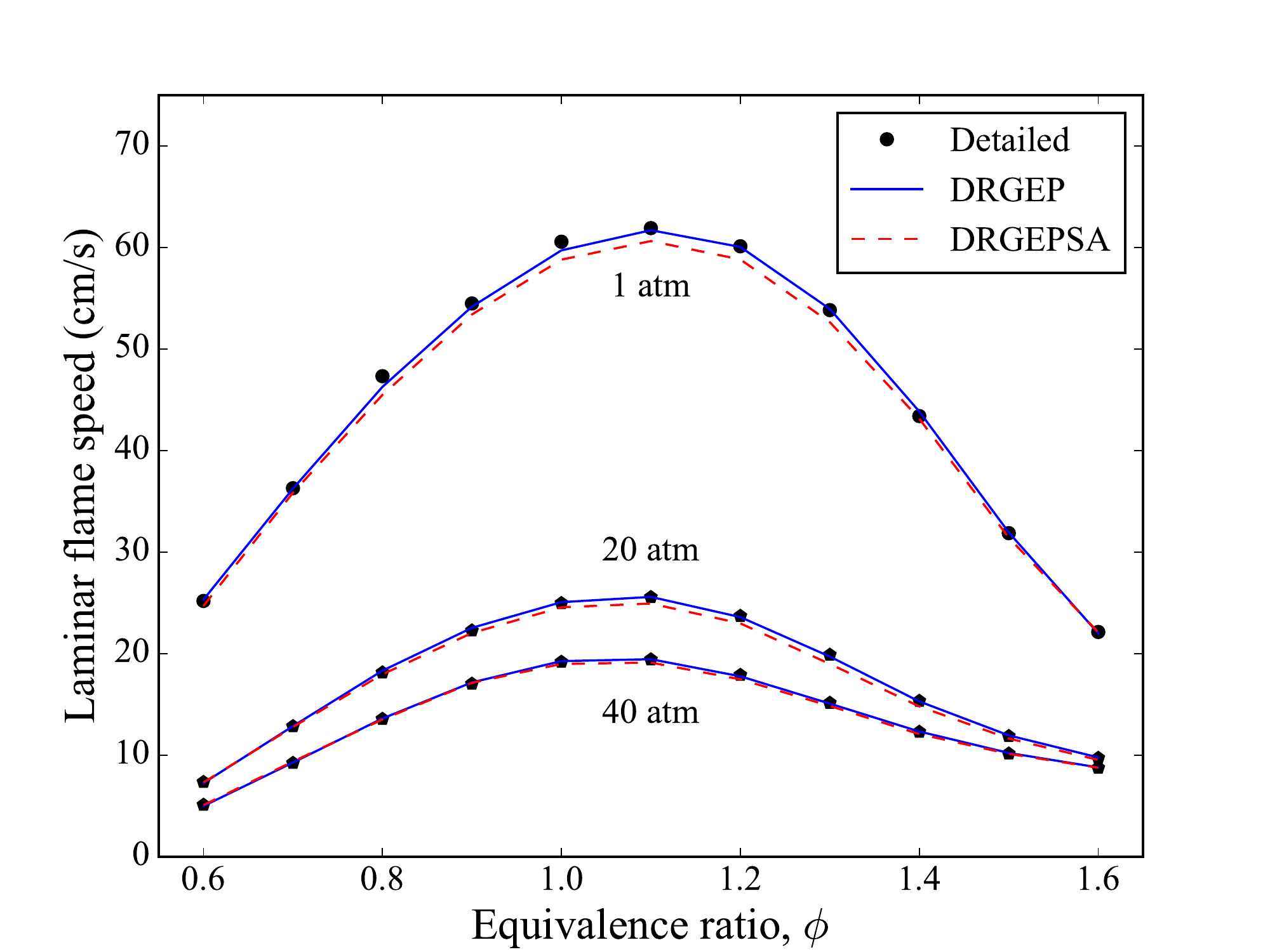}
        \caption{\textit{n}-butanol}
        \label{fig:Sarathy_nbutanol_skel_flame}
    \end{subfigure}
    ~
    \begin{subfigure}[b]{0.48\textwidth}
        \includegraphics[width=\textwidth]{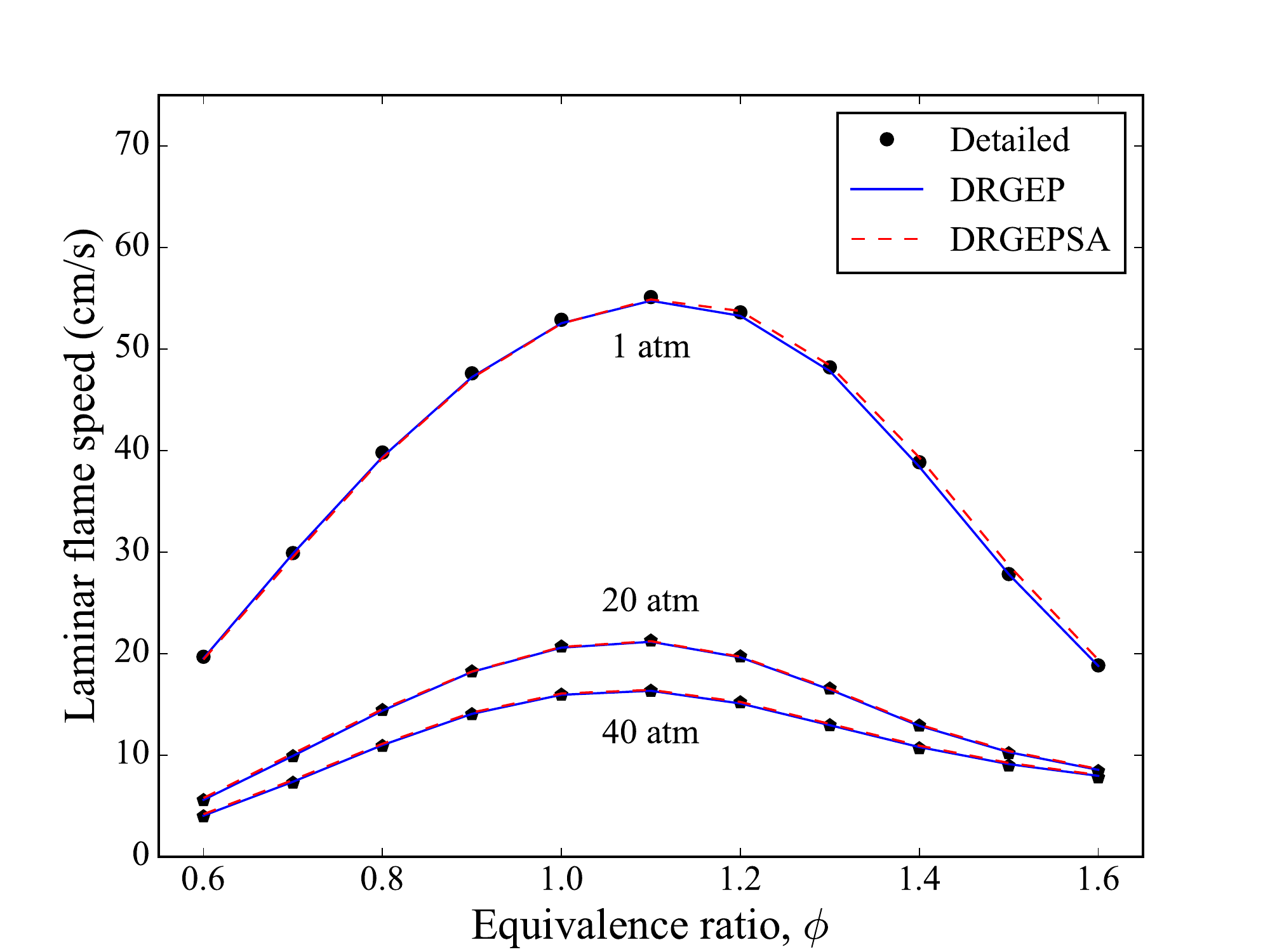}
        \caption{isobutanol}
        \label{fig:Sarathy_isobutanol_skel_flame}
    \end{subfigure}
    \\
    \begin{subfigure}[b]{0.48\textwidth}
        \includegraphics[width=\textwidth]{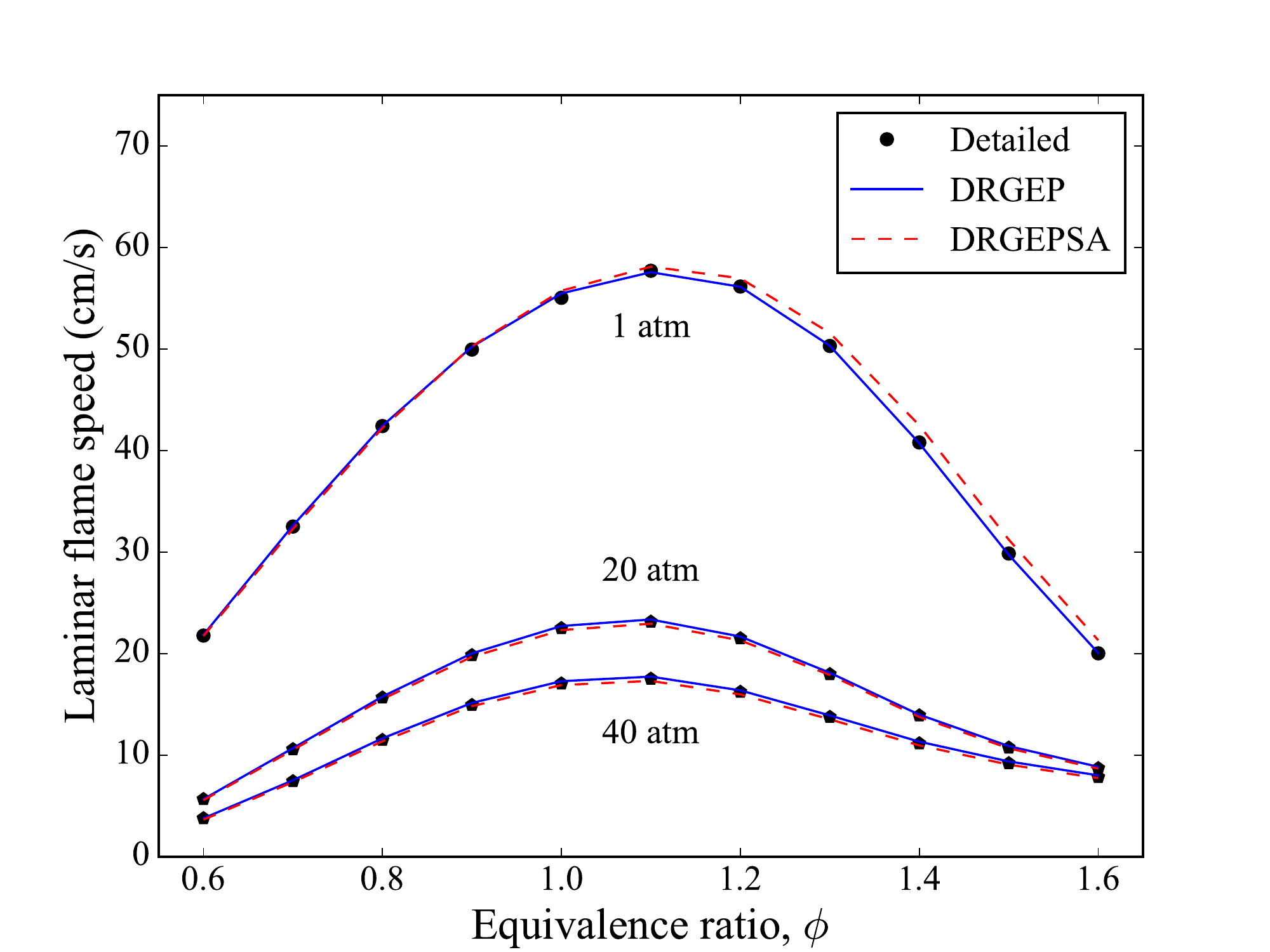}
        \caption{\textit{sec}-butanol}
        \label{fig:Sarathy_secbutanol_skel_flame}
    \end{subfigure}
    ~
    \begin{subfigure}[b]{0.48\textwidth}
        \includegraphics[width=\textwidth]{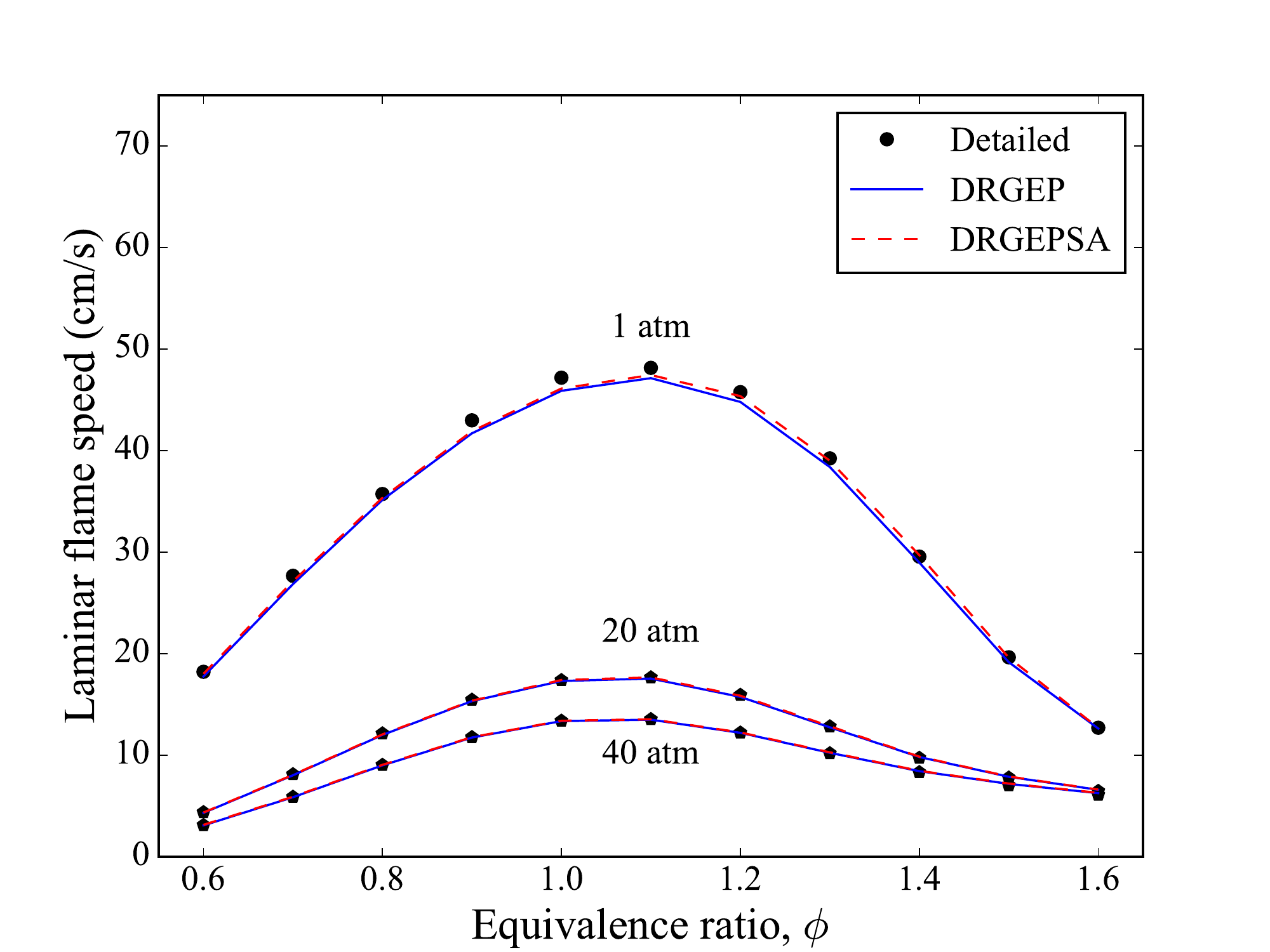}
        \caption{\textit{tert}-butanol}
        \label{fig:Sarathy_tertbutanol_skel_flame}
    \end{subfigure}
   \caption{Comparison of laminar flame speed predictions for butanol isomers as a function of equivalence ratio in air using Sarathy detailed and skeletal DRGEP and DRGEPSA mechanisms at pressures of \SIlist{1;20;40}{\atm} and an unburned mixture temperature of \SI{400}{\kelvin}.}
   \label{fig:Sarathy_flame_speed}
\end{figure}

\begin{figure}[htbp]
   \centering
    \begin{subfigure}[b]{0.48\textwidth}
         \includegraphics[width=\textwidth]{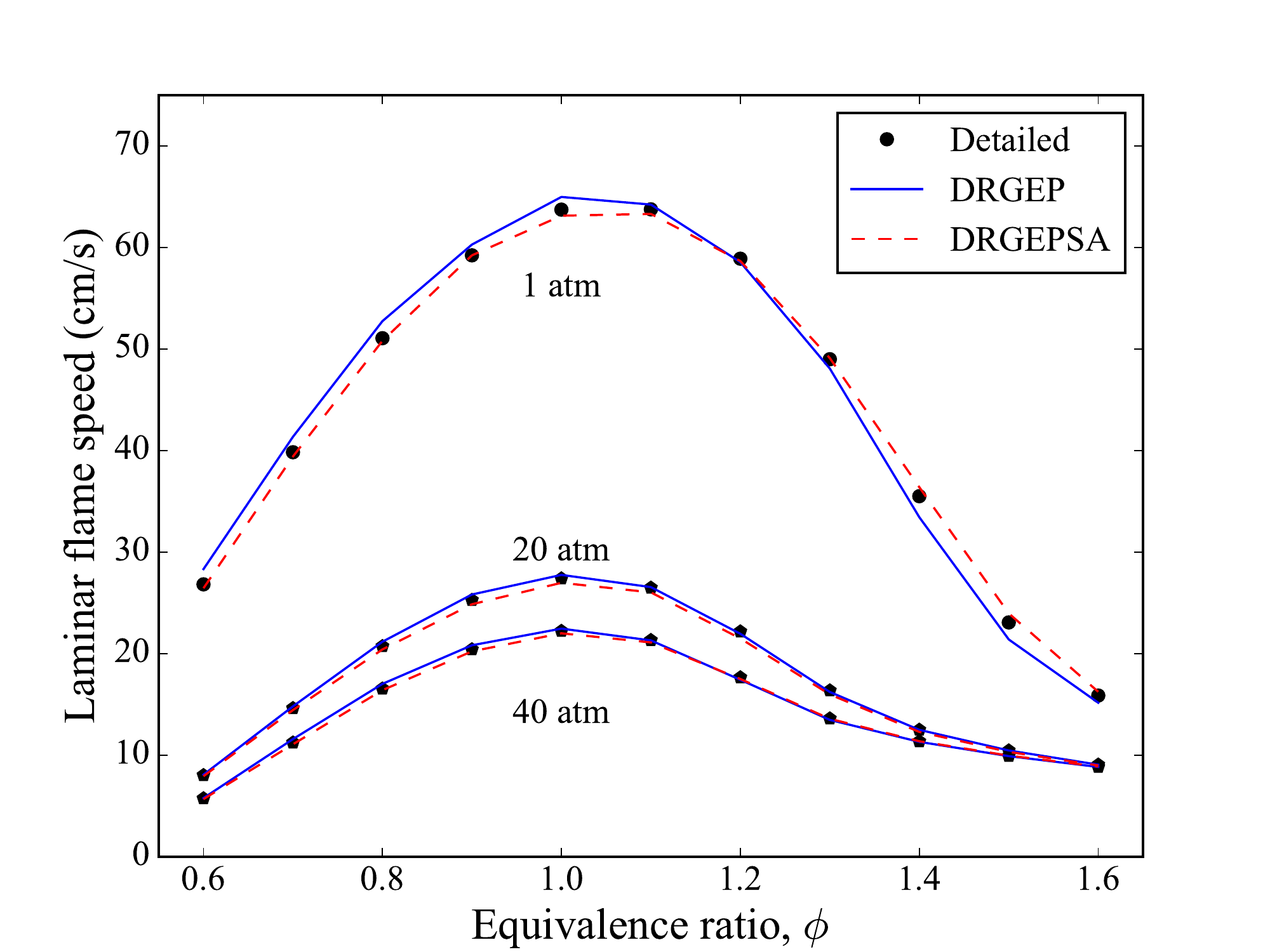}
         \caption{\textit{n}-butanol}
         \label{fig:Merchant_nbutanol_skel_flame}
     \end{subfigure}
     ~
     \begin{subfigure}[b]{0.48\textwidth}
         \includegraphics[width=\textwidth]{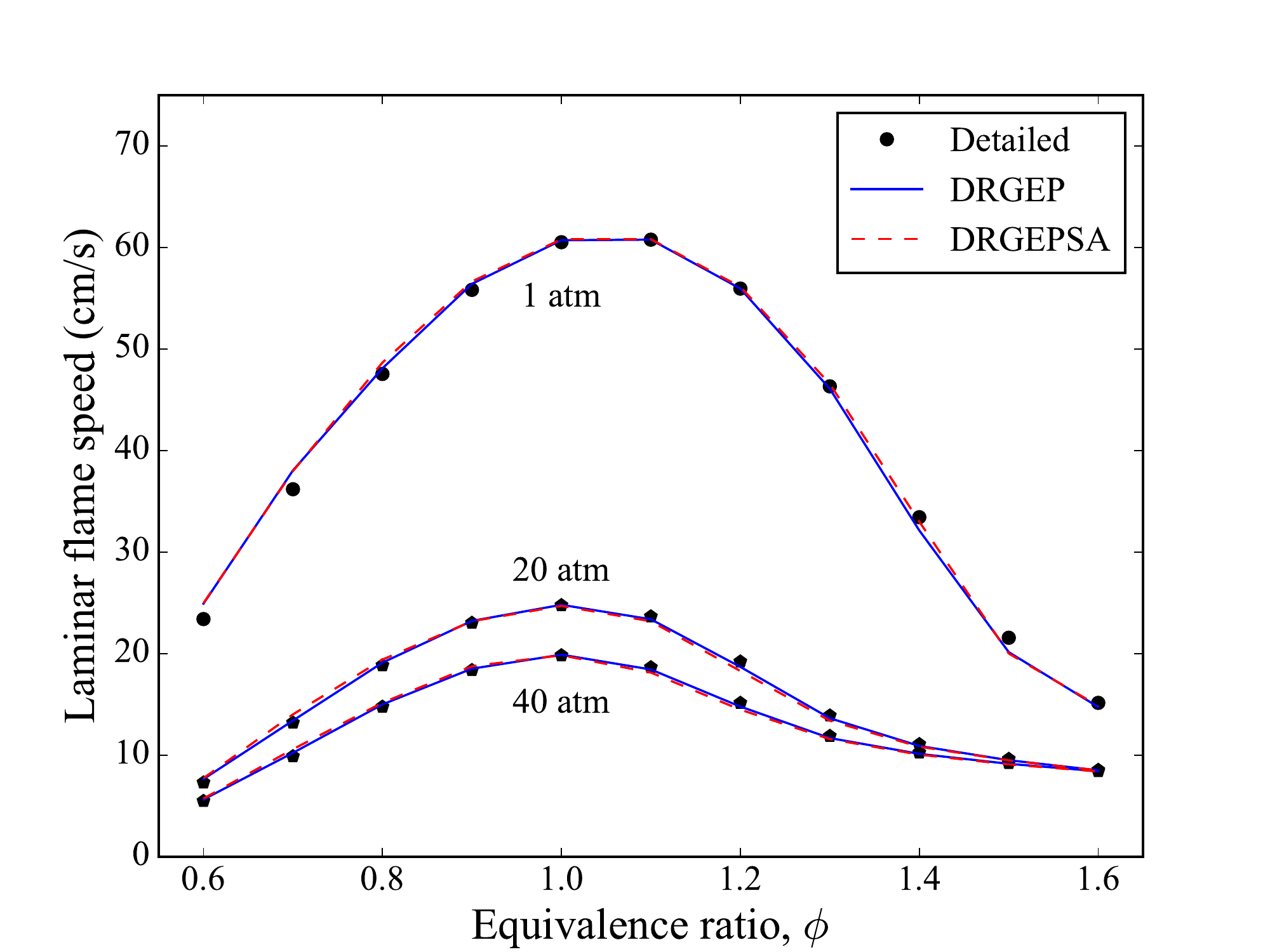}
         \caption{isobutanol}
         \label{fig:Merchant_isobutanol_skel_flame}
     \end{subfigure}
     \\
     \begin{subfigure}[b]{0.48\textwidth}
         \includegraphics[width=\textwidth]{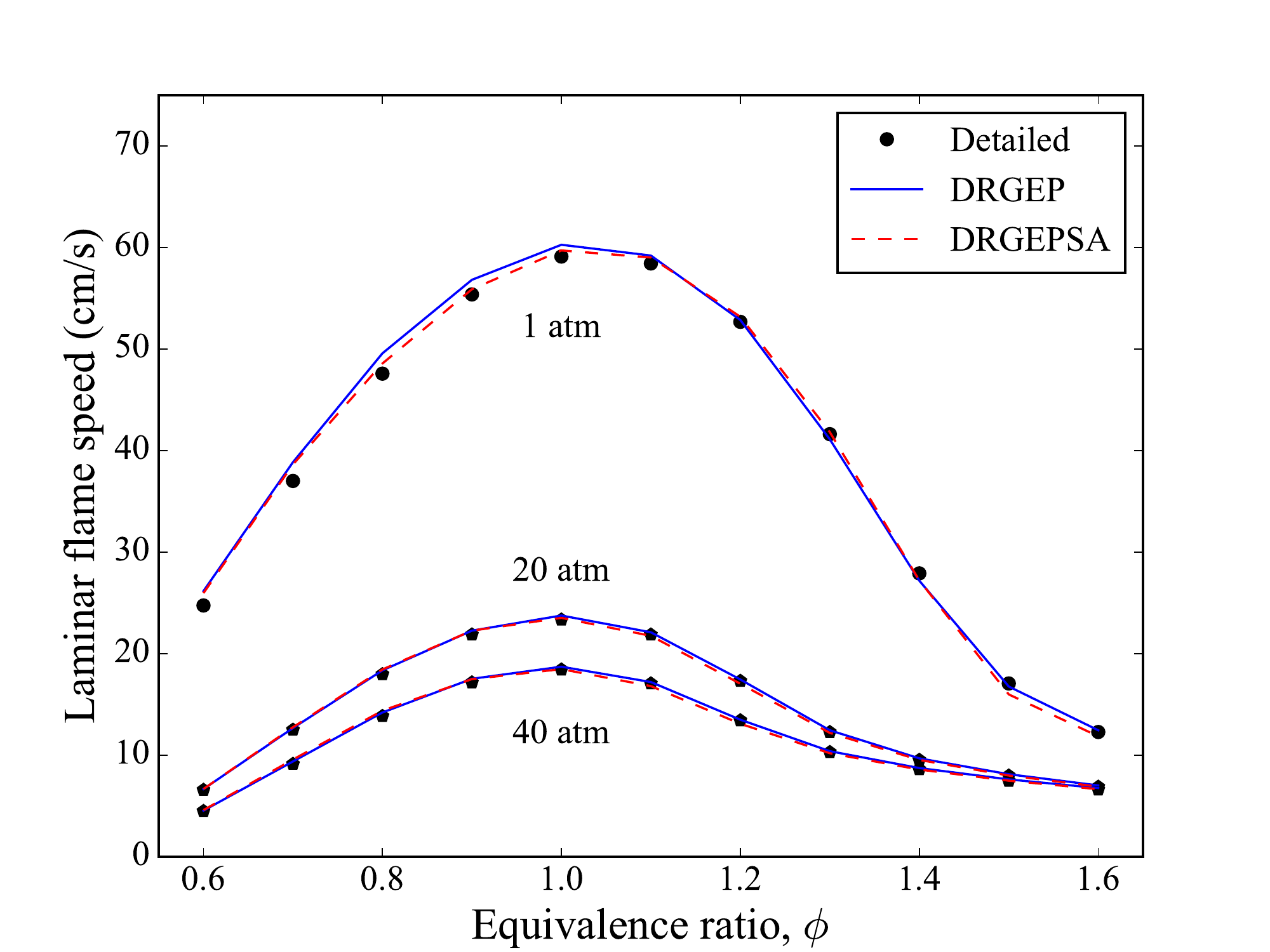}
         \caption{\textit{sec}-butanol}
         \label{fig:Merchant_secbutanol_skel_flame}
     \end{subfigure}
     ~
     \begin{subfigure}[b]{0.48\textwidth}
         \includegraphics[width=\textwidth]{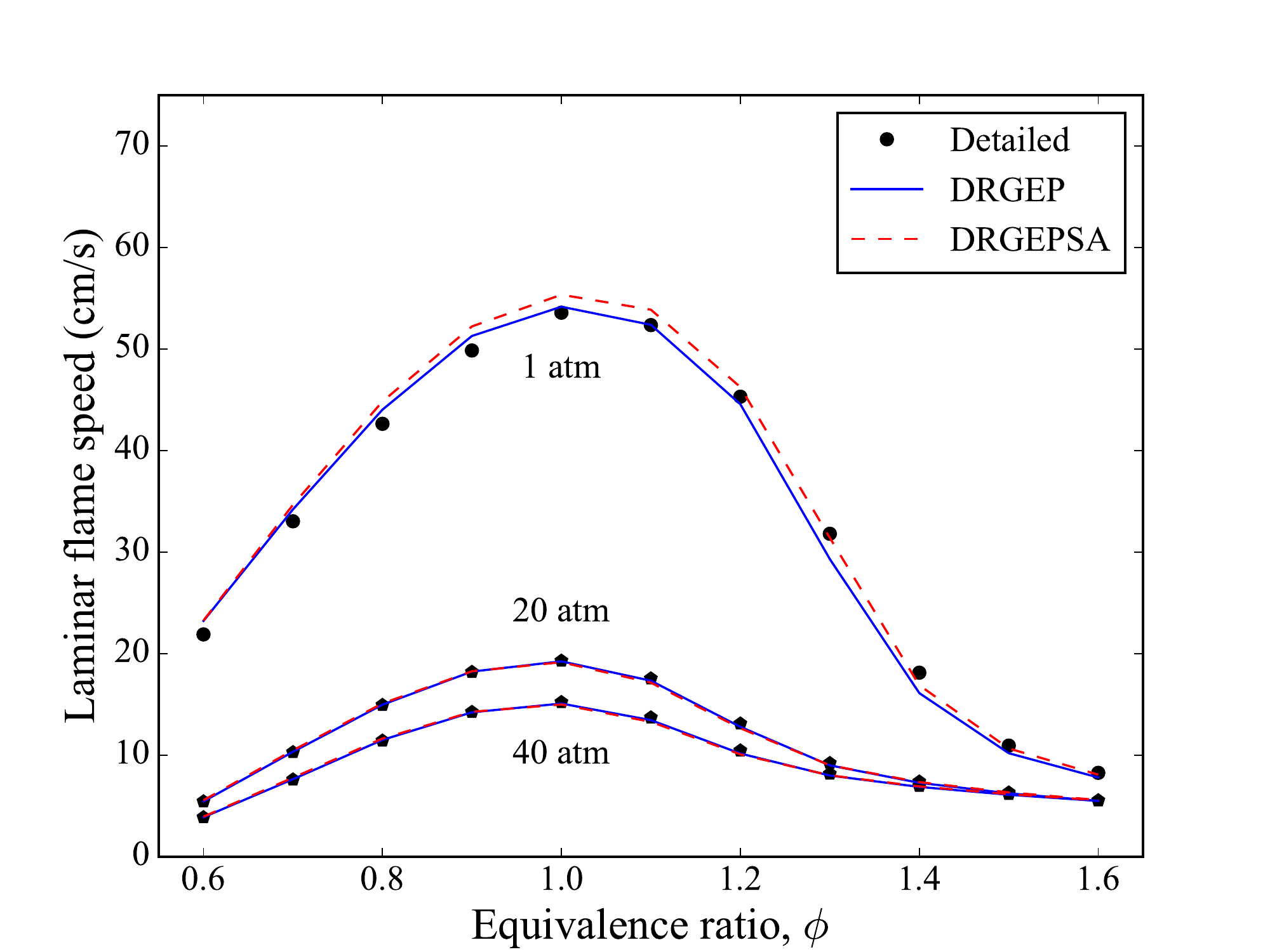}
         \caption{\textit{tert}-butanol}
         \label{fig:Merchant_tertbutanol_skel_flame}
     \end{subfigure}
   \caption{Comparison of laminar flame speed predictions for butanol isomers as a function of equivalence ratio in air using Merchant detailed and skeletal DRGEP and DRGEPSA mechanisms at pressures of \SIlist{1;20;40}{\atm} and an unburned mixture temperature of \SI{400}{\kelvin}.}
   \label{fig:Merchant_flame_speed}
\end{figure}

Figures~\ref{fig:Sarathy_flame_speed} and \ref{fig:Merchant_flame_speed} compare the simulated laminar flame speeds obtained by the detailed and skeletal mechanisms at an unburned mixture temperature of \SI{400}{\kelvin} and pressures of \SIlist{1;20;40}{\atm}.
The skeletal mechanisms generally closely match the detailed mechanisms for all butanol isomers with the Sarathy and Merchant mechanisms.
Average deviations from the detailed parent are \SI{0.88}{\percent} and \SI{1.40}{\percent} for DRGEP and DRGEPSA skeletal mechanisms of the Sarathy mechanism, respectively, and \SI{1.95}{\percent} and \SI{1.99}{\percent} for the Merchant mechanism.
Calculations with the final DRGEPSA skeletal mechanisms deviate at most \SI{6.4}{\percent} for \textit{sec}-butanol with the Sarathy mechanism and \SI{7.2}{\percent} for \textit{tert}-butanol with the Merchant mechanism, respectively.

\begin{figure}[htbp]
   \centering
   \includegraphics[width=0.75\linewidth]{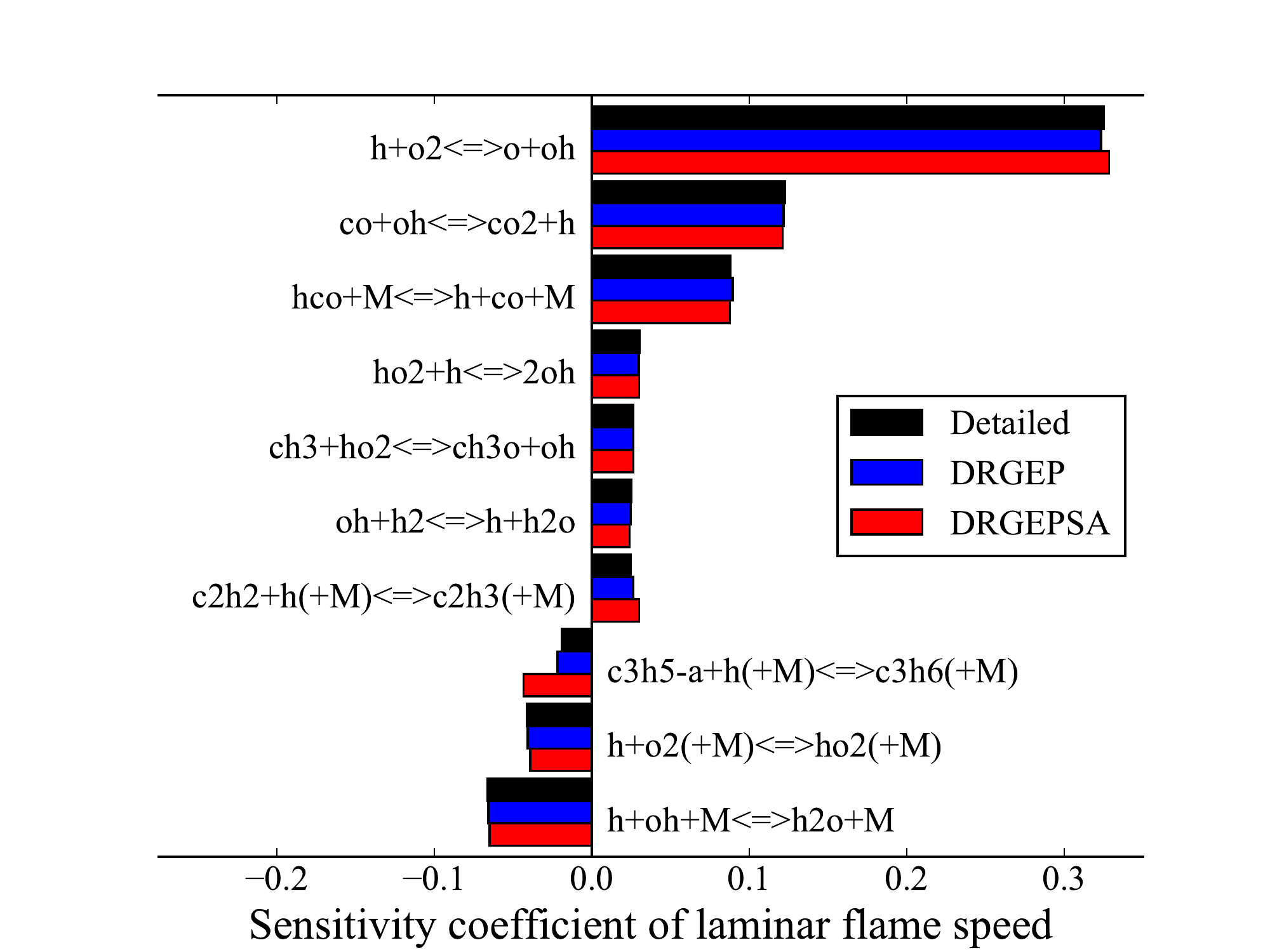}
   \caption{Sensitivity analysis of laminar flame speed for stoichiometric \textit{n}-butanol\slash air mixture using Sarathy detailed and skeletal DRGEP and DRGEPSA mechanisms at an unburned mixture temperature of \SI{400}{\kelvin} and pressure of \SI{1}{\atm}.}
   \label{fig:Sarathy_flame_sensitivity}
\end{figure}

For each mechabnism, we also performed sensitivity analyses of laminar flame speed at an unburned mixture temperature of \SI{400}{\kelvin}, pressure of \SI{1}{\atm}, and stoichiometric equivalence ratio.
Figure~\ref{fig:Sarathy_flame_sensitivity} shows the sensitivity of \textit{n}-butanol laminar flame speed with respect to pre-exponential factors using the Sarathy mechanism.
The laminar flame speed is most sensitive to reactions describing \ce{H2}\slash \ce{CO} chemistry with chain propagation\slash termination reactions involving \ce{H} and \ce{OH} radicals.
The skeletal mechanisms at both reduction levels produce similar sensitivities.
The close agreements in both laminar flame speed and sensitivities suggests that the skeletal mechanisms retain the dominant flame kinetics and important reactions through the reduction process.

\begin{figure}[htbp]
   \centering
   \includegraphics[width=0.65\linewidth]{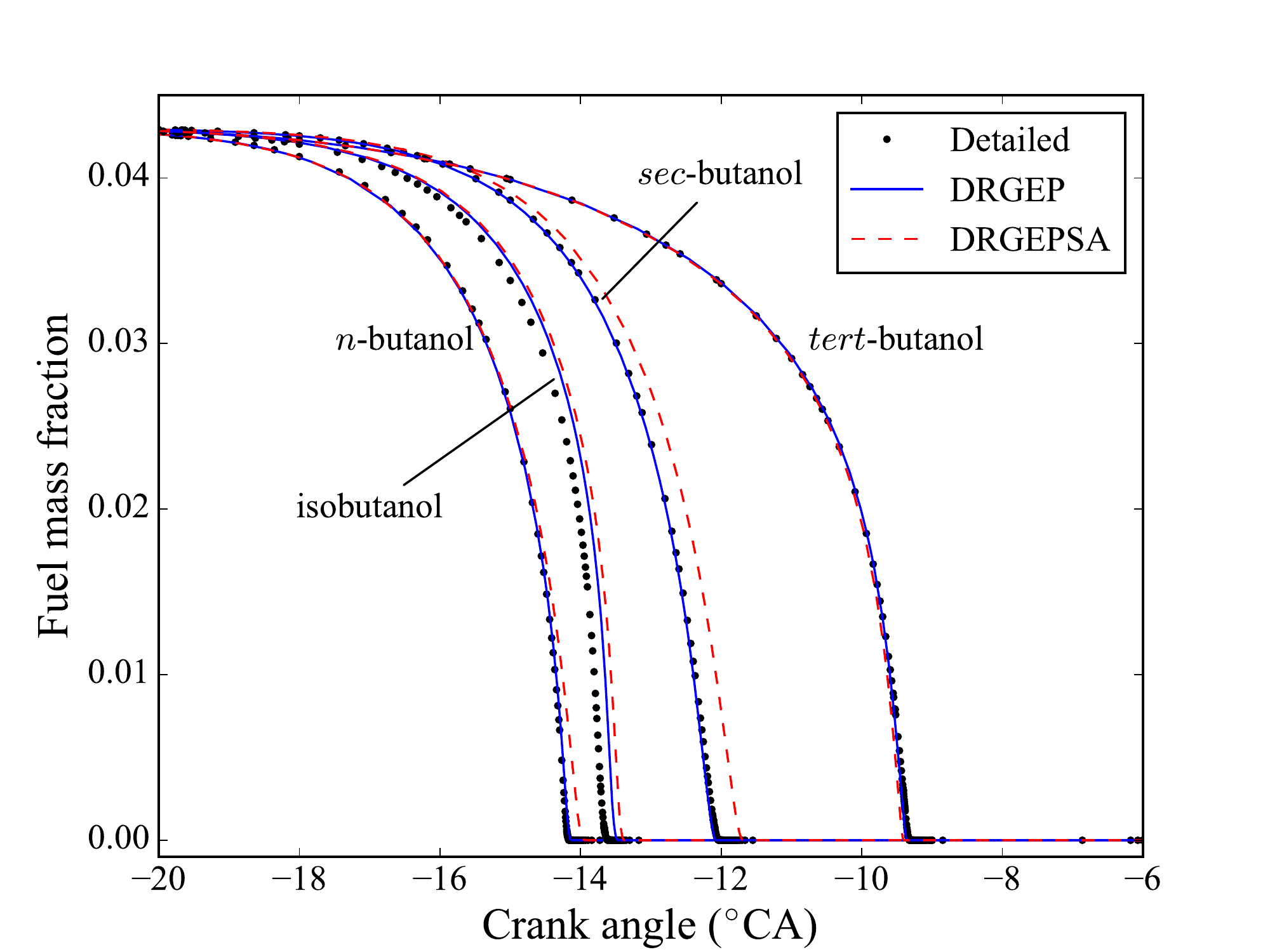}
   \caption{Mass fractions of butanol isomers in engine simulation using Sarathy detailed and skeletal DRGEP and DRGEPSA mechanisms at equivalence ratio of 0.5 in air, initial temperature of \SI{489}{\kelvin}, pressure of \SI{1.48}{\atm}, compression ratio of 14, and engine speed of \SI{1200}{\rpm}.}
   \label{fig:Sarathy_mass_fractions}
\end{figure}

\begin{figure}[htbp]
   \centering
   \includegraphics[width=0.65\linewidth]{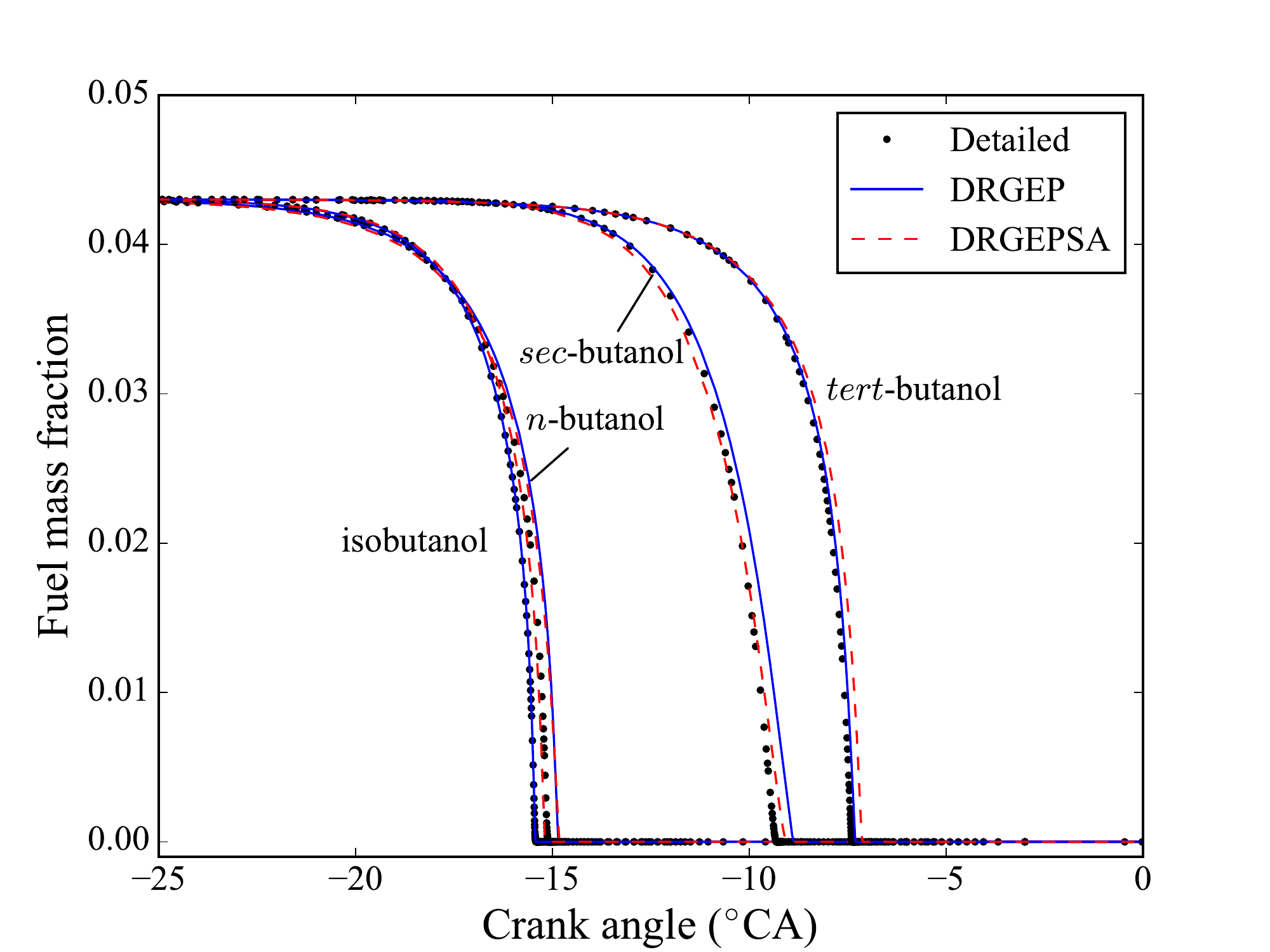}
   \caption{Mass fractions of butanol isomers in engine simulation using Merchant detailed and skeletal DRGEP and DRGEPSA mechanisms at equivalence ratio of 0.5 in air, initial temperature of \SI{489}{\kelvin}, pressure of \SI{1.48}{\atm}, compression ratio of 14, and engine speed of \SI{1200}{\rpm}.}
   \label{fig:Merchant_mass_fractions}
\end{figure}

While valuable, matching global combustion properties such as ignition delay times, extinction limits, and flame propagation does not necessarily guarantee fidelity of the skeletal mechanisms when simulating time-evolving combustion processes.
Therefore, further validation was performed for HCCI engine simulations by using an internal combustion engine simulator contained within CHEMKIN-PRO~\cite{chemkin:2012}.
The engine simulation was conducted at an equivalence ratio of 0.5, initial temperature of \SI{489}{\kelvin}, initial pressure of \SI{1.48}{\atm}, compression ratio of 14, and an engine speed of \SI{1200}{\rpm}.
Niemeyer and Sung~\cite{Niemeyer:2014} used the same conditions in a previous study of a gasoline surrogate.
Figures~\ref{fig:Sarathy_mass_fractions} and \ref{fig:Merchant_mass_fractions} show the mass fractions of butanol isomers as a function of engine crank angle (\textdegree CA) simulated by the Sarathy and Merchant mechanisms, respectively.
The simulations with the Sarathy- and Merchant-based skeletal mechanisms both agree well with the corresponding parent mechanism, both exhibiting maximum deviations of approximately \SI{0.5}{\textdegree CA} in the \textit{sec}-butanol case.

Comparing the four butanol isomers, the Sarathy mechanism predicts that \textit{n}-butanol exhibits the fastest consumption rate, and hence the shortest combustion duration (crank angles from \SI{90}{\percent} to \SI{10}{\percent} of fuel mass fraction), followed by iso, \textit{sec}-, and \textit{tert}-butanols.
The Merchant mechanism predicts that isobutanol and \textit{n}-butanol have the fastest consumption rates, followed by \textit{sec}- and \textit{tert}-butanols.
Compared to the other three isomers, the slower consumption rate of \textit{tert}-butanol predicted by both Sarathy and Merchant mechanisms can be attributed to its high activation energy; \textit{tert}-butanol's lower reactivity has been demonstrated in several experimental configurations including autoignition delay~\cite{Moss:2008bva,Stranic:2012jl,Weber:2013hs}, ignition temperature~\cite{Brady:2016dw}, and laminar flame speed~\cite{Gu:2010bo,Veloo:2011fr,Wu:2013et}.
It is also noted that \textit{tert}-butanol is currently used as an octane enhancer in gasoline.
Comparing the relative behavior of the butanol isomers between the Sarathy and Merchant mechanisms, \textit{n}-butanol exhibits a noticeably faster consumption rate than isobutanol in the Sarathy mechanism, but a somewhat slower consumption rate than isobutanol in the Merchant mechanism.
A close comparison of the experimental data on ignition delay and atmospheric laminar flame speed in Figs.~\ref{fig:ignition_validation} and \ref{fig:validation_laminar_flame}, respectively, shows that \textit{n}-butanol manifests faster ignition at near-atmospheric pressures and slightly higher laminar flame speeds than isobutanol; at high pressures, ignition delays for \textit{n}-butanol are comparable to those of isobutanol.
Previous studies on high temperature ignition delay~\cite{Stranic:2012jl,Moss:2008bva} and laminar flame speed~\cite{Gu:2010bo,Wu:2013et} showed that \textit{n}-butanol has the highest reactivity, followed by iso and \textit{sec}-butanols, and \textit{tert}-butanol has the lowest reactivity.
Other studies on forced ignition temperature~\cite{Brady:2016dw} and laminar flame speed~\cite{Veloo:2011fr} showed similar reactivity among \textit{n}-, iso, and \textit{sec}-butanols, while the reactivity of \textit{tert}-butanol is consistently lower than the other three isomers.\

\begin{figure}[htbp]
   \centering
   \includegraphics[width=0.65\linewidth]{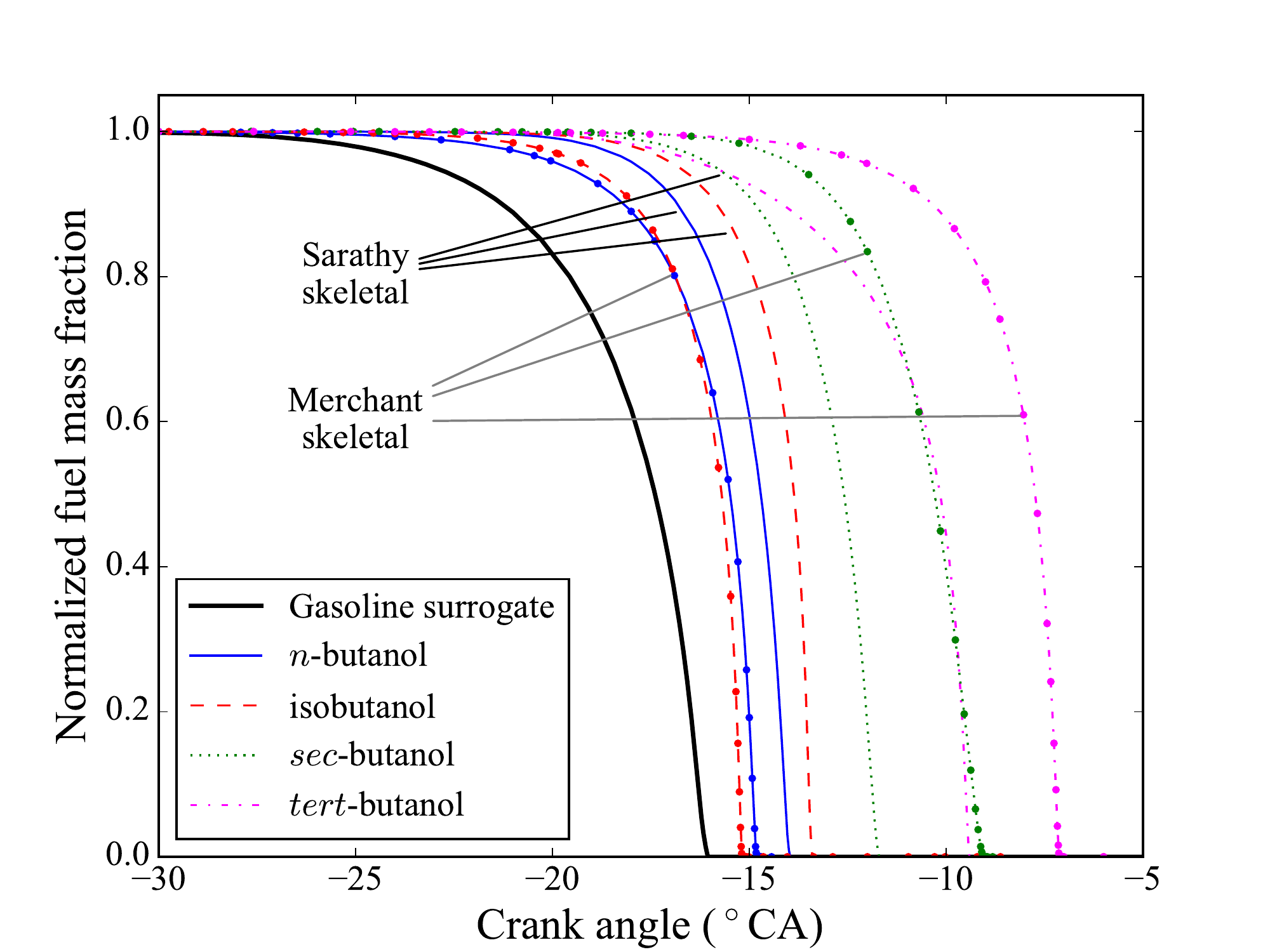}
   \caption{Comparison of normalized fuel mass fraction in engine simulations at equivalence ratio of 0.5 in air, initial temperature of \SI{489}{\kelvin}, pressure of \SI{1.48}{\atm}, compression ratio of 14, and engine speed of \SI{1200}{\rpm}. The thick black line indicates gasoline mass fraction and thinner lines indicate butanol isomer mass fractions, with symbols indicating results from the Merchant skeletal mechanisms (and lack of symbols for Sarathy skeletal mechanisms).}
   \label{fig:engine_fuel_mass}
\end{figure}

\begin{figure}[htbp]
   \centering
   \includegraphics[width=0.65\linewidth]{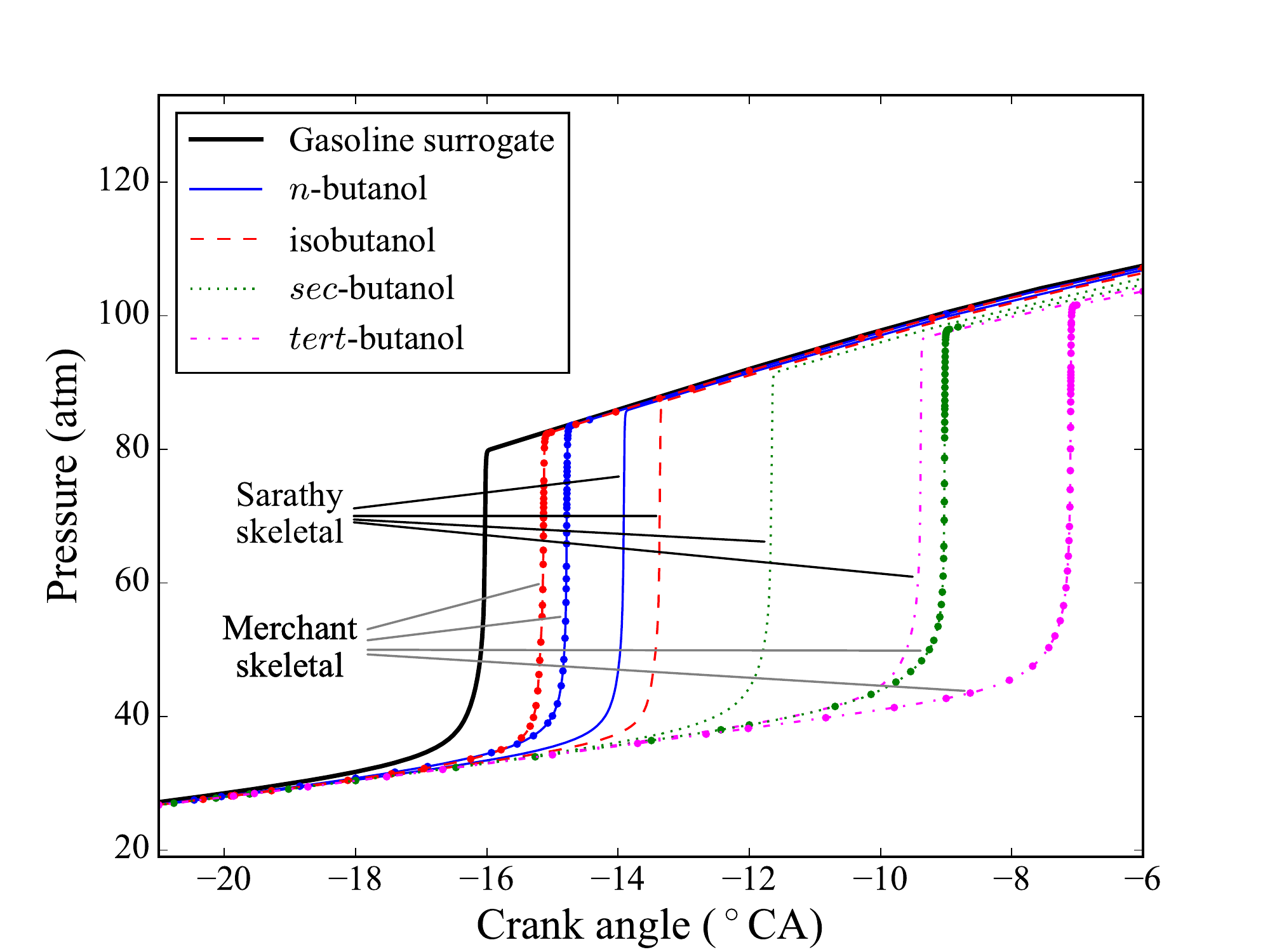}
   \caption{Comparison of pressure profiles in engine simulations at equivalence ratio of 0.5 in air, initial temperature of \SI{489}{\kelvin}, pressure of \SI{1.48}{\atm}, compression ratio of 14, and engine speed of \SI{1200}{\rpm}.
    The thick black line indicates the gasoline surrogate and thinner lines indicate butanol isomers, with symbols indicating results from Merchant skeletal mechanisms (and lack of symbols for Sarathy skeletal mechanisms).}
   \label{fig:engine_pressure}
\end{figure}

To compare the engine performance of butanol isomers with that of conventional gasoline, a skeletal mechanism~\cite{Niemeyer:2015wq} proposed for a gasoline surrogate~\cite{Mehl:2011cn,Mehl:2011jn} was used in our engine simulation for the same conditions used in the butanol isomer simulations.
This gasoline skeletal mechanism contains 97 species and 512 reactions for a gasoline surrogate consisting of \SI{48.8}{\percent} isooctane, \SI{15.3}{\percent} \emph{n}-heptane, \SI{30.6}{\percent} toluene, and \SI{5.3}{\percent} 2-pentene by mole, which has an anti-knock index of 87 and a sensitivity of 8~\cite{Mehl:2011jn}.
Figures~\ref{fig:engine_fuel_mass} and \ref{fig:engine_pressure} compare the normalized fuel mass fraction and pressure profiles of the gasoline surrogate with all four butanol isomers simulated by their respective skeletal mechanisms.
Both figures show that consumption of the gasoline surrogate initiates appreciably earlier than those of the butanol isomers, as predicted by both Sarathy and Merchant mechanisms.
Both mechanisms predict consumption of \textit{n}- and isobutanol closest to that of gasoline, although in different orders (\textit{n}-butanol and isobutanol are consumed first for the Sarathy and Merchant mechanism, respectively).
Both mechanisms predict earlier consumption of \textit{sec}-butanol than \textit{tert}-butanol, with nearly the same difference in consumption times between the two fuels.
The pressure profile comparison in Fig.~\ref{fig:engine_pressure} again shows that the gasoline surrogate ignites first, followed by iso and \textit{n}-butanols predicted by the Merchant mechanism, and then \textit{n}-, iso, \textit{sec}-, and \textit{tert}-butanols predicted by Sarathy mechanism.
The \textit{sec}-and \textit{tert}-butanols predicted by the Merchant mechanism ignite last.
In general, the observed trend somehow follows the octane rates of the gasoline surrogate~\cite{Mehl:2011jn} and butanol isomers~\cite{yano:2011,ef2010089}.
Overall, the results from both mechanisms suggest that \textit{n}-butanol and isobutanol more closely match the reactivity of the gasoline surrogate, and could thus more easily be combined with or replace gasoline in a compression-ignition engine.

\section{Concluding remarks}
\label{S:conclusions}

In the present study, skeletal mechanisms for each butanol isomer were generated by the DRGEP and DREGPSA reduction methods for the detailed mechanisms of Sarathy and coworkers~\cite{Sarathy:2012fj,Vasu:2013jc} and Merchant et al.~\cite{Merchant:2013kz}
The DRGEP method combined with an SA phase removed a significant proportion of the detailed mechanisms while maintaining their essential behaviors, resulting in final DRGEPSA skeletal mechanisms for each isomer ranging from \SIrange{18}{23}{\percent} of the size of their parent detailed mechanisms.
All of the resulting skeletal mechanisms are limited to high temperature conditions above \SI{1000}{\kelvin} and pressure up to \SI{40}{\atm}, and cover lean to rich equivalence ratios of \numrange{0.5}{1.5}.
The skeletal mechanisms generated in this study are included as supplemental material with this article.

Validation of the generated skeletal mechanisms demonstrated good matching with the detailed parent mechanisms in terms of ignition delay times, extinction residence times, and laminar flame speeds.
Further validation of the skeletal models in HCCI-like engine simulations also showed good agreement.
A comparison was also made of engine simulation results for the butanol isomers against a gasoline surrogate of an anti-knock index of 87 at an equivalence ratio of 0.5 in air, inlet temperature of \SI{489}{\kelvin} and pressure of \SI{1.48}{\atm}, with a compression ratio of 14 and engine speed of \SI{1200}{\rpm}.
While the gasoline surrogate is consumed faster than all butanol isomers for both skeletal mechanism sets, both sets predict that \textit{n}-butanol and isobutanol are consumed closest to the gasoline (albeit in different orders).
\textit{tert}-Butanol exhibits the slowest consumption rate amongst the four isomers regardless of the parent mechanisms.
For the same isomer, the Merchant mechanism predicts faster consumption rates of \textit{n}- and isobutanol than those predicted by the Sarathy mechanism, while the Sarathy mechanism predicts faster consumption of \textit{sec}- and \textit{tert}-butanol than the Merchant mechanism.
These results indicate that \textit{n}-butanol and isobutanol might be more suitable to supplement or replace gasoline than \textit{sec}- or \textit{tert}-butanol.

In addition to the skeletal mechanisms for butanol and comparisons with gasoline produced in this study, issues with pressure-dependent reactions expressed via logarithmic interpolation were also identified and discussed.
In particular, the rates of otherwise identical pressure-log reactions exhibit differences when expressed as  duplicate reactions or multiple expressions in single reaction.
Furthermore, issues can arise with negative pre-exponential Arrhenius factors in such reactions.
We therefore recommend that mechanism developers take care in selecting the formulation of pressure-log reactions.

\acknowledgement
The work at Beihang University was supported by the Project 51306010 of National Natural Science Foundation of China and Project 3152020 of Beijing Natural Science Foundation, while that at the University of Connecticut was supported as part of the Combustion Energy Frontier Research Center, an Energy Frontier Research Center funded by the U.S. Department of Energy, Office of Science, Office of Basic Energy Sciences under Award Number DE-SC0001198

\begin{suppinfo}
The butanol skeletal mechanisms generated from both the Sarathy and Merchant detailed mechanisms are available as supplementary material. 
The data and plotting scripts used to generate the figures in this paper are also available openly~\cite{Hui-data:2016}.
\end{suppinfo}

\bibliography{butanol-mechanism-paper}

\providecommand{\latin}[1]{#1}
\makeatletter
\providecommand{\doi}
  {\begingroup\let\do\@makeother\dospecials
  \catcode`\{=1 \catcode`\}=2\doi@aux}
\providecommand{\doi@aux}[1]{\endgroup\texttt{#1}}
\makeatother
\providecommand*\mcitethebibliography{\thebibliography}
\csname @ifundefined\endcsname{endmcitethebibliography}
  {\let\endmcitethebibliography\endthebibliography}{}
\begin{mcitethebibliography}{68}
\providecommand*\natexlab[1]{#1}
\providecommand*\mciteSetBstSublistMode[1]{}
\providecommand*\mciteSetBstMaxWidthForm[2]{}
\providecommand*\mciteBstWouldAddEndPuncttrue
  {\def\EndOfBibitem{\unskip.}}
\providecommand*\mciteBstWouldAddEndPunctfalse
  {\let\EndOfBibitem\relax}
\providecommand*\mciteSetBstMidEndSepPunct[3]{}
\providecommand*\mciteSetBstSublistLabelBeginEnd[3]{}
\providecommand*\EndOfBibitem{}
\mciteSetBstSublistMode{f}
\mciteSetBstMaxWidthForm{subitem}{(\alph{mcitesubitemcount})}
\mciteSetBstSublistLabelBeginEnd
  {\mcitemaxwidthsubitemform\space}
  {\relax}
  {\relax}

\bibitem[Nigam and Singh(2011)Nigam, and Singh]{Nigam:2011aa}
Nigam,~P.~S.; Singh,~A. Production of liquid biofuels from renewable resources.
  \emph{Prog. Energy Combust. Sci.} \textbf{2011}, \emph{37}, 52--68\relax
\mciteBstWouldAddEndPuncttrue
\mciteSetBstMidEndSepPunct{\mcitedefaultmidpunct}
{\mcitedefaultendpunct}{\mcitedefaultseppunct}\relax
\EndOfBibitem
\bibitem[Bergthorson and Thomson(2015)Bergthorson, and
  Thomson]{Bergthorson:2015dg}
Bergthorson,~J.~M.; Thomson,~M.~J. A review of the combustion and emissions
  properties of advanced transportation biofuels and their impact on existing
  and future engines. \emph{Renewable Sustainable Energy Rev.} \textbf{2015},
  \emph{42}, 1393--1417\relax
\mciteBstWouldAddEndPuncttrue
\mciteSetBstMidEndSepPunct{\mcitedefaultmidpunct}
{\mcitedefaultendpunct}{\mcitedefaultseppunct}\relax
\EndOfBibitem
\bibitem[Heufer \latin{et~al.}(2011)Heufer, Fernandes, Olivier, Beeckmann,
  R{\"o}hl, and Peters]{Heufer:2011jh}
Heufer,~K.~A.; Fernandes,~R.~X.; Olivier,~H.; Beeckmann,~J.; R{\"o}hl,~O.;
  Peters,~N. Shock tube investigations of ignition delays of \emph{n}-butanol
  at elevated pressures between 770 and 1250 {K}. \emph{Proc. Combust. Inst.}
  \textbf{2011}, \emph{33}, 359--366\relax
\mciteBstWouldAddEndPuncttrue
\mciteSetBstMidEndSepPunct{\mcitedefaultmidpunct}
{\mcitedefaultendpunct}{\mcitedefaultseppunct}\relax
\EndOfBibitem
\bibitem[Weber \latin{et~al.}(2011)Weber, Kumar, Zhang, and Sung]{Weber:2011fv}
Weber,~B.~W.; Kumar,~K.; Zhang,~Y.; Sung,~C.~J. Autoignition of
  \emph{n}-butanol at elevated pressure and low-to-intermediate temperature.
  \emph{Combust. Flame} \textbf{2011}, \emph{158}, 809--819\relax
\mciteBstWouldAddEndPuncttrue
\mciteSetBstMidEndSepPunct{\mcitedefaultmidpunct}
{\mcitedefaultendpunct}{\mcitedefaultseppunct}\relax
\EndOfBibitem
\bibitem[Stranic \latin{et~al.}(2012)Stranic, Chase, Harmon, Yang, Davidson,
  and Hanson]{Stranic:2012jl}
Stranic,~I.; Chase,~D.~P.; Harmon,~J.~T.; Yang,~S.; Davidson,~D.~F.;
  Hanson,~R.~K. Shock tube measurements of ignition delay times for the butanol
  isomers. \emph{Combust. Flame} \textbf{2012}, \emph{159}, 516--527\relax
\mciteBstWouldAddEndPuncttrue
\mciteSetBstMidEndSepPunct{\mcitedefaultmidpunct}
{\mcitedefaultendpunct}{\mcitedefaultseppunct}\relax
\EndOfBibitem
\bibitem[Weber and Sung(2013)Weber, and Sung]{Weber:2013hs}
Weber,~B.~W.; Sung,~C.~J. Comparative autoignition trends in butanol isomers at
  elevated pressure. \emph{Energy Fuels} \textbf{2013}, \emph{27},
  1688--1698\relax
\mciteBstWouldAddEndPuncttrue
\mciteSetBstMidEndSepPunct{\mcitedefaultmidpunct}
{\mcitedefaultendpunct}{\mcitedefaultseppunct}\relax
\EndOfBibitem
\bibitem[Moss \latin{et~al.}(2008)Moss, Berkowitz, Oehlschlaeger, Biet, Warth,
  Glaude, and Battin-Leclerc]{Moss:2008bva}
Moss,~J.~T.; Berkowitz,~A.~M.; Oehlschlaeger,~M.~A.; Biet,~J.; Warth,~V.;
  Glaude,~P.-A.; Battin-Leclerc,~F. An experimental and kinetic modeling study
  of the oxidation of the four isomers of butanol. \emph{J. Phys. Chem. A}
  \textbf{2008}, \emph{112}, 10843--10855\relax
\mciteBstWouldAddEndPuncttrue
\mciteSetBstMidEndSepPunct{\mcitedefaultmidpunct}
{\mcitedefaultendpunct}{\mcitedefaultseppunct}\relax
\EndOfBibitem
\bibitem[Brady \latin{et~al.}(2016)Brady, Hui, and Sung]{Brady:2016dw}
Brady,~K.~B.; Hui,~X.; Sung,~C.~J. Comparative study of the counterflow forced
  ignition of the butanol isomers at atmospheric and elevated pressures.
  \emph{Combust. Flame} \textbf{2016}, \emph{165}, 34--49\relax
\mciteBstWouldAddEndPuncttrue
\mciteSetBstMidEndSepPunct{\mcitedefaultmidpunct}
{\mcitedefaultendpunct}{\mcitedefaultseppunct}\relax
\EndOfBibitem
\bibitem[Brady \latin{et~al.}(2015)Brady, Hui, Sung, and
  Niemeyer]{Brady:2015fk}
Brady,~K.~B.; Hui,~X.; Sung,~C.~J.; Niemeyer,~K.~E. Counterflow ignition of
  n-butanol at atmospheric and elevated pressures. \emph{Combust. Flame}
  \textbf{2015}, \emph{162}, 3596--3611\relax
\mciteBstWouldAddEndPuncttrue
\mciteSetBstMidEndSepPunct{\mcitedefaultmidpunct}
{\mcitedefaultendpunct}{\mcitedefaultseppunct}\relax
\EndOfBibitem
\bibitem[Liu \latin{et~al.}(2011)Liu, Kelly, and Law]{Liu2011995}
Liu,~W.; Kelly,~A.~P.; Law,~C.~K. Non-premixed ignition, laminar flame
  propagation, and mechanism reduction of n-butanol, iso-butanol, and methyl
  butanoate. \emph{Proc. Combust. Inst.} \textbf{2011}, \emph{33},
  995--1002\relax
\mciteBstWouldAddEndPuncttrue
\mciteSetBstMidEndSepPunct{\mcitedefaultmidpunct}
{\mcitedefaultendpunct}{\mcitedefaultseppunct}\relax
\EndOfBibitem
\bibitem[Sarathy \latin{et~al.}(2009)Sarathy, Thomson, Togbe, Dagaut, Halter,
  and Mouna{\"\i}m-Rousselle]{Sarathy:2009js}
Sarathy,~S.~M.; Thomson,~M.~J.; Togbe,~C.; Dagaut,~P.; Halter,~F.;
  Mouna{\"\i}m-Rousselle,~C. An experimental and kinetic modeling study of
  \emph{n}-butanol combustion. \emph{Combust. Flame} \textbf{2009}, \emph{156},
  852--864\relax
\mciteBstWouldAddEndPuncttrue
\mciteSetBstMidEndSepPunct{\mcitedefaultmidpunct}
{\mcitedefaultendpunct}{\mcitedefaultseppunct}\relax
\EndOfBibitem
\bibitem[Broustail \latin{et~al.}(2011)Broustail, Seers, Halter, Mor{\'e}ac,
  and Mouna{\"\i}m-Rousselle]{Broustail:2011ez}
Broustail,~G.; Seers,~P.; Halter,~F.; Mor{\'e}ac,~G.;
  Mouna{\"\i}m-Rousselle,~C. Experimental determination of laminar burning
  velocity for butanol and ethanol iso-octane blends. \emph{Fuel}
  \textbf{2011}, \emph{90}, 1--6\relax
\mciteBstWouldAddEndPuncttrue
\mciteSetBstMidEndSepPunct{\mcitedefaultmidpunct}
{\mcitedefaultendpunct}{\mcitedefaultseppunct}\relax
\EndOfBibitem
\bibitem[Gu \latin{et~al.}(2010)Gu, Huang, Wu, and Li]{Gu:2010bo}
Gu,~X.; Huang,~Z.; Wu,~S.; Li,~Q. Laminar burning velocities and flame
  instabilities of butanol isomers--air mixtures. \emph{Combust. Flame}
  \textbf{2010}, \emph{157}, 2318--2325\relax
\mciteBstWouldAddEndPuncttrue
\mciteSetBstMidEndSepPunct{\mcitedefaultmidpunct}
{\mcitedefaultendpunct}{\mcitedefaultseppunct}\relax
\EndOfBibitem
\bibitem[Veloo and Egolfopoulos(2011)Veloo, and Egolfopoulos]{Veloo:2011fr}
Veloo,~P.~S.; Egolfopoulos,~F.~N. Flame propagation of butanol isomers/air
  mixtures. \emph{Proc. Combust. Inst.} \textbf{2011}, \emph{33},
  987--993\relax
\mciteBstWouldAddEndPuncttrue
\mciteSetBstMidEndSepPunct{\mcitedefaultmidpunct}
{\mcitedefaultendpunct}{\mcitedefaultseppunct}\relax
\EndOfBibitem
\bibitem[Wu and Law(2013)Wu, and Law]{Wu:2013et}
Wu,~F.; Law,~C.~K. An experimental and mechanistic study on the laminar flame
  speed, {Markstein} length and flame chemistry of the butanol isomers.
  \emph{Combust. Flame} \textbf{2013}, \emph{160}, 2744--2756\relax
\mciteBstWouldAddEndPuncttrue
\mciteSetBstMidEndSepPunct{\mcitedefaultmidpunct}
{\mcitedefaultendpunct}{\mcitedefaultseppunct}\relax
\EndOfBibitem
\bibitem[Agathou and Kyritsis(2012)Agathou, and Kyritsis]{Agathou:2012ey}
Agathou,~M.~S.; Kyritsis,~D.~C. Experimental investigation of bio-butanol
  laminar non-premixed flamelets. \emph{Appl. Energy} \textbf{2012}, \emph{93},
  296--304\relax
\mciteBstWouldAddEndPuncttrue
\mciteSetBstMidEndSepPunct{\mcitedefaultmidpunct}
{\mcitedefaultendpunct}{\mcitedefaultseppunct}\relax
\EndOfBibitem
\bibitem[Lefkowitz \latin{et~al.}(2012)Lefkowitz, Heyne, Won, Dooley, Kim,
  Haas, Jahangirian, Dryer, and Ju]{Lefkowitz:2012do}
Lefkowitz,~J.~K.; Heyne,~J.~S.; Won,~S.~H.; Dooley,~S.; Kim,~H.~H.;
  Haas,~F.~M.; Jahangirian,~S.; Dryer,~F.~L.; Ju,~Y. A chemical kinetic study
  of \emph{tertiary}-butanol in a flow reactor and a counterflow diffusion
  flame. \emph{Combust. Flame} \textbf{2012}, \emph{159}, 968--978\relax
\mciteBstWouldAddEndPuncttrue
\mciteSetBstMidEndSepPunct{\mcitedefaultmidpunct}
{\mcitedefaultendpunct}{\mcitedefaultseppunct}\relax
\EndOfBibitem
\bibitem[Sarathy \latin{et~al.}(2014)Sarathy, O{\ss}wald, Hansen, and
  Kohse-H{\"o}inghaus]{Sarathy:2014iq}
Sarathy,~S.~M.; O{\ss}wald,~P.; Hansen,~N.; Kohse-H{\"o}inghaus,~K. Alcohol
  combustion chemistry. \emph{Prog. Energy Combust. Sci.} \textbf{2014},
  \emph{44}, 40--102\relax
\mciteBstWouldAddEndPuncttrue
\mciteSetBstMidEndSepPunct{\mcitedefaultmidpunct}
{\mcitedefaultendpunct}{\mcitedefaultseppunct}\relax
\EndOfBibitem
\bibitem[No(2016)]{No:2016cq}
No,~S.-Y. Application of biobutanol in advanced {CI} engines {\textendash} A
  review. \emph{Fuel} \textbf{2016}, \emph{183}, 641--658\relax
\mciteBstWouldAddEndPuncttrue
\mciteSetBstMidEndSepPunct{\mcitedefaultmidpunct}
{\mcitedefaultendpunct}{\mcitedefaultseppunct}\relax
\EndOfBibitem
\bibitem[Szwaja and Naber(2010)Szwaja, and Naber]{Szwaja:2010db}
Szwaja,~S.; Naber,~J.~D. Combustion of \emph{n}-butanol in a spark-ignition
  {IC} engine. \emph{Fuel} \textbf{2010}, \emph{89}, 1573--1582\relax
\mciteBstWouldAddEndPuncttrue
\mciteSetBstMidEndSepPunct{\mcitedefaultmidpunct}
{\mcitedefaultendpunct}{\mcitedefaultseppunct}\relax
\EndOfBibitem
\bibitem[Gu \latin{et~al.}(2012)Gu, Huang, Cai, Gong, Wu, and Lee]{Gu:2012gh}
Gu,~X.; Huang,~Z.; Cai,~J.; Gong,~J.; Wu,~X.; Lee,~C.-f. Emission
  characteristics of a spark-ignition engine fuelled with
  gasoline-\emph{n}-butanol blends in combination with {EGR}. \emph{Fuel}
  \textbf{2012}, \emph{93}, 611--617\relax
\mciteBstWouldAddEndPuncttrue
\mciteSetBstMidEndSepPunct{\mcitedefaultmidpunct}
{\mcitedefaultendpunct}{\mcitedefaultseppunct}\relax
\EndOfBibitem
\bibitem[Al~Hasan and Al~Momany(2008)Al~Hasan, and Al~Momany]{AlHasan:2008cm}
Al~Hasan,~M.~I.; Al~Momany,~M. The effect of iso-butanol-diesel blends on
  engine performance. \emph{Transport} \textbf{2008}, \emph{23}, 306--310\relax
\mciteBstWouldAddEndPuncttrue
\mciteSetBstMidEndSepPunct{\mcitedefaultmidpunct}
{\mcitedefaultendpunct}{\mcitedefaultseppunct}\relax
\EndOfBibitem
\bibitem[Yao \latin{et~al.}(2010)Yao, Wang, Zheng, and Yue]{Yao:2010es}
Yao,~M.; Wang,~H.; Zheng,~Z.; Yue,~Y. Experimental study of n-butanol additive
  and multi-injection on {HD} diesel engine performance and emissions.
  \emph{Fuel} \textbf{2010}, \emph{89}, 2191--2201\relax
\mciteBstWouldAddEndPuncttrue
\mciteSetBstMidEndSepPunct{\mcitedefaultmidpunct}
{\mcitedefaultendpunct}{\mcitedefaultseppunct}\relax
\EndOfBibitem
\bibitem[Rakopoulos \latin{et~al.}(2011)Rakopoulos, Rakopoulos, Papagiannakis,
  and Kyritsis]{Rakopoulos:2011cp}
Rakopoulos,~D.~C.; Rakopoulos,~C.~D.; Papagiannakis,~R.~G.; Kyritsis,~D.~C.
  Combustion heat release analysis of ethanol or \emph{n}-butanol diesel fuel
  blends in heavy-duty {DI} diesel engine. \emph{Fuel} \textbf{2011},
  \emph{90}, 1855--1867\relax
\mciteBstWouldAddEndPuncttrue
\mciteSetBstMidEndSepPunct{\mcitedefaultmidpunct}
{\mcitedefaultendpunct}{\mcitedefaultseppunct}\relax
\EndOfBibitem
\bibitem[Saisirirat \latin{et~al.}(2011)Saisirirat, Togbe, Togbe, Chanchaona,
  Foucher, Foucher, Mounaim-Rousselle, Mouna{\"\i}m-Rousselle, Dagaut, and
  Dagaut]{Saisirirat:2011ci}
Saisirirat,~P.; Togbe,~C.; Togbe,~C.; Chanchaona,~S.; Foucher,~F.; Foucher,~F.;
  Mounaim-Rousselle,~C.; Mouna{\"\i}m-Rousselle,~C.; Dagaut,~P.; Dagaut,~P.
  Auto-ignition and combustion characteristics in {HCCI} and {JSR} using
  1-butanol/\emph{n}-heptane and ethanol/\emph{n}-heptane blends. \emph{Proc.
  Combust. Inst.} \textbf{2011}, \emph{33}, 3007--3014\relax
\mciteBstWouldAddEndPuncttrue
\mciteSetBstMidEndSepPunct{\mcitedefaultmidpunct}
{\mcitedefaultendpunct}{\mcitedefaultseppunct}\relax
\EndOfBibitem
\bibitem[He \latin{et~al.}(2013)He, Liu, Yuan, and Zhao]{He:2013bo}
He,~B.-Q.; Liu,~M.-B.; Yuan,~J.; Zhao,~H. Combustion and emission
  characteristics of a {HCCI} engine fuelled with n-butanol--gasoline blends.
  \emph{Fuel} \textbf{2013}, \emph{108}, 668--674\relax
\mciteBstWouldAddEndPuncttrue
\mciteSetBstMidEndSepPunct{\mcitedefaultmidpunct}
{\mcitedefaultendpunct}{\mcitedefaultseppunct}\relax
\EndOfBibitem
\bibitem[He \latin{et~al.}(2014)He, Yuan, Liu, and Zhao]{He:2014gl}
He,~B.-Q.; Yuan,~J.; Liu,~M.-B.; Zhao,~H. Combustion and emission
  characteristics of a \emph{n}-butanol {HCCI} engine. \emph{Fuel}
  \textbf{2014}, \emph{115}, 758--764\relax
\mciteBstWouldAddEndPuncttrue
\mciteSetBstMidEndSepPunct{\mcitedefaultmidpunct}
{\mcitedefaultendpunct}{\mcitedefaultseppunct}\relax
\EndOfBibitem
\bibitem[Dagaut \latin{et~al.}(2009)Dagaut, Sarathy, and
  Thomson]{Dagaut:2009kj}
Dagaut,~P.; Sarathy,~S.~M.; Thomson,~M.~J. A chemical kinetic study of
  \emph{n}-butanol oxidation at elevated pressure in a jet stirred reactor.
  \emph{Proc. Combust. Inst.} \textbf{2009}, \emph{32}, 229--237\relax
\mciteBstWouldAddEndPuncttrue
\mciteSetBstMidEndSepPunct{\mcitedefaultmidpunct}
{\mcitedefaultendpunct}{\mcitedefaultseppunct}\relax
\EndOfBibitem
\bibitem[Sarathy \latin{et~al.}(2012)Sarathy, Sarathy, Vranckx, Yasunaga, Mehl,
  O{\ss}wald, Metcalfe, Westbrook, Pitz, Kohse-H{\"o}inghaus, Fernandes, and
  Curran]{Sarathy:2012fj}
Sarathy,~S.~M.; Sarathy,~S.~M.; Vranckx,~S.; Yasunaga,~K.; Mehl,~M.;
  O{\ss}wald,~P.; Metcalfe,~W.~K.; Westbrook,~C.~K.; Pitz,~W.~J.;
  Kohse-H{\"o}inghaus,~K.; Fernandes,~R.~X.; Curran,~H.~J. A comprehensive
  chemical kinetic combustion model for the four butanol isomers.
  \emph{Combust. Flame} \textbf{2012}, \emph{159}, 2028--2055\relax
\mciteBstWouldAddEndPuncttrue
\mciteSetBstMidEndSepPunct{\mcitedefaultmidpunct}
{\mcitedefaultendpunct}{\mcitedefaultseppunct}\relax
\EndOfBibitem
\bibitem[Vasu and Sarathy(2013)Vasu, and Sarathy]{Vasu:2013jc}
Vasu,~S.~S.; Sarathy,~S.~M. On the high-temperature combustion of
  \emph{n}-butanol: shock tube data and an improved kinetic model. \emph{Energy
  Fuels} \textbf{2013}, \emph{27}, 7072--7080\relax
\mciteBstWouldAddEndPuncttrue
\mciteSetBstMidEndSepPunct{\mcitedefaultmidpunct}
{\mcitedefaultendpunct}{\mcitedefaultseppunct}\relax
\EndOfBibitem
\bibitem[Rosado-Reyes and Tsang(2012)Rosado-Reyes, and
  Tsang]{RosadoReyes:2012fl}
Rosado-Reyes,~C.~M.; Tsang,~W. Shock Tube Study on the Thermal Decomposition of
  \emph{n}-Butanol. \emph{J. Phys. Chem. A} \textbf{2012}, \emph{116},
  9825--9831\relax
\mciteBstWouldAddEndPuncttrue
\mciteSetBstMidEndSepPunct{\mcitedefaultmidpunct}
{\mcitedefaultendpunct}{\mcitedefaultseppunct}\relax
\EndOfBibitem
\bibitem[Frassoldati \latin{et~al.}(2012)Frassoldati, Grana, Faravelli, Ranzi,
  O{\ss}wald, and Kohse-H{\"o}inghaus]{Frassoldati:2012jn}
Frassoldati,~A.; Grana,~R.; Faravelli,~T.; Ranzi,~E.; O{\ss}wald,~P.;
  Kohse-H{\"o}inghaus,~K. Detailed kinetic modeling of the combustion of the
  four butanol isomers in premixed low-pressure flames. \emph{Combust. Flame}
  \textbf{2012}, \emph{159}, 2295--2311\relax
\mciteBstWouldAddEndPuncttrue
\mciteSetBstMidEndSepPunct{\mcitedefaultmidpunct}
{\mcitedefaultendpunct}{\mcitedefaultseppunct}\relax
\EndOfBibitem
\bibitem[Grana \latin{et~al.}(2010)Grana, Frassoldati, Faravelli, Niemann,
  Ranzi, Seiser, Cattolica, and Seshadri]{Grana:2010gk}
Grana,~R.; Frassoldati,~A.; Faravelli,~T.; Niemann,~U.; Ranzi,~E.; Seiser,~R.;
  Cattolica,~R.; Seshadri,~K. An experimental and kinetic modeling study of
  combustion of isomers of butanol. \emph{Combust. Flame} \textbf{2010},
  \emph{157}, 2137--2154\relax
\mciteBstWouldAddEndPuncttrue
\mciteSetBstMidEndSepPunct{\mcitedefaultmidpunct}
{\mcitedefaultendpunct}{\mcitedefaultseppunct}\relax
\EndOfBibitem
\bibitem[Van~Geem \latin{et~al.}(2010)Van~Geem, Pyl, Marin, Harper, and
  Green]{VanGeem:2010ca}
Van~Geem,~K.~M.; Pyl,~S.~P.; Marin,~G.~B.; Harper,~M.~R.; Green,~W.~H. Accurate
  High-Temperature Reaction Networks for Alternative Fuels: Butanol Isomers.
  \emph{Ind. Eng. Chem. Res.} \textbf{2010}, \emph{49}, 10399--10420\relax
\mciteBstWouldAddEndPuncttrue
\mciteSetBstMidEndSepPunct{\mcitedefaultmidpunct}
{\mcitedefaultendpunct}{\mcitedefaultseppunct}\relax
\EndOfBibitem
\bibitem[Merchant \latin{et~al.}(2013)Merchant, Zanoelo, Speth, Harper,
  Van~Geem, and Green]{Merchant:2013kz}
Merchant,~S.~S.; Zanoelo,~E.~F.; Speth,~R.~L.; Harper,~M.~R.; Van~Geem,~K.~M.;
  Green,~W.~H. Combustion and pyrolysis of \emph{iso}-butanol: Experimental and
  chemical kinetic modeling study. \emph{Combust. Flame} \textbf{2013},
  \emph{160}, 1907--1929\relax
\mciteBstWouldAddEndPuncttrue
\mciteSetBstMidEndSepPunct{\mcitedefaultmidpunct}
{\mcitedefaultendpunct}{\mcitedefaultseppunct}\relax
\EndOfBibitem
\bibitem[Lu and Law(2009)Lu, and Law]{Lu:2009gh}
Lu,~T.; Law,~C.~K. Toward accommodating realistic fuel chemistry in large-scale
  computations. \emph{Prog. Energy Combust. Sci.} \textbf{2009}, \emph{35},
  192--215\relax
\mciteBstWouldAddEndPuncttrue
\mciteSetBstMidEndSepPunct{\mcitedefaultmidpunct}
{\mcitedefaultendpunct}{\mcitedefaultseppunct}\relax
\EndOfBibitem
\bibitem[Tur{\'a}nyi and Tomlin(2014)Tur{\'a}nyi, and Tomlin]{Turanyi:2014aa}
Tur{\'a}nyi,~T.; Tomlin,~A.~S. \emph{Analysis of Kinetic Reaction Mechanisms};
  Springer-Verlag: Berlin Heidelberg, 2014\relax
\mciteBstWouldAddEndPuncttrue
\mciteSetBstMidEndSepPunct{\mcitedefaultmidpunct}
{\mcitedefaultendpunct}{\mcitedefaultseppunct}\relax
\EndOfBibitem
\bibitem[Lu and Law(2005)Lu, and Law]{Lu:2005ce}
Lu,~T.; Law,~C.~K. A directed relation graph method for mechanism reduction.
  \emph{Proc. Combust. Inst.} \textbf{2005}, \emph{30}, 1333--1341\relax
\mciteBstWouldAddEndPuncttrue
\mciteSetBstMidEndSepPunct{\mcitedefaultmidpunct}
{\mcitedefaultendpunct}{\mcitedefaultseppunct}\relax
\EndOfBibitem
\bibitem[Lu and Law(2006)Lu, and Law]{Lu:2006bb}
Lu,~T.; Law,~C.~K. Linear time reduction of large kinetic mechanisms with
  directed relation graph: \emph{n}-heptane and iso-octane. \emph{Combust.
  Flame} \textbf{2006}, \emph{144}, 24--36\relax
\mciteBstWouldAddEndPuncttrue
\mciteSetBstMidEndSepPunct{\mcitedefaultmidpunct}
{\mcitedefaultendpunct}{\mcitedefaultseppunct}\relax
\EndOfBibitem
\bibitem[Lu and Law(2006)Lu, and Law]{Lu:2006gi}
Lu,~T.; Law,~C.~K. On the applicability of directed relation graphs to the
  reduction of reaction mechanisms. \emph{Combust. Flame} \textbf{2006},
  \emph{146}, 472--483\relax
\mciteBstWouldAddEndPuncttrue
\mciteSetBstMidEndSepPunct{\mcitedefaultmidpunct}
{\mcitedefaultendpunct}{\mcitedefaultseppunct}\relax
\EndOfBibitem
\bibitem[Lu and Law(2008)Lu, and Law]{Lu:2008bi}
Lu,~T.; Law,~C.~K. Strategies for mechanism reduction for large hydrocarbons:
  \emph{n}-heptane. \emph{Combust. Flame} \textbf{2008}, \emph{154},
  153--163\relax
\mciteBstWouldAddEndPuncttrue
\mciteSetBstMidEndSepPunct{\mcitedefaultmidpunct}
{\mcitedefaultendpunct}{\mcitedefaultseppunct}\relax
\EndOfBibitem
\bibitem[Bendtsen \latin{et~al.}(2001)Bendtsen, Glarborg, and
  Dam-Johansen]{Bendtsen:2001vh}
Bendtsen,~A.; Glarborg,~P.; Dam-Johansen,~K. Visualization methods in analysis
  of detailed chemical kinetics modelling. \emph{Comput. Chem.} \textbf{2001},
  \emph{25}, 161--170\relax
\mciteBstWouldAddEndPuncttrue
\mciteSetBstMidEndSepPunct{\mcitedefaultmidpunct}
{\mcitedefaultendpunct}{\mcitedefaultseppunct}\relax
\EndOfBibitem
\bibitem[Pepiot-Desjardins and Pitsch(2008)Pepiot-Desjardins, and
  Pitsch]{Pepiot-Desjardins:2008}
Pepiot-Desjardins,~P.; Pitsch,~H. An efficient error-propagation-based
  reduction method for large chemical kinetic mechanisms. \emph{Combust. Flame}
  \textbf{2008}, \emph{154}, 67--81\relax
\mciteBstWouldAddEndPuncttrue
\mciteSetBstMidEndSepPunct{\mcitedefaultmidpunct}
{\mcitedefaultendpunct}{\mcitedefaultseppunct}\relax
\EndOfBibitem
\bibitem[Niemeyer \latin{et~al.}(2010)Niemeyer, Sung, and
  Raju]{Niemeyer:2010bt}
Niemeyer,~K.~E.; Sung,~C.~J.; Raju,~M.~P. Skeletal mechanism generation for
  surrogate fuels using directed relation graph with error propagation and
  sensitivity analysis. \emph{Combust. Flame} \textbf{2010}, \emph{157},
  1760--1770\relax
\mciteBstWouldAddEndPuncttrue
\mciteSetBstMidEndSepPunct{\mcitedefaultmidpunct}
{\mcitedefaultendpunct}{\mcitedefaultseppunct}\relax
\EndOfBibitem
\bibitem[Niemeyer and Sung(2014)Niemeyer, and Sung]{Niemeyer:2014}
Niemeyer,~K.~E.; Sung,~C.~J. Mechanism reduction for multicomponent surrogates:
  A case study using toluene reference fuels. \emph{Combust. Flame}
  \textbf{2014}, \emph{161}, 2752--2764\relax
\mciteBstWouldAddEndPuncttrue
\mciteSetBstMidEndSepPunct{\mcitedefaultmidpunct}
{\mcitedefaultendpunct}{\mcitedefaultseppunct}\relax
\EndOfBibitem
\bibitem[Niemeyer and Sung(2015)Niemeyer, and Sung]{Niemeyer:2015wq}
Niemeyer,~K.~E.; Sung,~C.~J. Reduced chemistry for a gasoline surrogate valid
  at engine-relevant conditions. \emph{Energy Fuels} \textbf{2015}, \emph{29},
  1172--1185\relax
\mciteBstWouldAddEndPuncttrue
\mciteSetBstMidEndSepPunct{\mcitedefaultmidpunct}
{\mcitedefaultendpunct}{\mcitedefaultseppunct}\relax
\EndOfBibitem
\bibitem[Sun \latin{et~al.}(2010)Sun, Chen, Gou, and Ju]{Sun:2010jf}
Sun,~W.; Chen,~Z.; Gou,~X.; Ju,~Y. A path flux analysis method for the
  reduction of detailed chemical kinetic mechanisms. \emph{Combust. Flame}
  \textbf{2010}, \emph{157}, 1298--1307\relax
\mciteBstWouldAddEndPuncttrue
\mciteSetBstMidEndSepPunct{\mcitedefaultmidpunct}
{\mcitedefaultendpunct}{\mcitedefaultseppunct}\relax
\EndOfBibitem
\bibitem[Bodenstein(1913)]{Bodenstein:1913tc}
Bodenstein,~M. Eine Theorie der photomechnischen reaktionsgeschwindigkeit.
  \emph{Z. Phys. Chem.} \textbf{1913}, \emph{85}, 329--397\relax
\mciteBstWouldAddEndPuncttrue
\mciteSetBstMidEndSepPunct{\mcitedefaultmidpunct}
{\mcitedefaultendpunct}{\mcitedefaultseppunct}\relax
\EndOfBibitem
\bibitem[Chapman and Underhill(1913)Chapman, and Underhill]{Chapman:1913dx}
Chapman,~D.; Underhill,~L. {LV}.---The interaction of chlorine and hydrogen.
  The influence of mass. \emph{J. Chem. Soc. Trans.} \textbf{1913}, \emph{103},
  496--508\relax
\mciteBstWouldAddEndPuncttrue
\mciteSetBstMidEndSepPunct{\mcitedefaultmidpunct}
{\mcitedefaultendpunct}{\mcitedefaultseppunct}\relax
\EndOfBibitem
\bibitem[Lam and Goussis(1988)Lam, and Goussis]{Lam:1988wc}
Lam,~S.-H.; Goussis,~D.~A. Understanding complex chemical kinetics with
  computational singular perturbation. \emph{Symp. (Int.) Combust.}
  \textbf{1988}, \emph{22}, 931--941\relax
\mciteBstWouldAddEndPuncttrue
\mciteSetBstMidEndSepPunct{\mcitedefaultmidpunct}
{\mcitedefaultendpunct}{\mcitedefaultseppunct}\relax
\EndOfBibitem
\bibitem[Lam(1993)]{Lam:1993ub}
Lam,~S.-H. Using {CSP} to understand complex chemical kinetics. \emph{Combust.
  Sci. Technol.} \textbf{1993}, \emph{89}, 375--404\relax
\mciteBstWouldAddEndPuncttrue
\mciteSetBstMidEndSepPunct{\mcitedefaultmidpunct}
{\mcitedefaultendpunct}{\mcitedefaultseppunct}\relax
\EndOfBibitem
\bibitem[Lam and Goussis(1994)Lam, and Goussis]{Lam:1994ws}
Lam,~S.-H.; Goussis,~D.~A. The {CSP} Method for Simplying Kinetics. \emph{Int.
  J. Chem. Kinet.} \textbf{1994}, \emph{26}, 461--486\relax
\mciteBstWouldAddEndPuncttrue
\mciteSetBstMidEndSepPunct{\mcitedefaultmidpunct}
{\mcitedefaultendpunct}{\mcitedefaultseppunct}\relax
\EndOfBibitem
\bibitem[Lu \latin{et~al.}(2001)Lu, Ju, and Law]{Lu:2001ve}
Lu,~T.; Ju,~Y.; Law,~C.~K. Complex {CSP} for chemistry reduction and analysis.
  \emph{Combust. Flame} \textbf{2001}, \emph{126}, 1445--1455\relax
\mciteBstWouldAddEndPuncttrue
\mciteSetBstMidEndSepPunct{\mcitedefaultmidpunct}
{\mcitedefaultendpunct}{\mcitedefaultseppunct}\relax
\EndOfBibitem
\bibitem[Niemeyer and Sung(2011)Niemeyer, and Sung]{Niemeyer:2011fe}
Niemeyer,~K.~E.; Sung,~C.~J. On the importance of graph search algorithms for
  {DRGEP}-based mechanism reduction methods. \emph{Combust. Flame}
  \textbf{2011}, \emph{158}, 1439--1443\relax
\mciteBstWouldAddEndPuncttrue
\mciteSetBstMidEndSepPunct{\mcitedefaultmidpunct}
{\mcitedefaultendpunct}{\mcitedefaultseppunct}\relax
\EndOfBibitem
\bibitem[Niemeyer(2016)]{MARS:2.3}
Niemeyer,~K.~E. MARS v2.3.0.
  \url{https://niemeyer-research-group.github.io/MARS/}, 2016\relax
\mciteBstWouldAddEndPuncttrue
\mciteSetBstMidEndSepPunct{\mcitedefaultmidpunct}
{\mcitedefaultendpunct}{\mcitedefaultseppunct}\relax
\EndOfBibitem
\bibitem[Lindemann \latin{et~al.}(1922)Lindemann, Arrhenius, Langmuir, Dhar,
  Perrin, and Lewis]{lindemann1922discussion}
Lindemann,~F.; Arrhenius,~S.; Langmuir,~I.; Dhar,~N.; Perrin,~J.; Lewis,~W.~M.
  Discussion on ``the radiation theory of chemical action''. \emph{Trans.
  Faraday Soc.} \textbf{1922}, \emph{17}, 598--606\relax
\mciteBstWouldAddEndPuncttrue
\mciteSetBstMidEndSepPunct{\mcitedefaultmidpunct}
{\mcitedefaultendpunct}{\mcitedefaultseppunct}\relax
\EndOfBibitem
\bibitem[Gilbert \latin{et~al.}(1983)Gilbert, Luther, and
  Troe]{gilbert1983theory}
Gilbert,~R.; Luther,~K.; Troe,~J. Theory of thermal unimolecular reactions in
  the fall-off range. II. Weak Collision Rate Constants. \emph{Ber. Bunsen Ges.
  Phys. Chem.} \textbf{1983}, \emph{87}, 169--177\relax
\mciteBstWouldAddEndPuncttrue
\mciteSetBstMidEndSepPunct{\mcitedefaultmidpunct}
{\mcitedefaultendpunct}{\mcitedefaultseppunct}\relax
\EndOfBibitem
\bibitem[Stewart \latin{et~al.}(1989)Stewart, Larson, and
  Golden]{stewart1989pressure}
Stewart,~P.; Larson,~C.; Golden,~D. Pressure and temperature dependence of
  reactions proceeding via a bound complex. 2. Application to 2{CH}$_3$
  $\rightarrow$ {C}$_2${H}$_5$ + {H}. \emph{Combust. Flame} \textbf{1989},
  \emph{75}, 25--31\relax
\mciteBstWouldAddEndPuncttrue
\mciteSetBstMidEndSepPunct{\mcitedefaultmidpunct}
{\mcitedefaultendpunct}{\mcitedefaultseppunct}\relax
\EndOfBibitem
\bibitem[{Reaction Design: San Diego}(2012)]{chemkin:2012}
{Reaction Design: San Diego}, {CHEMKIN-PRO} 15113. 2012\relax
\mciteBstWouldAddEndPuncttrue
\mciteSetBstMidEndSepPunct{\mcitedefaultmidpunct}
{\mcitedefaultendpunct}{\mcitedefaultseppunct}\relax
\EndOfBibitem
\bibitem[Goodwin \latin{et~al.}(2016)Goodwin, Moffat, and Speth]{cantera}
Goodwin,~D.~G.; Moffat,~H.~K.; Speth,~R.~L. {Cantera}: An object-oriented
  software toolkit for chemical kinetics, Thermodynamics, and Transport
  Processes. \url{http://www.cantera.org}, 2016; Version 2.2.1\relax
\mciteBstWouldAddEndPuncttrue
\mciteSetBstMidEndSepPunct{\mcitedefaultmidpunct}
{\mcitedefaultendpunct}{\mcitedefaultseppunct}\relax
\EndOfBibitem
\bibitem[Venkatesh \latin{et~al.}(1997)Venkatesh, Chang, Dean, Cohen, and
  Carr]{venkatesh1997parameterization}
Venkatesh,~P.~K.; Chang,~A.~Y.; Dean,~A.~M.; Cohen,~M.~H.; Carr,~R.~W.
  Parameterization of pressure-and temperature-dependent kinetics in multiple
  well reactions. \emph{AIChE J.} \textbf{1997}, \emph{43}, 1331--1340\relax
\mciteBstWouldAddEndPuncttrue
\mciteSetBstMidEndSepPunct{\mcitedefaultmidpunct}
{\mcitedefaultendpunct}{\mcitedefaultseppunct}\relax
\EndOfBibitem
\bibitem[Venkatesh(2000)]{Venkatesh:2000gj}
Venkatesh,~P.~K. Damped pseudospectral functional forms of the falloff behavior
  of unimolecular reactions. \emph{J. Phys. Chem. A} \textbf{2000}, \emph{104},
  280--287\relax
\mciteBstWouldAddEndPuncttrue
\mciteSetBstMidEndSepPunct{\mcitedefaultmidpunct}
{\mcitedefaultendpunct}{\mcitedefaultseppunct}\relax
\EndOfBibitem
\bibitem[Mehl \latin{et~al.}(2011)Mehl, Pitz, Westbrook, and
  Curran]{Mehl:2011cn}
Mehl,~M.; Pitz,~W.~J.; Westbrook,~C.~K.; Curran,~H.~J. Kinetic modeling of
  gasoline surrogate components and mixtures under engine conditions.
  \emph{Proc. Combust. Inst.} \textbf{2011}, \emph{33}, 193--200\relax
\mciteBstWouldAddEndPuncttrue
\mciteSetBstMidEndSepPunct{\mcitedefaultmidpunct}
{\mcitedefaultendpunct}{\mcitedefaultseppunct}\relax
\EndOfBibitem
\bibitem[Mehl \latin{et~al.}(2011)Mehl, Chen, Pitz, Sarathy, and
  Westbrook]{Mehl:2011jn}
Mehl,~M.; Chen,~J.-Y.; Pitz,~W.~J.; Sarathy,~S.~M.; Westbrook,~C.~K. An
  approach for formulating surrogates for gasoline with application toward a
  reduced surrogate mechanism for {CFD} engine modeling. \emph{Energy Fuels}
  \textbf{2011}, \emph{25}, 5215--5223\relax
\mciteBstWouldAddEndPuncttrue
\mciteSetBstMidEndSepPunct{\mcitedefaultmidpunct}
{\mcitedefaultendpunct}{\mcitedefaultseppunct}\relax
\EndOfBibitem
\bibitem[Yanowitz \latin{et~al.}(2011)Yanowitz, Christensen, and
  McCormick]{yano:2011}
Yanowitz,~J.; Christensen,~E.; McCormick,~R.~L. Utilization of renewable
  oxygenates as gasoline blending components. Technical Report
  NREL/TP-5400-50791, 2011; \url{https://dx.doi.org/10.2172/1024518}\relax
\mciteBstWouldAddEndPuncttrue
\mciteSetBstMidEndSepPunct{\mcitedefaultmidpunct}
{\mcitedefaultendpunct}{\mcitedefaultseppunct}\relax
\EndOfBibitem
\bibitem[Christensen \latin{et~al.}(2011)Christensen, Yanowitz, Ratcliff, and
  McCormick]{ef2010089}
Christensen,~E.; Yanowitz,~J.; Ratcliff,~M.; McCormick,~R.~L. Renewable
  oxygenate blending effects on gasoline properties. \emph{Energy Fuels}
  \textbf{2011}, \emph{25}, 4723--4733\relax
\mciteBstWouldAddEndPuncttrue
\mciteSetBstMidEndSepPunct{\mcitedefaultmidpunct}
{\mcitedefaultendpunct}{\mcitedefaultseppunct}\relax
\EndOfBibitem
\bibitem[Hui \latin{et~al.}(2016)Hui, Niemeyer, Brady, and Sung]{Hui-data:2016}
Hui,~X.; Niemeyer,~K.~E.; Brady,~K.~B.; Sung,~C.-J. Data and plotting scripts
  for ``Reduced chemistry for butanol isomers at engine-relevant conditions''.
  figshare, 2016; \url{https://dx.doi.org/10.6084/m9.figshare.3505799}\relax
\mciteBstWouldAddEndPuncttrue
\mciteSetBstMidEndSepPunct{\mcitedefaultmidpunct}
{\mcitedefaultendpunct}{\mcitedefaultseppunct}\relax
\EndOfBibitem
\end{mcitethebibliography}

\end{document}